\documentclass[12pt]{article}
\pdfoutput=1

\usepackage{draft}
\usepackage{putex} 
\usepackage[all,cmtip]{xy}
\usepackage{fancybox}
\usepackage{subcaption}
\usepackage{graphicx}
\usepackage{cite}
\usepackage{mciteplus}
\usepackage{skak}
\usepackage{amsmath}
\usepackage{multirow}
\DeclareFontFamily{OT1}{pzc}{}
\DeclareFontShape{OT1}{pzc}{m}{it}{<-> s * [1.10] pzcmi7t}{}
\DeclareMathAlphabet{\mathpzc}{OT1}{pzc}{m}{it}

\def\({\left(}
\def\){\right)}
\newcommand{\eqnDiag}[1]{ \vcenter{\hbox{ #1}} }

\usepackage{tikz-cd}
\makeatletter
\DeclareFontFamily{OMX}{MnSymbolE}{}
\DeclareSymbolFont{MnLargeSymbols}{OMX}{MnSymbolE}{m}{n}
\SetSymbolFont{MnLargeSymbols}{bold}{OMX}{MnSymbolE}{b}{n}
\DeclareFontShape{OMX}{MnSymbolE}{m}{n}{
    <-6>  MnSymbolE5
   <6-7>  MnSymbolE6
   <7-8>  MnSymbolE7
   <8-9>  MnSymbolE8
   <9-10> MnSymbolE9
  <10-12> MnSymbolE10
  <12->   MnSymbolE12
}{}
\DeclareFontShape{OMX}{MnSymbolE}{b}{n}{
    <-6>  MnSymbolE-Bold5
   <6-7>  MnSymbolE-Bold6
   <7-8>  MnSymbolE-Bold7
   <8-9>  MnSymbolE-Bold8
   <9-10> MnSymbolE-Bold9
  <10-12> MnSymbolE-Bold10
  <12->   MnSymbolE-Bold12
}{}

\let\llangle\@undefined
\let\rrangle\@undefined
\DeclareMathDelimiter{\llangle}{\mathopen}%
                     {MnLargeSymbols}{'164}{MnLargeSymbols}{'164}
\DeclareMathDelimiter{\rrangle}{\mathclose}%
                     {MnLargeSymbols}{'171}{MnLargeSymbols}{'171}
\makeatother

\usepackage{latexsym}
\usepackage{stmaryrd}

\usepackage{cite}
\usepackage{framed}
\definecolor{shadecolor}{rgb}{0.95,0.95,0.97}
\definecolor{refkey}{rgb}{0.5,0.5,0}
\definecolor{labelkey}{rgb}{0.5,0.5,0}
\definecolor{citekey}{rgb}{0.5,0.5,0}
\definecolor{darkgreen}{rgb}{0,0.5,0}
\definecolor{darkblue}{cmyk}{0.9,0.9,0,0}
\definecolor{darkred}{rgb}{0.6,0,0.3}
\colorlet{mydarkblue}{blue!50!black}
\colorlet{myred}{red!65!black}
\usepackage{ifpdf}
\ifpdf
\usepackage{graphicx} 
\usepackage{epsfig}
\usepackage[setpagesize=false,pagebackref=false, linktocpage, bookmarksopen=true, colorlinks=true, linkcolor=blue,citecolor=blue,urlcolor=blue]{hyperref}
\usepackage{hyperref}

\else
\usepackage{graphicx}

\fi

\newcommand{\ep}{\epsilon}
\newcommand{\ve}{\varepsilon}

\renewcommand{\Re}{{\rm Re}}

\def\fn#1{\footnote{#1}}
\def\nn{\nonumber}
\def\eqref#1{(\ref{#1})}
\def\comma{\,,}
\def\period{\,.}

\def\bm#1{\text{\boldmath $#1$}}

\def\XXint#1#2#3{{\setbox0=\hbox{$#1{#2#3}{\int}$}
		\vcenter{\hbox{$#2#3$}}\kern-.5\wd0}}

\newcommand{\beq}{\begin{equation}}
\newcommand{\eeq}{\end{equation}}

\def\nullify#1{}

\makeatletter
\def\section{\@startsection {section}{1}{\z@}{-3.5ex plus -1ex minus 
		-.2ex}{2.3ex plus .2ex}{\large\bf}}
\makeatother
\makeatletter
\def\subsection{\@startsection {subsection}{1}{\z@}{-3.5ex plus -1ex minus 
		-.2ex}{2.3ex plus .2ex}{\normalsize\bf}}
\makeatother

\begin{document}
	
	\preprint{CERN-TH-2023-089}
	
	\institution{CERN}{Department of Theoretical Physics, CERN, 1211 Meyrin, Switzerland}
	\institution{NYU}{Center for Cosmology and Particle Physics, New York University, New York, NY 10003, USA}
	
	\title{
	Large Charge 't Hooft Limit of\\ $\mathcal{N}=4$ Super-Yang-Mills
	}

	\authors{Jo\~ao Caetano\worksat{\CERN}, Shota Komatsu\worksat{\CERN}    and Yifan Wang\worksat{\NYU}}

		\abstract
	{
	The planar integrability of $\mathcal{N}=4$ super-Yang-Mills (SYM) is the cornerstone for numerous exact observables. We show that the large charge sector of the ${\rm SU}(2)$ $\mathcal{N}=4$ SYM provides another interesting solvable corner which exhibits striking similarities despite being far from the planar limit. We study non-BPS operators obtained by small deformations of half-BPS operators with $R$-charge $J$ in the limit $J\to\infty$ with $\lambda_{J}\equiv g_{\rm YM}^2 J/2$ fixed. The dynamics in this {\it large charge 't Hooft limit} is constrained by a centrally-extended $\mathfrak{psu}(2|2)^2$ symmetry that played a crucial role for the planar integrability. To the leading order in $1/J$, the spectrum is fully fixed by this symmetry, manifesting the magnon dispersion relation familiar from the planar limit, while it is constrained up to a few constants at the next order. We also determine the structure constant of two large charge operators and the Konishi operator, revealing a rich structure interpolating between the perturbative series at weak coupling and the worldline instantons at strong coupling. In addition we compute heavy-heavy-light-light (HHLL) four-point functions of half-BPS operators in terms of resummed conformal integrals and recast them into an integral form reminiscent of the hexagon formalism in the planar limit. For general ${\rm SU}(N)$ gauge groups, we study integrated HHLL correlators by supersymmetric localization and identify a dual matrix model of size $J/2$ that reproduces our large charge result at $N=2$. Finally we discuss a relation to the physics on the Coulomb branch and explain how the dilaton Ward identity emerges from a limit of the conformal block expansion. We comment on generalizations including the large spin 't Hooft limit, the combined large $N$-large $J$ limits, and applications to general $\mathcal{N}=2$ superconformal field theories.
}
	\date{}

	\maketitle
	
	\tableofcontents
	\newpage
	\section{Introduction}
	\subsection{Introduction and outline of the paper}
	A system with a large number of degrees of freedom often exhibits emergent properties which are difficult to deduce from its Lagrangian. Among multiple ways of introducing a large number of degrees of freedom to a given system, the two approaches often discussed in the literature are
	\begin{enumerate}
	\item Consider a family of theories parametrized by a parameter $N$ which quantifies the number of degrees of freedom, and take the limit $N\to \infty$.
	\item Consider a state in a given theory in which a large number of degrees of freedom are excited, such as a state with a large density of particles.
	\end{enumerate}
	A prominent example of the first is the large $N_c$ limit: gauge theories with a large number of colors $N_c$ admit a double-scaling limit in which $N_c$ is sent to infinity while the 't Hooft coupling $\lambda =g_{\rm YM}^2 N_c$ is kept finite. As pointed out by 't Hooft \cite{tHooft:1973alw}, Feynman diagrams contributing to this limit can be classified by the two-dimensional topology: the leading large $N_c$ answer is given by diagrams that can be drawn on a two-dimensional sphere while the subleading corrections come from diagrams that can be drawn on higher-genus Riemann surfaces. This ``empirically-observed'' connection to two-dimensional surfaces was promoted to a full-fledged duality after the discovery of AdS/CFT correspondence \cite{Maldacena:1997re}, which relates a special class of large $N_c$ gauge theories to string theory in AdS spacetime and provides a physical interpretation of the two-dimensional surfaces that show up in the large $N_c$ expansion.
			\begin{figure}[t]
    \centering 
           \includegraphics[width=0.98\textwidth]{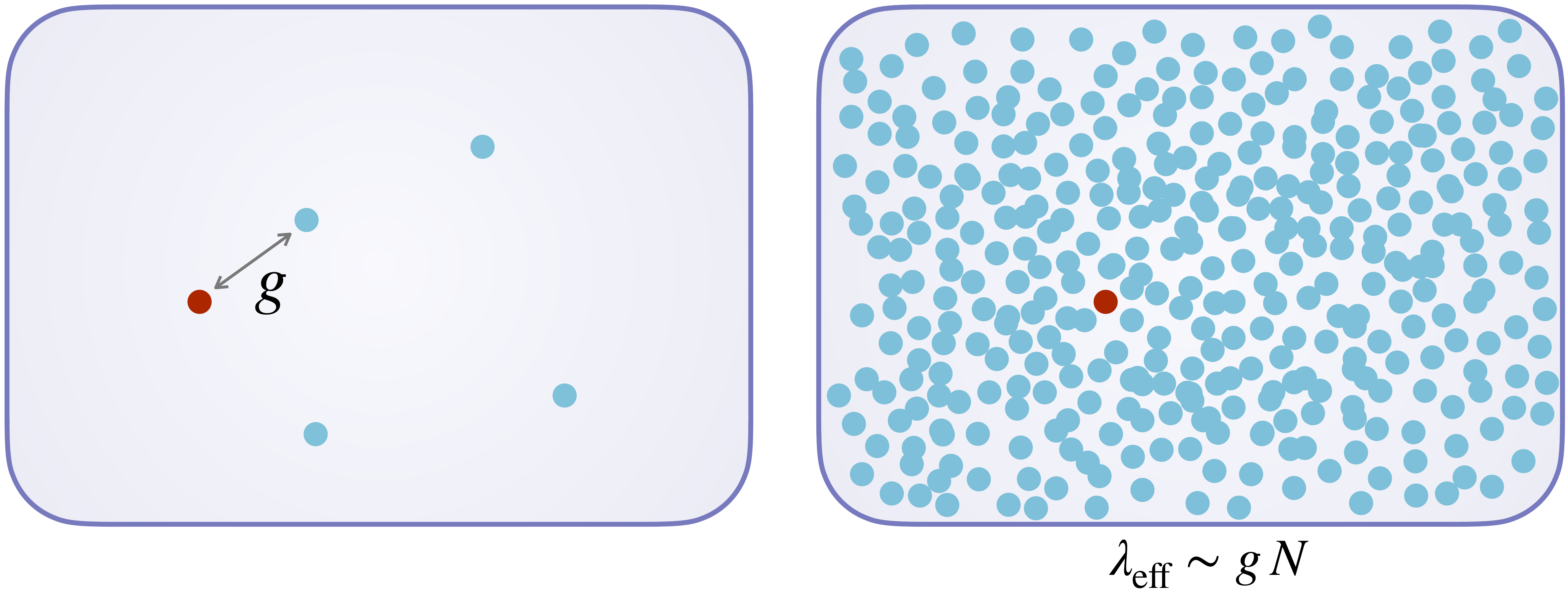} 
\caption{The figure represents a system of $N$ particles interacting with a pairwise coupling constant $g$ for small $N$ (left panel) and large $N$ (right panel). For a small number of particles $N\sim \mathcal{O}(1)$ and weak interaction $g \ll 1$, the system is approximately free and can be studied within the standard perturbation theory. However, when the number of particles $N$ becomes large, the interaction strength is roughly enhanced to $\lambda_{\rm eff}\sim g N$. In the double-scaling limit where $\lambda_{\rm eff}$ is kept fixed, one can replace the two-particle interaction by an effective one-particle potential.}\label{fig:particles}
\end{figure}
	
	On the other hand, examples of the second kind abound in condensed matter and statistical physics, and understanding their emergent properties is one of the central goals in these fields. To appreciate its importance in a simple setup, let us consider a system of particles with two-particle interaction strength $g$. If both the number of particles and the interaction strength are small, the system can be studied by perturbation theory around a free particle system. However, if the system contains a large number of particles $N$, the effective interaction strength gets enhanced to 
	\beq
	\lambda_{\rm eff}\sim g N\,, \label{eq:effectiveg}
	\eeq
	simply because the probability of a given particle to interact with another is proportional to $N$ (see Figure \ref{fig:particles}). Admittedly, the relation \eqref{eq:effectiveg} is based on a rather crude estimate, and in actual physical systems, there can be various other effects to be taken into account. Nonetheless, \eqref{eq:effectiveg} already highlights several important features of a system with a large number of degrees of freedom:
	\begin{itemize}
	\item Even if the fundamental interaction is weakly-coupled ($g<1$), a sector with a large number of degrees of freedom can be driven to a strongly-coupled phase.
	\item It suggests the existence of a double-scaling limit in which $N$ is sent to infinity with $\lambda_{\rm eff}$ fixed. In the simple example discussed above, this is a limit in which the system is well-described by a mean-field approximation and the two-particle interaction can be replaced with an effective one-particle potential.
	\end{itemize} 
	At least formally, this double-scaling limit is reminiscent of the 't Hooft limit of the large $N_c$ gauge theory, and it is tempting to ask whether and how the physics of the two setups exhibits similarities on a more concrete level.
	
	Partial answers to this question were given thanks to vigorous studies in the past years on the large charge sectors of conformal field theories (CFTs) with global symmetry, which were initiated in \cite{Hellerman:2015nra}. In a series of works \cite{Hellerman:2015nra,Monin:2016jmo,Alvarez-Gaume:2016vff} (see also a review \cite{Gaume:2020bmp}),  operators with large global charge $J\gg 1$ in generic (non-supersymmetric) CFTs have been studied using the effective field theory (EFT) techniques, and universal predictions on the conformal dimensions and the structure constants of large-charge operators have been obtained, which were later borne out by the direct large $N$ analysis \cite{Alvarez-Gaume:2019biu,Orlando:2019hte,Giombi:2020enj,Dondi:2021buw,Orlando:2021usz,Moser:2021bes} and by the lattice simulation \cite{Banerjee:2017fcx,Banerjee:2019jpw,Banerjee:2021bbw,Singh:2022akp,Cuomo:2023mxg}. The EFT approach was subsequently generalized to supersymmetric field theories \cite{Hellerman:2017veg,Hellerman:2017sur,Bourget:2018obm,Hellerman:2018xpi,Beccaria:2018xxl,Grassi:2019txd,Hellerman:2020sqj,Sharon:2020mjs,Hellerman:2021yqz,Hellerman:2021duh}, for which the relevant EFT is a low-energy EFT on the Coulomb branch. One of the main outcomes in the application of this EFT to supersymmetric theories is the determination of the asymptotic behavior of two-point functions of chiral and anti-chiral operators (equivalently the extremal correlators) in rank-one $\mathcal{N}=2$ superconformal field theories (SCFTs), which, for theories with a Lagrangian description, reproduces the exact results obtained by supersymmetric localization \cite{Gerchkovitz:2016gxx}. 
	
	The results based on EFTs are valid irrespective of the interaction strength and are applicable to theories without a weak-coupling description. The flip side of this universality is that it is difficult to capture intricate dynamics which are theory-dependent. One approach that could overcome this limitation is to consider CFTs with a weak-coupling parameter $g$ and take a double-scaling limit in which one sends the charge to infinity while keeping a product $g^{\#}J$ fixed, as was first found in \cite{Bourget:2018obm}. The physics in this limit --- which we call the {\it large charge 't Hooft limit} in this paper --- is different from the standard large charge limit reviewed above and exhibits certain similarities with the large $N_c$ 't Hooft limit. Much like the large $N_c$ 't Hooft limit, this double-scaling limit selects a certain class of Feynman diagrams and the observables in this limit can be computed by a resummation of such diagrams \cite{Arias-Tamargo:2019xld,Arias-Tamargo:2019kfr}. Alternatively, the limit can be studied by the semiclassical analysis as was demonstrated in \cite{Badel:2019oxl,Watanabe:2019pdh}. This parallels the fact that the large $N_c$ limit of the gauge theory corresponds to a classical limit of the holographic dual. In fact, the analogy goes even further: it was shown in \cite{Giombi:2018qox,Giombi:2018hsx,Grassi:2019txd,Giombi:2020amn,Beccaria:2020azj,Giombi:2021zfb,Giombi:2022anm} that various results obtained from supersymmetric localization in the large charge limit can be recast into an ``emergent'' matrix model of size of order $J$, for which the large charge 't Hooft limit corresponds to the standard 't Hooft limit.\footnote{Such an ``emergent'' matrix model was obtained first for the generalized cusp anomalous dimension of the BPS Wilson loop in $\mathcal{N}=4$ SYM by Gromov and Sever using the integrability approach. See \cite{Gromov:2012eu} and \cite{Giombi:2018qox,Gromov:2013qga,Sizov:2013joa}.} In addition, it was shown in the analysis of the $O(N)$ model \cite{Dondi:2021buw} that the $1/J$ perturbative series in the large charge 't Hooft limit is asymptotic and the coefficients at large order grow {\it double}-factorially,\fn{Namely, the coefficient at $\ell$-loop grows as $(2\ell)!$ unlike in standard perturbation theory in quantum field theory, in which the coefficient grows as $\ell!$.} much like in the $1/N_c$ expansion of the large $N_c$ gauge theory. 
	
	Finally, already back in 2012, Polchinski and Silverstein \cite{Polchinski:2012nh} argued that the large charge limit alone (without taking a conventional large $N$ limit) can lead to a holographic dual description. The most notable example is a system of a few NS5 branes and a large number of fundamental strings, which is known to be dual to type IIB string theory on ${\rm AdS}_3\times {\rm S}^3\times T^4$ \cite{Maldacena:1998bw,Maldacena:2000hw}. When viewed from the fundamental strings, this leads to a conventional large $N$ theory in two dimensions while, when viewed from the NS5 branes, this corresponds to a large charge state in six dimensions. This suggests that the large charge limit and the large $N$ limit are sometimes dual descriptions of the same system.
	
	In this work, we study the large charge 't Hooft limit of the four-dimensional $\mathcal{N}=4$ supersymmetric Yang-Mills theory (SYM) with the gauge group SU$(2)$ and reveal yet another similarity between the large charge 't Hooft limit and the standard 't Hooft limit. In particular, we demonstrate that the observables in the large charge 't Hooft limit are highly constrained by the maximally-centrally-extended $\mf{psu}(2|2)^2$ symmetry (see around \eqref{univcentral}), which played a crucial role in the integrability approach to the standard 't Hooft limit \cite{Beisert:2005tm,Beisert:2006qh}. Specifically, we analyze the spectrum of non-BPS operators obtained by small deformations of the half-BPS operators with large ${\rm U}(1)_R$ charge $J$,\footnote{This is the universal ${\rm U}(1)_R$ symmetry for general $\cN=2$ SCFTs.} in the limit $J$ is sent to infinity while keeping $\lambda_J \equiv g_{\rm YM}^2 J/2$ fixed.\fn{We choose this definition of $\lambda_J$ (rather than $\lambda_J =g_{\rm YM}^2 J$) since the localization result for the integrated heavy-heavy-light-light (HHLL) four-point function can be recast into a matrix integral of size $J/2$ (instead of size $J$). See Section~\ref{subsec:MM} for details.} At the leading order in the $1/J$ expansion, we show that the spectrum is given by a gauge-invariant combination of ``magnons'' satisfying the following ``dispersion relation'':
	\begin{align}
	E_n=\sqrt{n^2+16g^2}\comma\qquad\qquad  \left(g^2\equiv \frac{\lambda_J}{16\pi^2}=\frac{g_{\rm YM}^2J}{32\pi^2}\comma\quad  n\in \mathbb{N}\right)\period
	\label{dispersion}
	\end{align}
	\begin{figure}[t]
    \centering 
           \includegraphics[width=0.56\textwidth]{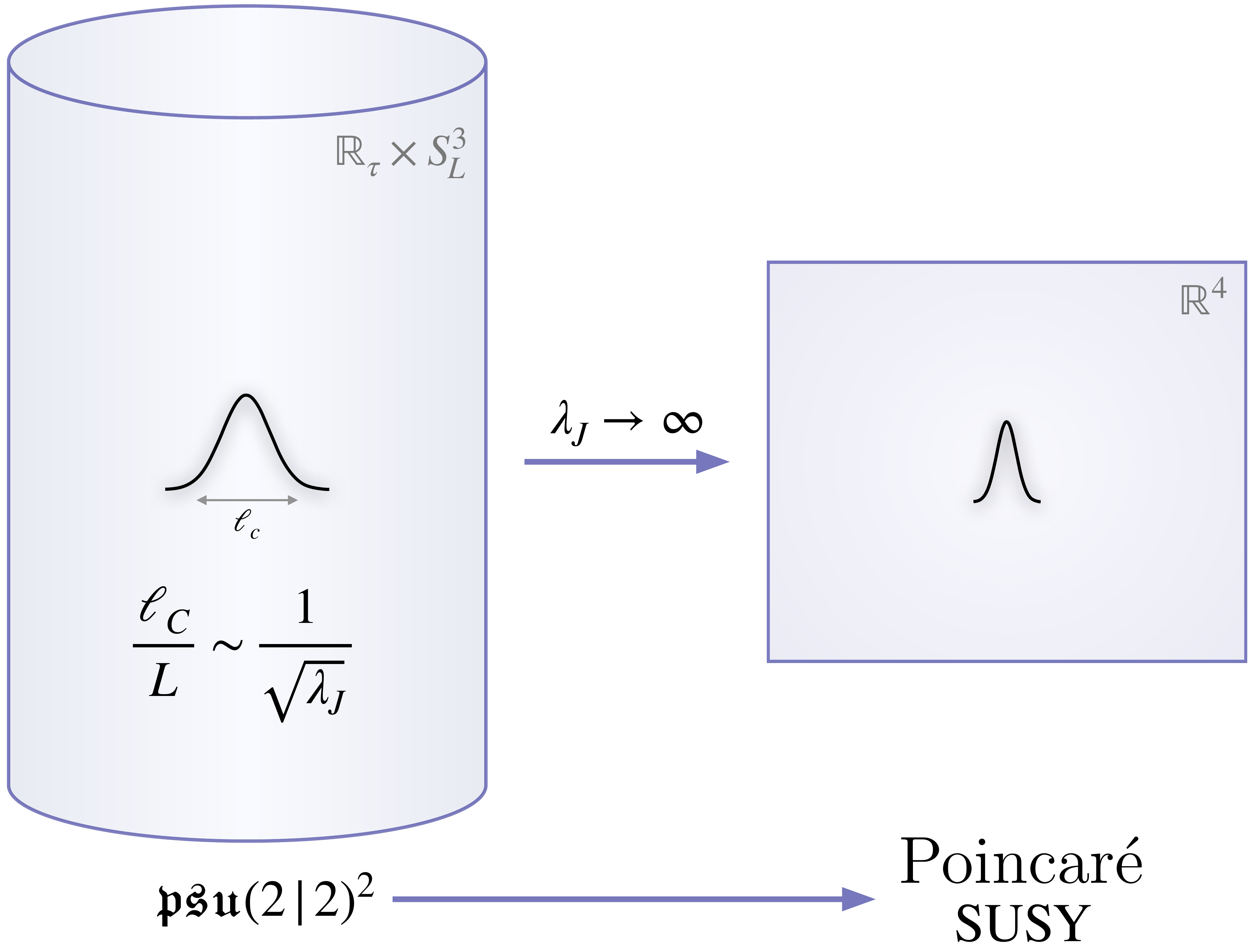} 
\caption{The Compton wavelength  $\ell_c$ of a particle in a short representation of $\mf{psu}(2|2)^2$ scales inversely with the large charge 't Hooft coupling $\sqrt{\lambda_{J}}$ in units of the ${\rm S}^3$ radius $L$. The standard large charge limit (fixed $g_{\rm YM}$  and $J\to\infty$) corresponds to $\lambda_J\to\infty$. In this limit, the sphere flattens out, resulting in a BPS particle protected by the flat space Poincar\'{e} supersymmetry.}\label{poincare} 
\end{figure}
Interestingly, this takes the same form as the dispersion relation of magnons with momentum $p=\pm \pi$ in the large $N_c$ limit \cite{Beisert:2005tm},
	\beq
	E_n(p)=\sqrt{n^2+16 \tilde{g}^2\sin^2 \frac{p}{2}}\comma\qquad \left(\tilde{g}^2\equiv \frac{\lambda}{16\pi^2}=\frac{g_{\rm YM}^2N_c}{16\pi^2}\comma\quad  n\in \mathbb{N}\right)\comma
	\eeq
	where the integer $n$ signifies the $n$-th bound state of fundamental magnons. As we explain in this paper, this is not a coincidence but is a direct consequence of the maximally-centrally-extended $\mf{psu}(2|2)^2$ symmetry present in the large charge 't Hooft limit. We also show that the spectrum at order $1/J$ is constrained by the symmetry up to a few overall constants, some of which can be determined by a straightforward semiclassical analysis around a BPS background with large charge. 
	
	Moreover we establish that in the standard large charge limit (fixed $g_{\rm YM}$ and $J\to\infty$), the maximally-centrally-extended $\mf{psu}(2|2)^2$ symmetry undergoes an Lie algebra contraction, morphing into the centrally-extended Poincar\'{e} supersymmetry. Correspondingly the short representations of the former symmetry become the BPS particle representations of the contracted symmetry. This group-theoretical understanding provides a solid foundation for the relationship between the large charge limit of SCFTs and the dynamics on the Coulomb branch discussed in the literature. For more explanation, see Figure~\ref{poincare} and Section~\ref{subsubsec:Poincare}.
		\begin{figure}[t]
    \centering 
    \begin{minipage}{0.48\textwidth}
        \centering
        \includegraphics[width=0.96\textwidth]{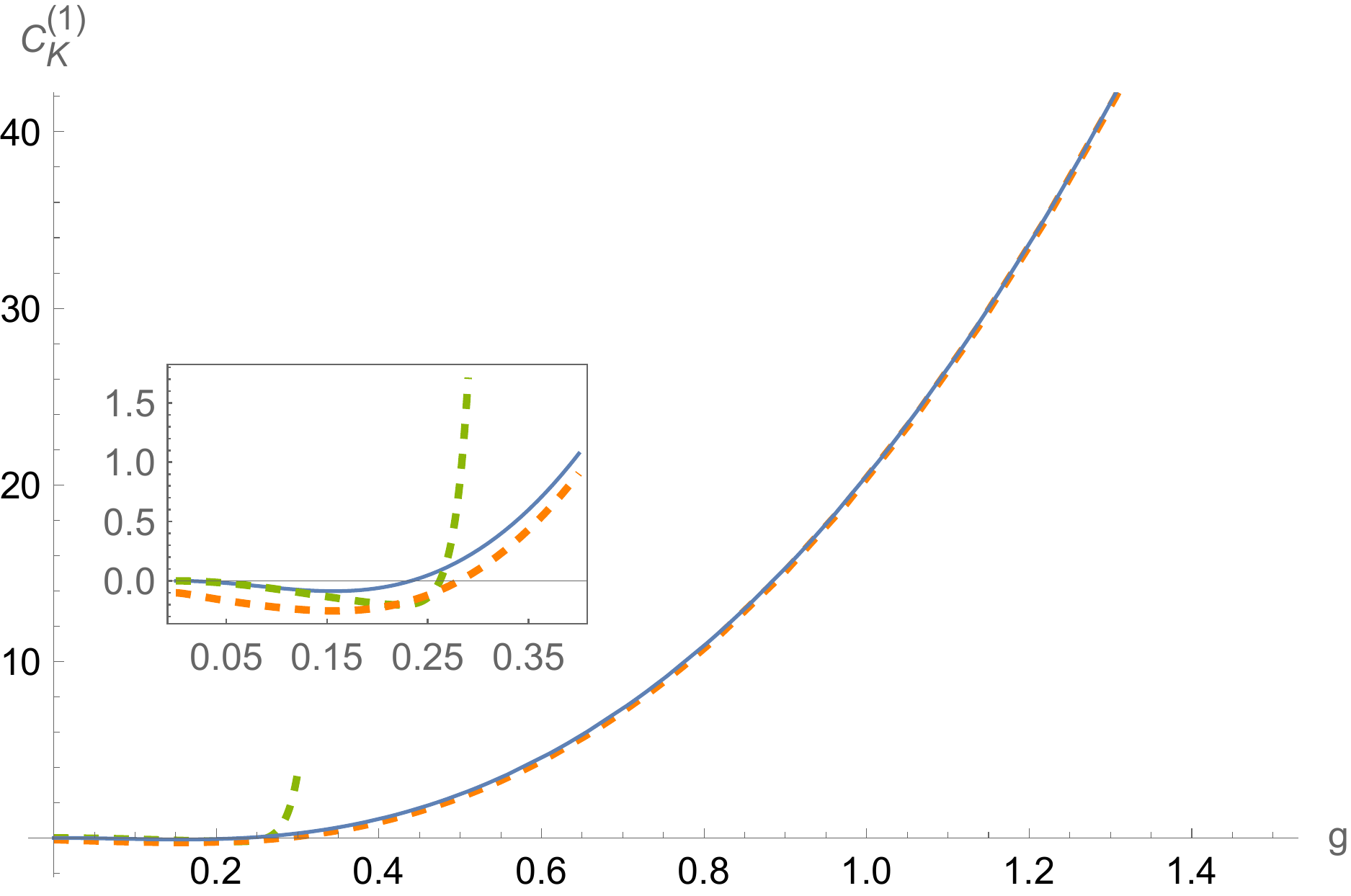} 
    \end{minipage}\hfill
    \begin{minipage}{0.48\textwidth}
        \centering
        \includegraphics[width=0.96\textwidth]{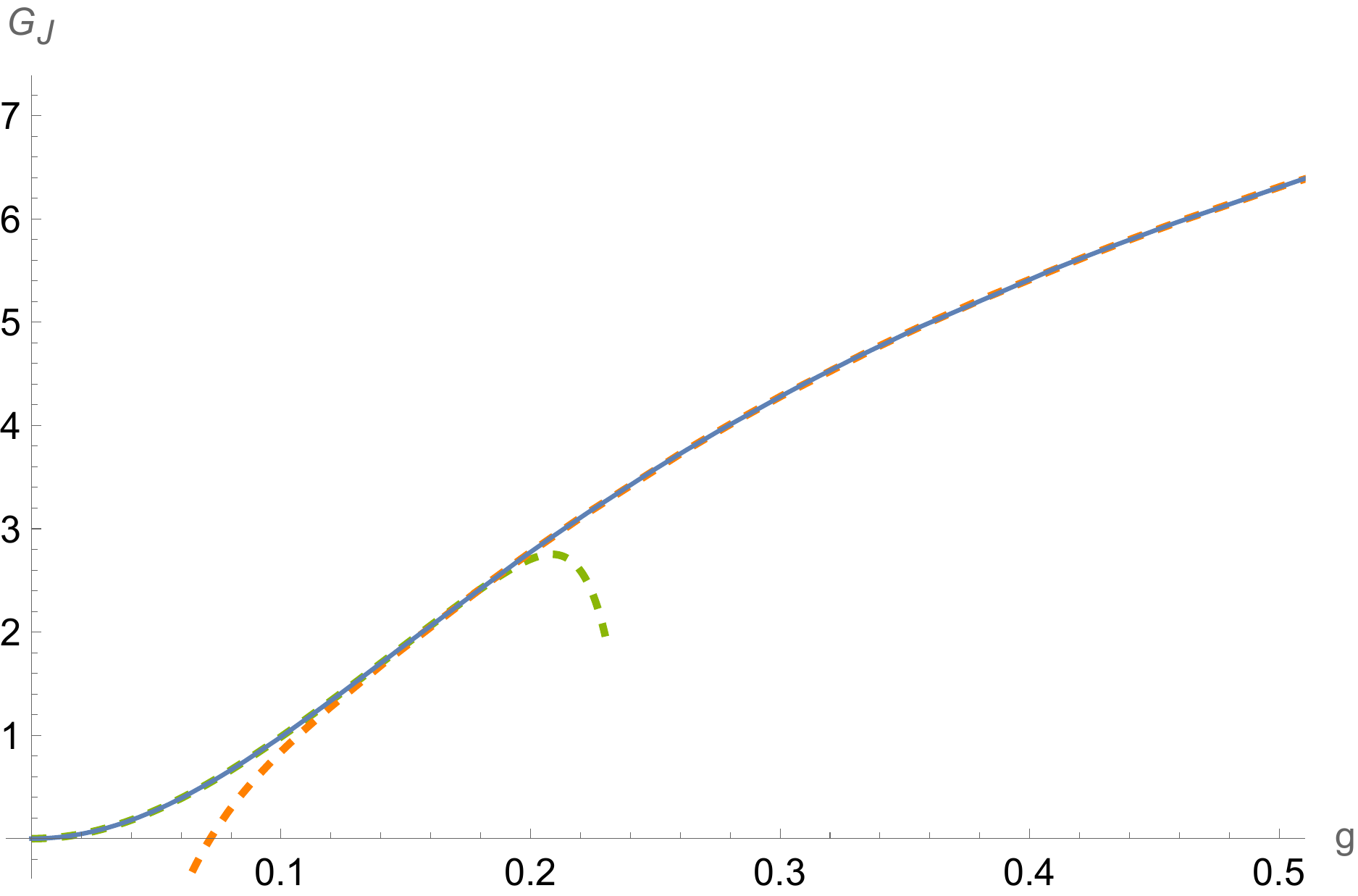} 
    \end{minipage}
 \caption{The three-point coefficient $C_K^{(1)}$ of the Konishi operator and two large-charge BPS operators (on the left) and the integrated  HHLL four-point function $G_J$ (on the right) are depicted as a function of the large charge 't Hooft coupling.
  The blue lines are the exact results. The green and orange dashed lines correspond to the weak and strong coupling  expansions respectively. The weak coupling expansion is truncated to the first ten terms, while the strong coupling expansion incorporates a single worldline instanton contribution. Remarkably, the latter yields a highly accurate result (especially for the three-point coefficient), even at small coupling. The zoomed-in box in the left picture highlights the weak coupling region.
}\label{interpolplots}
\end{figure}

	In addition to determining  the large charge spectrum, we compute the three-point function of the Konishi operator and two large-charge BPS operators, and heavy-heavy-light-light (HHLL) four-point functions of two large-charge BPS operators and two light BPS operators. The three-point function is given by a simple integral involving the Bessel function, 
	\beq
	C_K^{(1)}= -8g^2+4g\int_0^{\infty}dw \frac{4gw-J_{1}(8g w)}{\sinh^2(w)}\comma
	\eeq
	which nevertheless exhibits a rich structure interpolating between the perturbative series at weak coupling ($\lambda_J\ll 1$) and the worldline instantons at strong coupling ($\lambda_J\gg 1$).\footnote{See \eqref{dispersion} for the relation between $\lambda_J$ and $g$.} See equation \eqref{eq:KonishiCnext} for the explicit result and also Figure~\ref{interpolplots} for a summary. The HHLL four-point functions are given in terms of a resummation of conformal ladder integrals  \cite{Broadhurst:2010ds},\footnote{The same resummation of the conformal ladder integrals shows up also in the large charge double-scaling limit of the critical $O(N)$ model \cite{Giombi:2020enj}.} 
	\beq
	t(z,\bar{z})=\sum_{k=0}^{\infty}(-4g^2)^{k}(1-z)(1-\bar{z})F^{(k)}(z,\bar{z})\comma
	\eeq
	where $z$ and $\bar{z}$ are conformal cross ratios and $F^{(k)}$ is a $k$-loop conformal ladder integral  which can be represented as
	  \begin{align}
       & \eqnDiag{\includegraphics[scale=0.2]{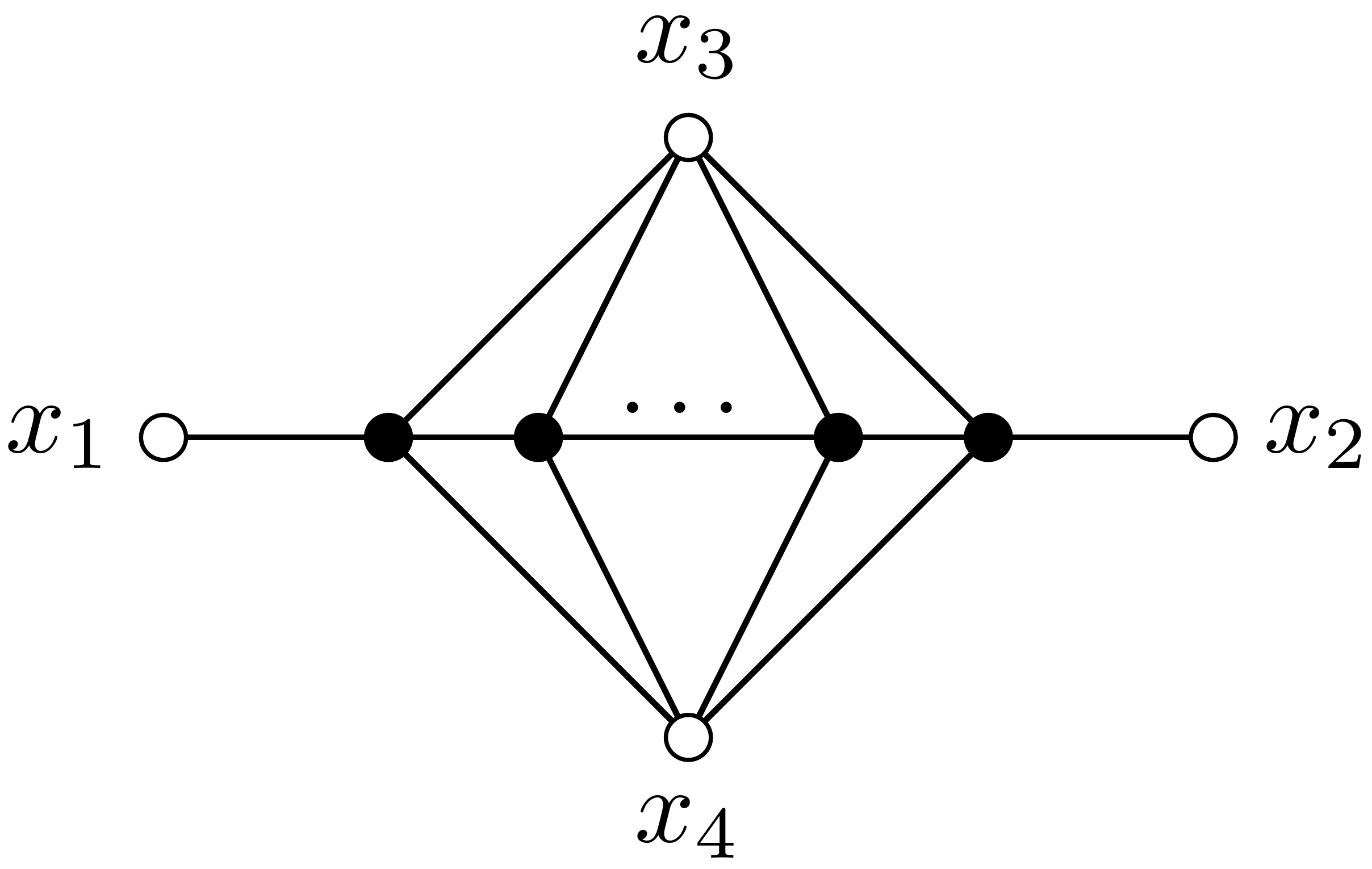}} = \frac{\pi^{2k}}{ x_{12}^2 x_{34}^{2k}}(1-z)(1-\bar{z})F^{(k)}(z,\bar{z})\period \label{eq:ladder}
  \end{align}
Here the black dots are being integrated. The points $x_{3,4}$ are the locations of the large charge operators  and the horizontal line represents the propagation of a scalar field in this background from $x_1$ to $x_2$ (see Section~\ref{sec:higher} for details).

	\begin{figure}[t]
    \centering 
           \includegraphics[width=0.26\textwidth]{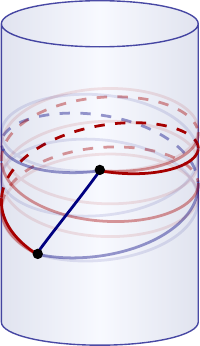} 
\caption{An illustration of the worldline instanton representation (\ref{eq:propinstant}) of the HHLL four-point function from summing massive propagators. The blue and red curves depict the two distinct terms, representing the worldline instanton's wrapping of the cylinder in opposite directions. The varying tones of blue and red correspond to different winding numbers ($n=0,1,2$) for each respective direction.}\label{helix}
\end{figure}
		 Similar but different resummations of conformal ladder integrals show up in the so-called hexagon approach \cite{Basso:2015zoa} to the four-point functions in the planar limit \cite{Fleury:2016ykk,Basso:2017jwq,Coronado:2018ypq,Coronado:2018cxj,Basso:2018cvy}. Taking inspiration from it, we express the resummed conformal ladder integrals as a sum over magnons, 
		 \beq
t(z,\bar{z})=\frac{(1-z)(1-\bar{z})}{\sqrt{z\bar{z}}}\sum_{a=1}^{\infty}\frac{ae^{-\sigma \sqrt{a^2+16g^2}}}{\sqrt{a^2+16g^2}}\frac{\sin (a\varphi)}{\sin(\varphi)}\comma
\eeq
with $e^{-\sigma}=\sqrt{z\bar{z}}$ and $e^{i\varphi}=\sqrt{z/\bar{z}}$,
	 and use this to read off the operator-product-expansion (OPE) data. In addition, we derive a strong coupling expansion of the resummed ladder integrals,
	 \beq
	 \begin{aligned} \label{eq:propinstant}
	& t(z,\bar{z})=\frac{(1-z)(1-\bar{z})}{\sqrt{z\bar{z}}} \,  \sum_{n=0}^{\infty}{\color{mydarkblue}{W(\varphi+2\pi n)}}+{\color{myred}{W(2\pi -\varphi+2\pi n)}} \comma\\
	 &W(x)=\frac{4g xK_1(4g \sqrt{x^2+\sigma^2})}{\sin (x)\sqrt{x^2+\sigma^2}}\period
	 \end{aligned}
	 \eeq
	 This expression has two remarkable features,
	 \begin{enumerate}
	 \item It provides an exact rewriting of the resummed conformal ladder integrals in terms of the worldline instanton contributions $W(x)$ depending on the orientation (blue versus red) and the winding number $n$ of the instanton. See Figure~\ref{helix}.
	 \item The worldline instantons have the same functional form as the massive propagator in flat space (i.e. Bessel $K_1$). This provides a concrete link between the large charge limit in the CFT and the physics on the Coulomb branch.
	 \end{enumerate}
	 See Section~\ref{subsec:resummedladder} for further discussions on these points.
	 	 
	 We also compute the integrated four-point functions for arbitrary ${\rm SU}(N)$ gauge groups by recasting results from supersymmetric localization into an ``emergent'' matrix integral of size $J/2$.  The matrix integrals can be evaluated by saddle-point techniques and the answers are in precise agreement with the results in the literature obtained by other methods \cite{Paul:2023rka,Brown:2023why}. For $N=2$, we derive the same answer from our un-integrated HHLL four-point function by explicitly carrying out the integral over the conformal cross-ratios $z,\bar z$, thus providing a nontrivial consistency check of our formulae.
	 See Figure~\ref{interpolplots} for an illustrative summary.  Note that the large charge 't Hooft limit corresponds to a standard 't Hooft limit of this matrix integral, thereby offering another evidence for the similarity between the two limits.
	
	The rest of this paper is organized as follows. In {\bf Section}~\ref{subsec:generalities}, we explain generalities of the large charge limit and the large charge 't Hooft limit in (S)CFTs and discuss similarities and differences.  In {\bf Section}~\ref{sec:weak}, we compute the non-BPS spectrum in the large charge 't Hooft limit by perturbation theory in $\lambda_J$. For the explicit computation, we use the dilatation operator at weak coupling, which was determined fully at the non-planar level in \cite{Beisert:2004ry}. We find that the resulting spectrum at leading order in the $1/J$ expansion is consistent with an interpretation that the non-BPS operators are composites made of ``magnons'', each carrying a definite energy, subject to parity and gauge singlet constraints. We consolidate this observation by examining the partition function of free $\mathcal{N}=4$ SYM and taking the large charge limit. In {\bf Section}~\ref{sec:leading}, we explain the structure of the large $J$ spectrum from the point of view of the maximally-centrally-extended $\mf{psu}(2|2)^2$ symmetry. We first give a physics derivation of the existence of the centrally-extended symmetry in the large charge 't Hooft limit. We also provide a brief explanation on how this symmetry is related to the centrally-extended Poincar\'e supersymmetry on the Coulomb branch of the $\mathcal{N}=4$ SYM. We then use this symmetry to constrain the spectrum at leading order in the large $J$ limit. This allows us to fix the energy of individual magnons non-perturbatively as a function of $\lambda_J$. We then check the results against the semiclassical analysis around a large-charge BPS background. It is in principle possible to determine the full $\lambda_J$-dependence just from the semi-classics, but as we will see, the actual semiclassical computation is rather complicated: even in the leading large $J$ limit, the computation involves diagonalizing the unconventional kinetic terms since the background is space-time dependent. The centrally-extended symmetry provides a useful guiding principle for organizing the computation and also allows us to promote the results obtained in a scalar subsector to sectors involving fermions and gauge fields. In {\bf Section}~\ref{sec:subleading}, we analyze the $1/J$ correction to the spectrum. For simplicity, we focus on the SU$(2|2)$ subsector and show that the spectrum is constrained by the centrally extended symmetry up to a few overall constants. We then compute some of these constants by a direct semiclassical analysis. In {\bf Section}~\ref{sec:higher}, we analyze a sample of higher-point functions. First we study the three-point function of the Konishi operator and two large-charge BPS operators up to order $1/J$ and obtain an integral representation exact in $\lambda_J$. We then analyze the weak- and strong-coupling limits making contact with the perturbation theory and the worldline instantons. We next study HHLL four-point functions of two large-charge BPS operators and two small-charge BPS operators in the large charge 't Hooft limit. The results are given by a resummed conformal integral which we recast into an ordinary contour integral. Using this representation, we read off the OPE data. We also explore a connection to the standard large charge limit and study a relation to the physics on the Coulomb branch. In particular, we discuss how the form factor expansion of the two-point function in the Coulomb branch arises as a limit of the conformal block expansion in the heavy-light channel. We then derive an ``emergent'' matrix integral which computes the integrated four-point functions. Using the matrix model representation, we analyze the large charge 't Hooft limit and obtain the exact answer in $\lambda_J$.  Finally in {\bf  Section}~\ref{sec:generalization}, we conclude and discuss possible generalizations including extensions to less-supersymmetric large charge states, theories with higher-rank gauge groups, more general three-point functions, the combination of large $N$ and large $J$ limits, and applications to general $\mathcal{N}=2$ SCFTs.
	\subsection{Large charge limit vs.~large charge 't Hooft limit\label{subsec:generalities}} Before delving into the details of the computation, here we review a physical picture of the large charge limit and the large charge 't Hooft limit, and discuss their similarities and differences. Along the way, we also mention several open questions.
	
	\paragraph{Basics of large charge limit.}The simplest way to discuss the large charge limit in conformal field theory in $d$ dimensions is to consider on flat space $\mR^{d}$  a two-point function of operators  that have the minimal conformal dimension $\Delta_{\rm min}(J)$ for a given large charge $J $. We then map it to a cylinder $\mR_\tau \times {\rm S}^{d-1}$, where the radius of the sphere is taken  to be $L$. Under this mapping, the large charge operators are mapped to states with the following energy and charge:
	\beq \label{enconf}
	E_{\rm state}=\frac{\Delta_{\rm min}(J)}{L}\comma\qquad J_{\rm state}=J\period
	\eeq
	The factor $1/L$ in the expression for the energy comes simply from the dimensional analysis. These relations translate to the energy and charge densities ($\epsilon_{\rm state}$ and $j_{\rm state}$) of the states as
	\beq
	\epsilon_{\rm state}\sim \frac{\Delta_{\rm min}(J)}{L^{d}}\comma\qquad j_{\rm state}\sim \frac{J}{L^{d-1}}\period
	\eeq
	We then take a scaling limit in which we send both $J$ and $L$ to infinity while keeping $j_{\rm state}$ to be finite. As a result, we obtain the lowest energy state in flat space with a given charge density $j_{\rm state}$. The physics of such states depends on the behavior of a theory in flat space:
	\begin{enumerate}
	\item In generic (non-supersymmetric) conformal field theories, we expect that a state with a non-zero charge density in flat space is a highly excited state with an $O(1)$ energy density. By requiring $\epsilon_{\rm state}\sim j_{\rm state}\sim O(1)$, we arrive at the famous scaling relation of the large charge operator 
	\beq\label{eq:largechargerel}
	\Delta_{\rm min}(J)\sim J^{\frac{d}{d-1}}\period
	\eeq
	\item In theories with a moduli space of vacua, the lowest energy state with a charge density $j$ not necessarily zero  in flat space has exactly zero energy $\epsilon=0$. In this case, the scaling relation \eqref{eq:largechargerel} is violated since $\epsilon_{\rm state}$ asymptotes to zero in the large $L$ limit while $j_{\rm state}$ can approach a finite nonzero value. This is in particular the case for large classes of superconformal field theories, for which there exist the BPS operators  satisfying a linear relation between the dimension and the charge, $\Delta_{\rm min}(J) \sim \# J$.
	\end{enumerate}
	Before proceeding, let us make two side comments regarding the second scenario. First, in order to have a state with $j\neq 0$ and $\epsilon=0$ in flat space, it is enough to have an operator whose dimension scales as $\Delta_{\rm min}(J)\sim J^{\alpha}$ with $\alpha$ strictly smaller than $\frac{d}{d-1}$. However in all the examples known in the literature, the relation between the minimal dimension and the charge is linear, suggesting that this might be a universal feature of CFTs with a vacuum manifold. In fact, there is an interesting result on condensed matter systems \cite{Tasaki:2021jmi}, which proved that the off-diagonal long range order---a hallmark of the spontaneous symmetry breaking present in the moduli space of vacua---implies a linear energy-charge relation. It would be interesting to try to adapt the proof to CFT. Second, the paper \cite{Aharony:2021mpc} proposed\footnote{More recently, counter-examples were found in \cite{Sharon:2023drx} for the original formulation of the conjecture, but even in those counter-examples, the convexity still holds in the large charge limit.} that the weak gravity conjecture in AdS implies a convexity of conformal dimensions of charged operators in CFTs.  When applied to the large charge sector, this means that the exponent $\alpha$ in the dimension-charge relation must satisfy $\alpha\geq 1$. This is indeed satisfied in all the known examples. Thus, combining the two statements, we are led to the following conjecture:
	\begin{enumerate}
	\item[] {\bf Conjecture:}\fn{{\bf 1.}~The upper bound $\alpha\leq \frac{d}{d-1}$ comes from requiring that the finite charge density state in flat space has finite energy density, which is a physically well-motivated assumption. {\bf 2.}~The results in the literature are consistent with a stronger version of the conjecture, which states that $\alpha$ must be either $\frac{d}{d-1}$ or $1$.} In any CFTs in $d>2$, the dimension-charge relation $\Delta_{\rm min}(J)\sim J^{\alpha}$ satisfies $1\leq \alpha \leq \frac{d}{d-1}$. The lower bound is saturated ($\alpha=1$), if and only if the CFT comes with a moduli space of vacua.
	\end{enumerate} 
It is an interesting open problem to establish this claim using non-perturbative techniques\footnote{In a recent paper \cite{Orlando:2023ljh}, it was shown that the convexity $(\alpha\geq 1)$ follows from the consistency of the large charge EFT under the assumption that the large charge limit is described by the EFT that consists of a single Goldstone boson. It would be interesting to prove the statement without making such assumptions (i.e.~purely based on the conformal bootstrap).} such as the conformal bootstrap.
\paragraph{Large charge limit of SCFT and large charge 't Hooft limit.} Let us now take a closer look into the large charge sector of SCFTs. To be concrete, we consider $\mathcal{N}=2$ SCFTs in four dimensions and discuss Coulomb branch operators\fn{They are half-BPS  scalar primary operators and elements of the Coulomb branch chiral ring whose vacuum expectation values are (anti)holomorphic  functions on the Coulomb branch of the vacuum manifold. The ``rank'' of an $\cN=2$ SCFT refers to the complex dimension of its Coulomb branch.} with large ${\rm U}(1)_R$ charge $J$ and scaling dimension $\Delta=|J|$. Roughly speaking, the insertions of these operators effectively introduce (nonlinear) source terms for the scalar fields in the vector multiplets, and the path integral will be dominated by a configuration with a nontrivial profile for these scalar fields.\footnote{This is most obvious for Lagrangian SCFTs (e.g. $\cN=2$ superconformal QCD). For non-Lagrangian SCFTs (e.g. the Argyres-Douglas SCFT \cite{Argyres:1995jj}), these vector multiplets come from the Coulomb branch effective action (see \cite{Hellerman:2017sur}).}
This is physically similar to considering theories on the Coulomb branch in flat space, which is described by an effective Lagrangian of axion-dilaton (and also other massless scalars if ${\rm rank}>1$) together with super-partners at low energy. As demonstrated in \cite{Hellerman:2017veg,Hellerman:2017sur} for the rank-1 SCFTs, this intuitive picture can be turned into a computational framework with the help of the EFT techniques. Namely, by writing down all the interactions allowed by the symmetry and organizing them according to the derivative expansion, one can compute observables in the large charge limit as a function of the coefficients in the effective Lagrangian. In particular, it turns out that some of the observables depend only on universal properties of the theories such as the conformal anomalies. Thus one can make a solid prediction from EFT and test it against the exact results from supersymmetric localization. 

In theories without exactly marginal parameters such as the Argyres-Douglas theories \cite{Argyres:1995jj,Argyres:1995xn}, this EFT approach is perhaps the best one can do. In contrast, in theories with tunable couplings (to be denoted by $g_{\rm YM}$), one can better understand inner workings of the large charge limit. In these theories, the large charge operators give a vacuum expectation value to scalar fields in the vector multiplet of order $\langle \phi \rangle \sim g_{\rm YM}\sqrt{J}$. As a result, BPS W-bosons acquire mass of order $m_{\rm W}\sim g_{\rm YM}\sqrt{J}$ while BPS monopoles acquire mass of order $m_{\rm monopole}\sim \sqrt{J}/g_{\rm YM}$. In the standard large charge limit in which $J$ is sent to infinity with $g_{\rm YM}$ fixed, both particles become infinitely massive in units of the radius of ${\rm S}^3$. By integrating them out, we generate higher-derivative contact interactions of the axion-dilaton fields that are suppressed by inverse powers of $J$. This is quite analogous to what happens when we integrate out a propagator of a heavy particle in flat space:
\beq
\frac{1}{p^2-M^2} =- \frac{1}{M^2}-\frac{p^2}{M^4} +\cdots\period 
\eeq
In theories with a Lagrangian description, it should be possible to perform this integrating-out procedure explicitly  and derive the EFT Lagrangian from first principles although it has not been demonstrated in the literature to the best of our knowledge.

To better understand the physics of these heavy particles, it is instructive to take a slightly different limit in which $J$ is sent to infinity while the {\it large charge 't Hooft coupling} $\lambda_J\equiv g_{\rm YM}^2J/2$ is kept finite. The physics in this {\it large charge 't Hooft limit} differs from that of the standard large charge limit since the mass of BPS W-bosons $m_{\rm W}\sim g_{\rm YM}\sqrt{J}$ remains finite and they contribute  to observables even in  the $J\to\infty$ limit. Furthermore, the very notion of ``BPS particles'' is modified in this limit (see Figure~\ref{poincare}): in the standard large charge limit, the Compton wavelengths of the particles are much smaller than the size of the sphere, allowing one to classify them by BPS representations of the Poincar\'e supersymmetry in flat space. On the other hand the Compton wavelengths are finite in the large charge 't Hooft limit and the particles ``feel'' the curvature of ${\rm S}^3$. Thus, a priori, the classification based on the BPS representations in flat space cannot be justified. One of the punchlines of our work is that the correct symmetry algebra governing this limit is the maximally-centrally-extended $\mf{psu}(2|2)^2$ symmetry, which was first introduced by Beisert in the analysis of the large $N$ $\mathcal{N}=4$ SYM \cite{Beisert:2005tm,Beisert:2006qh}. This symmetry turns out to be powerful enough to fully determine the leading large $J$ spectrum as well as constrain the structure of $1/J$ corrections. Now, in order to relate this to the standard large charge limit, one needs to take the ``strong coupling limit'', i.e. $\lambda_J=g_{\rm YM}^2J /2\to \infty$. As we will see, in this further limit this centrally-extended superconformal symmetry contracts to the centrally-extended Poincar\'e supersymmetry and the short representations of former become the BPS representations of the latter, providing a concrete link between the two centrally-extended supersymmetries studied in different contexts.
	\section{Weak Coupling Analysis\label{sec:weak}}
	In this section, we analyze the spectrum in the large charge 't Hooft limit at weak coupling (in $\lambda_J$) by directly diagonalizing the dilatation operator $\hat D$ in the limit. The result suggests that non-BPS operators in this limit are made out of fundamental excitations with definite energy which we call ``magnons'', subject to certain constraints. We focus on simple closed subsectors of the theory, namely the so-called SU(2), SU(3) and SL(2) sectors. As we will see below, despite their simplicity, these sectors exhibit crucial features of the spectrum in the large charge 't Hooft limit such as the relation between operators and magnons, the parity constraint, and magnon bound states. 
	 We also provide evidence for this interpretation by analyzing the large-charge limit of the partition function of free $\mathcal{N}=4$ SYM. 
	 
	 We denote the three complex adjoint scalar fields of the ${\rm SU}(2)$ SYM by $X,Y,Z$ and choose the ${\rm U}(1)_R$ generator such that $X,Y$ has R-charge $J=0$ and $Z$ has R-charge $J=1$. In the following, we will consider gauge invariant  operators built out of these scalar fields and their derivatives. 
	  
	\subsection{SU(2) sector: operators and magnons}
	We first consider the so-called SU(2) sector; namely gauge invariant operators which consist only of two complex scalar fields $Z$ and $X$. As shown\fn{This property is discussed mostly in the planar limit but it also holds in theories at finite $N_c$.} in \cite{Beisert:2003tq}, they form a closed subsector under the action of the perturbative dilatation operator. 
	The full dilatation operator $\hat D$ at one and two loops in this sector is given by\fn{See \cite{Beisert:2003tq}. A more explicit expression can be found in \cite{Kristjansen:2010kg}. See also the thesis by Beisert \cite{Beisert:2004ry}, in which various aspects of perturbative dilatation operators are discussed.}
\beq\label{eq:def1loop}
\begin{aligned}
\hat D_{1}=&-\frac{g_{\rm YM}^2}{8\pi^2}:{\rm Tr}[X,Z][\check{X},\check{Z}]:\comma \qquad \check{Z}_{ab}=\frac{d}{d Z_{ba}} 
\\
\hat D_{2} =& -\frac{1}{2}\left(\frac{g_{\rm YM}^2}{8\pi^2}\right)^2\bigl( :{\rm Tr}\left[[X,Z],\check{Z} \right]\left[ [\check{X},\check{Z}],Z\right]:
\\
+&:{\rm Tr}\left[[X,Z],\check{X} \right]\left[ [\check{X},\check{Z}],X\right]:
\\
+&:{\rm Tr}\left[[X,Z],T^{a} \right]\left[ [\check{X},\check{Z}],T^{a}\right]: \bigr)\comma
\end{aligned}
\eeq
where $T^{a}$ are the SU(2) generators.
Here the normal ordering symbol $:\bullet :$ means that operators $\check{Z}$ and $\check{X}$ do not act on fields inside $\hat D_1$ and $\hat D_2$, \eqref{eq:def1loop}. 

In the large $N_c$ limit, single-trace operators provide basic building blocks of gauge-invariant operators and the action of the dilatation operators on them was found to be isomorphic to Hamiltonians of an integrable spin chain \cite{Minahan:2002ve}. In this paper, we are interested in the opposite limit, namely in $\mathcal{N}=4$ SYM with the SU(2) gauge group. In such theories, most of single-trace operators can be rewritten as multi-trace operators of shorter lengths using the trace relations of SU(2),
\beq
\begin{aligned} \label{tracerels}
&Z^2=\frac{{\rm Tr}Z Z}{2}{\bf 1}_{2}\comma\qquad X^2=\frac{{\rm Tr}XX}{2}{\bf 1}_{2}\comma\qquad {\rm Tr}X={\rm Tr}Z=0\comma\\
&\Tr\, Z X Z X =-\frac{1}{2} \Tr\, Z Z \,\Tr\, XX+\Tr\, Z X  \, \Tr\, Z X\period
\end{aligned}
\eeq
(They can be verified by explicitly writing down components of the matrices.) Using these, one can express all gauge-invariant operators as products of the following single traces:
\beq
\Tr\,ZZ \,, \quad \Tr\,ZX\,, \quad \Tr\,XX\,.
\eeq

To study the large-charge sector using this basis, we consider a background of a large multi-trace of the form $(\Tr \,ZZ)^{J/2}$ with excitations consisting of any of the remaining two letters. This provides a complete basis for all excited states in this sector.
To simplify the discussion we will assign names to these letters,
\beq \label{particles}
\mathfrak{a} := \Tr \,ZZ \quad  \mathfrak{b}:=  \Tr\,ZX \quad  \mathfrak{c}:= \Tr\,XX
\eeq
We will be dealing with operators of the form
\beq \label{eq:basis}
| \mathfrak{a}^{\frac{J-n_{\mathfrak{b}}}{2}} \, \mathfrak{b}^{n_{\mathfrak{b}}}\, \mathfrak{c}^{n_{\mathfrak{c}}} \rangle\,.
\eeq
Here we shifted the exponent of $\mathfrak{a}$ so that the total $R$-charge in the $Z$-direction is $J$. We then consider the limit where $J \gg n_{\mathfrak{b},\mathfrak{c}}\sim O(1)$. The goal now is to determine the action of the Hamiltonian for the operators in this basis.
\paragraph{Simplification at large $J$.} Before delving into the actual computation, let us discuss briefly how the action of the dilatation operator simplifies in the large $J$ limit. The one-loop dilatation operator $D_1$ contains two derivatives $\check{X}$ and $\check{Z}$ (see \eqref{eq:def1loop}). The derivative $\check{X}$ always acts on the excitations, namely $\mathfrak{b}$ or $\mathfrak{c}$, while $\check{Z}$  can act either on the background $\mathfrak{a}^{(J-n_{\mathfrak{a}})/2}$ or the $\mathfrak{b}$-excitations. In the large charge 't Hooft limit, the leading contribution comes from the action of $\check{Z}$ on the background
\beq
\check{Z}_{ab}\mathfrak{a}^{\frac{J-n_{\mathfrak{b}}}{2}}\overset{J\gg 1}{\sim}J \, \mathfrak{a}^{\frac{J-n_{\mathfrak{b}}}{2}-1} Z_{ab}\comma
\eeq
which gives an $O(1)$ answer after multiplied with the coupling constant $g_{\rm YM}^2=2\lambda_J/J$. By contrast, if $\hat{Z}$ acts on the excitations, the result will be suppressed by $1/J$. A similar simplification takes place at two loops. Among the three terms in \eqref{eq:def1loop} for $D_2$, only the first term in \eqref{eq:def1loop} survives in the leading large $J$ limit since only that term contains two $\check{Z}$ and can produce $g_{\rm YM}^{4}J^2=4\lambda_{J}^2$. Once again, by acting with any of the two $\check{Z}$ on the excitations one produces terms that are subleading in the large $J$ expansion. We conclude that 
the leading large $J$ action of the dilatation operator on the basis of states (\ref{eq:basis}) can be easily obtained at each order by choosing the terms of the dilatation operator with the maximal number of $\check{Z}$ and act with all of them on the background. We obtain 
\beq 
\begin{aligned}
\hat D_1 \,|\mathfrak{a}^{\frac{J-n_{\mathfrak{b}}}{2}}\, \mathfrak{b}^{n_{\mathfrak{b}}}\,\mathfrak{c}^{n_{\mathfrak{c}}} \rangle &= 16  g^2n_{\mathfrak{c}} \left( | \mathfrak{a}^{\frac{J-n_{\mathfrak{b}}}{2}}\, \mathfrak{b}^{n_{\mathfrak{b}}}\,\mathfrak{c}^{n_{\mathfrak{c}}} \rangle -| \mathfrak{a}^{\frac{J-n_{\mathfrak{b}}}{2}-1}\, \mathfrak{b}^{n_{\mathfrak{b}}+2}\,\mathfrak{c}^{n_{\mathfrak{c}}-1}\rangle \right) \\
\hat D_2 \,|\mathfrak{a}^{\frac{J-n_{\mathfrak{b}}}{2}}\, \mathfrak{b}^{n_{\mathfrak{b}}}\,\mathfrak{c}^{n_{\mathfrak{c}}} \rangle &= -128  g^4n_{\mathfrak{c}} \left( | \mathfrak{a}^{\frac{J-n_{\mathfrak{b}}}{2}}\, \mathfrak{b}^{n_{\mathfrak{b}}}\,\mathfrak{c}^{n_{\mathfrak{c}}} \rangle -| \mathfrak{a}^{\frac{J-n_{\mathfrak{b}}}{2}-1}\, \mathfrak{b}^{n_{\mathfrak{b}}+2}\,\mathfrak{c}^{n_{\mathfrak{c}}-1}\rangle \right)\,.
\end{aligned}
\eeq
Given a number $n_{X} $ of $X$s, we can construct $\lfloor\frac{n_{X}}{2} \rfloor+1$ eigenstates from this dilatation operator and the corresponding spectrum of anomalous dimensions $\gamma$ is given by
\beq \label{su2lead}
\gamma_{k,n_{X}}= k\times2\, \epsilon\quad {\rm{with}}\quad \epsilon = 8 g^2-32 g^4\,,\quad k=0,\dots, \left\lfloor \frac{n_{X}}{2} \right\rfloor\,.
\eeq
In addition, because of the structure of basic single-trace operators \eqref{particles}, $n_X$ needs to be even (odd) if $J$ is even (odd).

Let us make two observations on this spectrum. First, \eqref{su2lead} can be reproduced by a collection of non-interacting ``magnons'' of the following kind,
\begin{itemize}
\item ``$0$'' (or massless) magnons whose anomalous dimension is 0,
\item ``$+$'' and ``$-$'' massive magnons whose anomalous dimensions are $\epsilon$,
\end{itemize}
subject to the constraints that
\begin{itemize}
\item the total numbers of $+$ and $-$ magnons are equal,
\item the total number of magnons need to be even if $J$ is even and odd if $J$ is odd.
\end{itemize}
The first condition can be interpreted as the consequence of the U(1) gauge invariance unbroken by the large charge background (for details see Section~\ref{sec:leading}). We will see in the next subsection that the second constraint will be replaced with a certain parity condition for more general operators.

Using this physical picture, the energy \eqref{su2lead} can be interpreted as the energy of a state with $k$ $\pm$ magnons and $(n_X-k)$ $0$ magnons. 
Second, the expansion of $\epsilon$ coincides with the expansion of the ``magnon energy'' mentioned in the introduction,
\beq
\sqrt{1+16g^2}=1+8g^2-16g^4+\ldots\period
\eeq
at weak coupling. (The leading term can be interpreted as the bare dimension of the field $X$.) Since we only have the leading two terms in the expansion, this is more like numerology at this point. However, as we will see later, the symmetry around the large charge vacuum fixes the energy of the excitation to be precisely of this form. 

 In the large $N$ limit of $\mathcal{N}=4$ SYM a similar magnon picture also emerges for single trace operators carrying a large charge $J$. In this limit, the single trace operators can be thought of as periodic ferromagnetic spin chains with asymptotically large length. The states are obtained by exciting magnons without restrictions on their number and the total energy amounts to add the corresponding individual energies. This picture holds up to finite size effects which lead to exponentially suppressed corrections in the length  and modify the asymptotic spin chain interpretation.  
As will be discussed in the next section, in the large $J$ limit, we will also find some one-dimensional system describing these magnons although there is no really underlying lattice like in the large $N$ limit.

\paragraph{Higher orders.}
A natural way of writing the dilatation operator is to use creation and annihilation operators for the letters (\ref{particles}),
\beq
{\hat{\mathfrak{a}}}^{\dagger}|0\rangle = |\mathfrak{a} \rangle\,,\quad  {\hat{\mathfrak{b}}}^{\dagger}|0\rangle = |\mathfrak{b} \rangle\,,\quad  {\hat{\mathfrak{c}}}^{\dagger}|0\rangle = |\mathfrak{c} \rangle\,,
\eeq
with the standard commutation rules $[{\hat{\mathfrak{a}}},{\hat{\mathfrak{a}}}^{\dagger}] =[{\hat{\mathfrak{b}}},{\hat{\mathfrak{b}}}^{\dagger}] =[{\hat{\mathfrak{c}}},{\hat{\mathfrak{c}}}^{\dagger}] =1$ and vanishing commutators involving operators of different species. 
This can be easily obtained by studying how the dilatation operator acts on individual letters.
In terms of such operators, we can write 
\beq
\begin{aligned}
\hat D_1 &= -2\, \frac{g_{\rm YM}^2}{8\pi^2} \left( {\hat{\mathfrak{a}}}^{\dagger}\,  {\hat{\mathfrak{b}}}\,{\hat{\mathfrak{b}}}\, {\hat{\mathfrak{c}}}^{\dagger}+4\,{\hat{\mathfrak{a}}}\, {\hat{\mathfrak{b}}}^{\dagger}\,{\hat{\mathfrak{b}}}^{\dagger}\, {\hat{\mathfrak{c}}}- 4\, \hat{\mathfrak{n}}_{\mathfrak{c}}\,\hat{\mathfrak{n}}_{\mathfrak{a}}+\hat{\mathfrak{n}}_{\mathfrak{b}}-\hat{\mathfrak{n}}_{\mathfrak{b}}^2  \right)\,,
 \\
\hat D_2 &= -2\, \frac{g_{\rm YM}^2}{8\pi^2}  (\hat{\mathfrak{n}}_{\mathfrak{a}}+\hat{\mathfrak{n}}_{\mathfrak{b}}+\hat{\mathfrak{n}}_{\mathfrak{c}}) \,\hat D_1
\end{aligned}
\eeq
where we have defined the number operator for each letter $\hat{\mathfrak{n}}_{\chi}=\hat{\chi}^{\dagger}\hat{\chi}$. Their action on states  is easier to compute now, for example we have for $\hat D_1$
\beq
\begin{aligned}
\hat D_1 \,|\mathfrak{a}^{\frac{J-n_{\mathfrak{b}}}{2}}\, \mathfrak{b}^{n_{\mathfrak{b}}}\,\mathfrak{c}^{n_{\mathfrak{c}}} \rangle &=16 \,g^2\left(1- \frac{n_{\mathfrak{b}}}{J}\right) n_{\mathfrak{c}} \left( |\mathfrak{a}^{\frac{J-n_{\mathfrak{b}}}{2}}\, \mathfrak{b}^{n_{\mathfrak{b}}}\,\mathfrak{c}^{n_{\mathfrak{c}}} \rangle-|\mathfrak{a}^{\frac{J-n_{\mathfrak{b}}}{2}-1}\, \mathfrak{b}^{n_{\mathfrak{b}}+2}\,\mathfrak{c}^{n_{\mathfrak{c}}-1} \rangle\right)\\
&+\frac{8\, n_{\mathfrak{b}}(n_{\mathfrak{b}}-1)}{J} g^2 \left(  |\mathfrak{a}^{\frac{J-n_{\mathfrak{b}}}{2}}\, \mathfrak{b}^{n_{\mathfrak{b}}}\,\mathfrak{c}^{n_{\mathfrak{c}}} \rangle -  |\mathfrak{a}^{\frac{J-n_{\mathfrak{b}}}{2}+1}\, \mathfrak{b}^{n_{\mathfrak{b}}-2}\,\mathfrak{c}^{n_{\mathfrak{c}}+1} \rangle\right)\,.
\end{aligned}
\eeq

The dilatation operator can be easily diagonalized and we obtain the spectrum which we write in a concise form as 
\beq \label{datasu2}
\gamma_{k,n_{X}}=k  \left(  2\, \epsilon -\frac{128\, g^4 \,n_X}{J}  \right) \left( 1+\frac{1}{J}(1-2k +n_X)\right)\,,\quad k=0,\dots, \left\lfloor \frac{n_{X}}{2} \right\rfloor\,.
\eeq
We want  to interpret the $1/J$ correction of this formula in light of the magnon picture advocated in the previous section. Accordingly, let us write the total number of magnons as $n_X = 2 k+ n_0$ where $2k$ is the number of massive magnons and $n_0$ is the number of massless magnons.
Let us focus on the leading correction in $1/J$ and write it as
\beq \label{eigensu2}
\gamma_{k,n_{X}}\Big\rvert_{\frac{1}{J}} = 2k \, n_0 \,\epsilon_{0} + 2k \,\epsilon_{1}+  \binom{2k}{2} \, \epsilon_{2}\,,
\eeq
with
\beq\label{eq:defe012}
\epsilon_0 =\epsilon_1= 8 g^2-96 g^4\comma\qquad \epsilon_2=-128 g^4 \period
\eeq
This formula makes it clear that  at order $1/J$, magnons interact at most pairwise. The contribution $\epsilon_0$ arises from the pairwise interaction between massless and massive magnons whereas $\epsilon_2$ has its origin in the pairwise interaction of massive magnons. Finally, the $\epsilon_1$ term corresponds to the correction to the mass of the magnons.  At order $1/J^2$, we find terms that scale with $k^3$ which can be interpreted as arising from a  three-body interaction. 
This is the main piece of data of this section and in what follows we will  justify both qualitatively and quantitatively the leading and subleading terms of the $1/J$ expansion of the  formula (\ref{datasu2}).
	\subsection{SU(3) sector: parity projection} \label{su3parity}
	We next consider the so-called SU(3) sector involving all three complex scalars $X,Y,Z$. This sector is closed under the action of the dilatation operator at one-loop, but at higher orders they generally mix with operators that contain fermions. The full one-loop dilatation operator involving all SO(6) scalar fields is known from \cite{Beisert:2003tq}, and when restricted to the three complex scalars it reads
	\beq \label{su3dila}
	\hat D_1=-\left(\frac{g^{2}_{{\rm{YM}}}}{8 \pi^2} \right)\left(:\Tr [X,Y][\check{X},\check{Y}]:+:\Tr[X,Z][\check{X},\check{Z}]:+:\Tr[Y,Z][\check{Y},\check{Z}] :\right)\,.
	\eeq 
As in the previous SU(2) sector, using the relations (\ref{tracerels}) together with 
\beq
\Tr X YZ = -\Tr XZY\,,
\eeq
one can express all gauge-invariant operators in terms of the basic letters
\beq
\begin{aligned}
&\mathfrak{a}:=\Tr ZZ \quad \mathfrak{b}:= \Tr ZX \quad \mathfrak{c}:=\Tr X X \\
 &\mathfrak{d}:=\Tr Z Y \quad \mathfrak{e}:=\Tr Y Y \quad \mathfrak{f}:=\Tr X Y  \quad \mathfrak{g}:=\Tr X Y Z\,.
\end{aligned}
\eeq
As before, the states we will be considering are of the form
\beq
|\mathfrak{a}^{\frac{J-n_{\mathfrak{b}}-n_{\mathfrak{d}}-n_{\mathfrak{g}}}{2}} \mathfrak{b}^{n_{\mathfrak{b}}}  \mathfrak{c}^{n_{\mathfrak{c}}} \mathfrak{d}^{n_{\mathfrak{d}}}  \mathfrak{e}^{n_{\mathfrak{e}}} \mathfrak{f}^{n_{\mathfrak{f}}}  \mathfrak{g}^{n_{\mathfrak{g}}} \rangle\,,
\eeq
and the leading $J$ contribution is extracted from the two last terms of (\ref{su3dila}) by acting with $\check{Z}$ on the background fields.

\paragraph{Leading order.} At leading order in the large $J$ expansion, we find that the energies of states are always given by integer multiples of $8g^2$ as was the case with the SU(2) sector (\ref{su2lead}). However, it turns out that the detailed spectrum and the degeneracy show a rather intricate pattern and depend on whether the charge $J$ is even or odd. See Table~\ref{tabstatesbrutef} for the full spectrum for states with a small total number of $X$ and $Y$ scalars (which we denote by $N$).
\begin{table}[t]
\centering
\scriptsize
$\begin{array}[t]{|c|c|c|c|}\hline
 {N} &J& \rm{eigenstate}&\rm{\gamma} \\\hline\hline
\multirow{7}{*}{2} &\multirow{6}{*}{{\rm{even}}} & |\mathfrak{a}^{\frac{J}{2}-1} \mathfrak{b}\, \mathfrak{d}\,\rangle & 0 \\
& &|\mathfrak{a}^{\frac{J}{2}-1} \mathfrak{d}^2\rangle &0\\
& &|\mathfrak{a}^{\frac{J}{2}-1} \mathfrak{b}^2\rangle &0\\
& &|\mathfrak{a}^{\frac{J}{2}-1} \mathfrak{b}\, \mathfrak{d}\rangle-|\mathfrak{a}^{\frac{J}{2}} \mathfrak{f}\rangle &16g^2+\frac{16g^2}{J}\\
& &|\mathfrak{a}^{\frac{J}{2}}\mathfrak{e}\rangle-|\mathfrak{a}^{\frac{J}{2}-1}\mathfrak{b}^2 \rangle &16g^2+\frac{16g^2}{J}\\
& &|\mathfrak{a}^{\frac{J}{2}}\mathfrak{c}\rangle-|\mathfrak{a}^{\frac{J}{2}-1}\mathfrak{d}^2\rangle &16g^2+\frac{16g^2}{J}\\
\cline{2-4}
 &{\rm{odd}} & |\mathfrak{a}^{\frac{J-1}{2}}\mathfrak{g}\rangle& 16g^2+\frac{32g^2}{J}  \\
 \hline
 \multirow{12}{*}{3} &\multirow{2}{*}{{\rm{even}}} &|\mathfrak{a}^{\frac{J}{2}-1} \mathfrak{b}\,\mathfrak{g}\rangle &16g^2+\frac{48 g^2}{J} \\
 & &|\mathfrak{a}^{\frac{J}{2}-1}\mathfrak{d}\,\mathfrak{g}\rangle & 16 g^2+\frac{48 g^2}{J} \\
 \cline{2-4}
 &\multirow{10}{*}{{\rm{odd}}} &|\mathfrak{a}^{\frac{J-3}{2}} \mathfrak{b}\, \mathfrak{d}^2\rangle & 0 \\
  & &|\mathfrak{a}^{\frac{J-3}{2}}\mathfrak{d}^3\rangle  & 0 \\
  & &|\mathfrak{a}^{\frac{J-3}{2}} \mathfrak{b}^2 \mathfrak{d}\rangle  & 0 \\
  &&|\mathfrak{a}^{\frac{J-3}{2}}\mathfrak{b}^3\rangle &0\\
  &&|\mathfrak{a}^{\frac{J-1}{2}}\mathfrak{d}\, \mathfrak{f}\rangle-|\mathfrak{a}^{\frac{J-3}{2}}\mathfrak{b}\,\mathfrak{d}^2\rangle &16 g^2+ \frac{32 g^2}{J}\\
  && |\mathfrak{a}^{\frac{J-1}{2}} \mathfrak{d}\,\mathfrak{e}\rangle-|\mathfrak{a}^{\frac{J-3}{2}} \mathfrak{d}^3\rangle  &16 g^2+ \frac{32 g^2}{J}\\
  && |\mathfrak{a}^{\frac{J-1}{2}} \mathfrak{c}\,\mathfrak{d}\rangle-|\mathfrak{a}^{\frac{J-3}{2}} \mathfrak{b}^2\,\mathfrak{d}\rangle  &16 g^2+ \frac{32 g^2}{J}\\
  && |\mathfrak{a}^{\frac{J-1}{2}} \mathfrak{b}\,\mathfrak{f}\rangle-|\mathfrak{a}^{\frac{J-3}{2}} \mathfrak{b}^2\,\mathfrak{d}\rangle  &16 g^2+ \frac{32 g^2}{J}\\
   && |\mathfrak{a}^{\frac{J-1}{2}} \mathfrak{b}\,\mathfrak{e}\rangle-|\mathfrak{a}^{\frac{J-3}{2}} \mathfrak{b}\,\mathfrak{d}^2\rangle  &16 g^2+ \frac{32 g^2}{J}\\
   && |\mathfrak{a}^{\frac{J-1}{2}} \mathfrak{b}\,\mathfrak{c}\rangle-|\mathfrak{a}^{\frac{J-3}{2}} \mathfrak{b}^3 \rangle  &16 g^2+ \frac{32 g^2}{J}\\
\hline
\end{array}  \quad
\begin{array}[t]{|c|c|c|c|}\hline
 {N} &J& \rm{eigenstate}&\rm{\gamma}\\\hline\hline
\multirow{26}{*}{4} &\multirow{20}{*}{{\rm{even}}} &|\mathfrak{a}^{\frac{J}{2}-2} \mathfrak{b}^3 \mathfrak{d}\rangle & 0 \\
& & | \mathfrak{a}^{\frac{J}{2}-2} \mathfrak{b} \mathfrak{d}^3\rangle &0\\
& &| \mathfrak{a}^{\frac{J}{2}-2} \mathfrak{b}^2 \mathfrak{d}^2\rangle & 0\\
& &| \mathfrak{a}^{\frac{J}{2}-2} \mathfrak{b}^4\rangle & 0 \\
& &| \mathfrak{a}^{\frac{J}{2}-2} \mathfrak{d}^4\rangle & 0 \\
& & |\mathfrak{a}^{\frac{J}{2}-1} \mathfrak{b}^2 \mathfrak{f}\rangle-|\mathfrak{a}^{\frac{J}{2}-2} \mathfrak{b}^3 \mathfrak{d}\rangle & 32 g^2+\frac{32 g^2}{J}\\
&&+|\mathfrak{a}^{\frac{J}{2}-1} \mathfrak{b} \mathfrak{c} \mathfrak{d}\rangle-|\mathfrak{a}^{J/2} \mathfrak{c} \mathfrak{f}\rangle  & \\
& & |\mathfrak{a}^{\frac{J}{2}-1} \mathfrak{b} \mathfrak{d} \mathfrak{e}\rangle-|\mathfrak{a}^{\frac{J}{2}-2} \mathfrak{b} \mathfrak{d}^3\rangle &32 g^2+ \frac{32 g^2}{J}\\
&& +|\mathfrak{a}^{\frac{J}{2}-1} \mathfrak{d}^2 \mathfrak{f}\rangle-|\mathfrak{a}^{J/2} \mathfrak{e} \mathfrak{f}\rangle &\\
& &\frac{1}{4} |\mathfrak{a}^{\frac{J}{2}-1} \mathfrak{b}^2 \mathfrak{e}\rangle -\frac{3}{4} |\mathfrak{a}^{\frac{J}{2}-2} \mathfrak{b}^2 \mathfrak{d}^2\rangle+|\mathfrak{a}^{\frac{J}{2}-1} \mathfrak{b} \mathfrak{d} \mathfrak{f}\rangle &32 g^2+ \frac{32 g^2}{J}\\
&&+\frac{1}{4} |\mathfrak{a}^{\frac{J}{2}-1} \mathfrak{c} \mathfrak{d}^2\rangle-\frac{1}{4} |\mathfrak{a}^{J/2} \mathfrak{c} \mathfrak{e}\rangle-\frac{1}{2} |\mathfrak{a}^{J/2} \mathfrak{f}^2\rangle & \\
& &|\mathfrak{a}^{\frac{J}{2}-1} \mathfrak{d}^2 \mathfrak{e}\rangle -\frac{1}{2} |\mathfrak{a}^{\frac{J}{2}-2} \mathfrak{d}^4\rangle-\frac{1}{2} |\mathfrak{a}^{J/2} \mathfrak{e}^2\rangle & 32 g^2+ \frac{32 g^2}{J}\\
& &|\mathfrak{a}^{\frac{J}{2}-1} \mathfrak{b}^2 \mathfrak{c}\rangle -\frac{1}{2} |\mathfrak{a}^{\frac{J}{2}-2} \mathfrak{b}^4\rangle-\frac{1}{2} |\mathfrak{a}^{J/2} \mathfrak{c}^2\rangle & 32 g^2+ \frac{32 g^2}{J}\\
& & |\mathfrak{a}^{\frac{J}{2}-1} \mathfrak{b} \mathfrak{c} \mathfrak{d}\rangle-|\mathfrak{a}^{\frac{J}{2}-1} \mathfrak{b}^2 \mathfrak{f}\rangle &16 g^2+ \frac{48 g^2}{J}\\
& & |\mathfrak{a}^{\frac{J}{2}-1} \mathfrak{b} \mathfrak{d} \mathfrak{e}\rangle-|\mathfrak{a}^{\frac{J}{2}-1} \mathfrak{d}^2 \mathfrak{f}\rangle &16 g^2+ \frac{48 g^2}{J}\\
& & |\mathfrak{a}^{\frac{J}{2}-1} \mathfrak{d}^2 \mathfrak{f} \rangle-|\mathfrak{a}^{\frac{J}{2}-2} \mathfrak{b} \mathfrak{d}^3 \rangle &16 g^2+ \frac{48 g^2}{J}\\
& & |\mathfrak{a}^{\frac{J}{2}-1} \mathfrak{b}^2 \mathfrak{f} \rangle-|\mathfrak{a}^{\frac{J}{2}-2} \mathfrak{b}^3 \mathfrak{d} \rangle &16 g^2+ \frac{48 g^2}{J}\\
& & |\mathfrak{a}^{\frac{J}{2}-1} \mathfrak{b} \mathfrak{d} \mathfrak{f}\rangle-|\mathfrak{a}^{\frac{J}{2}-2} \mathfrak{b}^2 \mathfrak{d}^2\rangle &16 g^2+ \frac{48 g^2}{J}\\
& & |\mathfrak{a}^{\frac{J}{2}-1} \mathfrak{b}^2 \mathfrak{e} \rangle-|\mathfrak{a}^{\frac{J}{2}-2} \mathfrak{b}^2 \mathfrak{d}^2\rangle &16 g^2+ \frac{48 g^2}{J}\\
& & |\mathfrak{a}^{\frac{J}{2}-1} \mathfrak{d}^2 \mathfrak{e} \rangle-|\mathfrak{a}^{\frac{J}{2}-2} \mathfrak{d}^4 \rangle &16 g^2+ \frac{48 g^2}{J}\\
& & |\mathfrak{a}^{\frac{J}{2}-1} \mathfrak{c}  \mathfrak{d}^2\rangle-|\mathfrak{a}^{\frac{J}{2}-1} \mathfrak{b}^2 \mathfrak{e} \rangle &16 g^2+ \frac{48 g^2}{J}\\
& & |\mathfrak{a}^{\frac{J}{2}-1} \mathfrak{c}  \mathfrak{d}^2\rangle-|\mathfrak{a}^{\frac{J}{2}-1} \mathfrak{b}^2 \mathfrak{e} \rangle &16 g^2+ \frac{48 g^2}{J}\\
& & |\mathfrak{a}^{\frac{J}{2}-1} \mathfrak{b}^2  \mathfrak{c}\rangle-|\mathfrak{a}^{\frac{J}{2}-2} \mathfrak{b}^4 \rangle &16 g^2+ \frac{48 g^2}{J}\\
& &|\mathfrak{a}^{\frac{J}{2}-1} \mathfrak{b} \mathfrak{d} \mathfrak{f}\rangle -\frac{1}{2} |\mathfrak{a}^{\frac{J}{2}-1} \mathfrak{b}^2 \mathfrak{e}\rangle-\frac{1}{2} |\mathfrak{a}^{\frac{J}{2}-1} \mathfrak{c} \mathfrak{d}^2\rangle &32 g^2+ \frac{80 g^2}{J} \\
&&+\frac{1}{2} |\mathfrak{a}^{J/2} \mathfrak{c} \mathfrak{e}\rangle -\frac{1}{2} |\mathfrak{a}^{J/2} \mathfrak{f}^2\rangle &\\
 \cline{2-4}
 &\multirow{6}{*}{{\rm{odd}}} &|\mathfrak{a}^{\frac{J-1}{2}} \mathfrak{f}\, \mathfrak{g} \rangle-|\mathfrak{a}^{\frac{J-3}{2}} \mathfrak{b}\, \mathfrak{d}\, \mathfrak{g} \rangle & 32 g^2+\frac{64 g^2}{J} \\
 & &|\mathfrak{a}^{\frac{J-3}{2}} \mathfrak{d}^2\, \mathfrak{g} \rangle-|\mathfrak{a}^{\frac{J-1}{2}} \mathfrak{e}\,  \mathfrak{g} \rangle & 32 g^2+\frac{64 g^2}{J} \\
  & &|\mathfrak{a}^{\frac{J-3}{2}} \mathfrak{b}^2\, \mathfrak{g} \rangle-|\mathfrak{a}^{\frac{J-1}{2}} \mathfrak{c}\,  \mathfrak{g} \rangle & 32 g^2+\frac{64 g^2}{J} \\
    & &|\mathfrak{a}^{\frac{J-3}{2}} \mathfrak{b} \, \mathfrak{d}\, \mathfrak{g} \rangle  & 16 g^2+\frac{64 g^2}{J} \\
    & &|\mathfrak{a}^{\frac{J-3}{2}} \mathfrak{d}^2 \, \mathfrak{g} \rangle  & 16 g^2+\frac{64 g^2}{J} \\
    & &|\mathfrak{a}^{\frac{J-3}{2}} \mathfrak{b}^2 \, \mathfrak{g} \rangle  & 16 g^2+\frac{64 g^2}{J} 
 \\
\hline
\end{array}$ 
\qquad
\caption{This table contains the spectrum of states of the dilatation operator at one-loop in the large charge 't Hooft coupling and up to the first $1/J$ correction in a sector involving two complex scalars. Note in particular that the spectrum depends on the parity of $J$.}
\label{tabstatesbrutef}
\end{table}
For instance, the spectrum of states with $N=2$ is
\beq\label{eq:spectrumsu31}
\begin{aligned}
N=2\qquad &\text{Even $J$:}\qquad  \{0,0,0,16g^2,16g^2,16g^2\}\comma\\
&\text{Odd $J$:}\qquad  \{16g^2\}\comma
\end{aligned}
\eeq 
while, for states with $N=3$, we have
\beq\label{eq:spectrumsu32}
\begin{aligned}
N=3\qquad &\text{Even $J$:}\qquad  \{16g^2,16g^2\}\comma\\
&\text{Odd $J$:} \qquad \{0,0,0,0,16g^2,16g^2,16g^2,16g^2,16g^2,16g^2\}\comma
\end{aligned}
\eeq 

Although the spectrum and the degeneracy look rather complicated, we find that there is a simple rule that reproduces the spectrum from the magnon picture that we discussed in the previous subsection:
\begin{enumerate}
\item Both $X$ and $Y$ magnons come with three different types; ``massless'', $+$ and $-$ types discussed in the previous subsection.
\item \textbf{U(1) gauge invariance}: The total numbers of $+$ and $-$ excitations are equal.
\item \textbf{Parity}: Consider the ``parity'' transformation which multiply $-1$ both to magnons and background $Z$ fields, and swaps $+$ and $-$ magnons. Under this transformation, the states need to be even. This is equivalent to saying that combinations of magnons need to be even (odd) under the parity if $J$ is even (odd).
\end{enumerate}
Postponing the explanation of the origin of these rules to Section~\ref{sec:leading}, let us see below how these rules reproduce the spectrum given in \eqref{eq:spectrumsu31}. For $N=2$, the sets of magnons that satisfy the second condition above are
\beq
\begin{aligned}
&|\phi_1^{0}\phi_1^{0}\rangle\comma\quad |\phi_1^{0}\phi_2^{0}\rangle\comma\quad |\phi_2^{0}\phi_2^{0}\rangle\comma\quad |\phi_1^{+}\phi_1^{-}\rangle\comma\quad |\phi_1^{+}\phi_2^{-}\rangle \comma\quad |\phi_2^{+}\phi_1^{-}\rangle\comma\quad |\phi_2^{+}\phi_2^{-}\rangle\period
\end{aligned}
\eeq
Here $\phi_{1}$ ($\phi_2$) denotes the $X$ ($Y$) scalar 
and the superscripts $0$ and $\pm$ signify the particle types. We next impose the parity condition and find, for even $J$,
\beq\label{eq:2particlesu3even}
\begin{aligned}
|\phi_1^{0}\phi_1^{0}\rangle\comma\quad |\phi_1^{0}\phi_2^{0}\rangle\comma\quad |\phi_2^{0}\phi_2^{0}\rangle\comma\quad |\phi_1^{+}\phi_1^{-}\rangle\comma\quad \frac{1}{\sqrt{2}}\left(|\phi_1^{+}\phi_2^{-}\rangle + |\phi_2^{+}\phi_1^{-}\rangle\right)\comma\quad |\phi_2^{+}\phi_2^{-}\rangle
\end{aligned}
\eeq
The first three states have zero energy while the last three states have the energy $16g^2$, reproducing the spectrum \eqref{eq:spectrumsu31} at leading large $J$.
For odd $J$, the only state that survives the parity projection is
\beq
 \frac{1}{\sqrt{2}}\left(|\phi_1^{+}\phi_2^{-}\rangle - |\phi_2^{+}\phi_1^{-}\rangle\right)\comma
\eeq
which has the energy $16g^2$ in agreement with \eqref{eq:spectrumsu31}. Performing a similar analysis for $N=3$, one can also reproduce the spectrum \eqref{eq:spectrumsu32}.
\paragraph{Subleading order.}
At order $1/J$, the magnons begin to interact similarly to the picture suggested by the previous results. By an explicit diagonalization of the dilatation operator, we find that the anomalous dimensions in this sector depend on whether $N+J$ is even or odd. We have generated  data for a large set of states involving several magnons and observed that it can be accommodated by the following expression
\beq
\begin{aligned} \label{su3eigens}
\gamma_{k, N,n}\Big\rvert_{\frac{1}{J}}  = 
2k \, n_0 \,\epsilon_{0} + 2k \,\epsilon_{1}+  \binom{2k}{2} \, \epsilon_{2}+2n\, (2k-2n+1) \,  \epsilon_3
\end{aligned}
\eeq
with $n_0=N-2k$ and
\beq
\begin{aligned}
n&=0,\dots ,\left\lfloor\frac{k}{2}\right\rfloor\,\quad \quad {\text{for }}\quad N+J {\text{ even}}\,, \\
n&=\left\lfloor\frac{k}{2}\right\rfloor+1,\dots ,k \,\quad \quad {\text{for }}\quad N+J {\text{ odd}}\,
\end{aligned}
\eeq	
and as before $k=1,\dots, \left\lfloor \frac{N}{2} \right\rfloor$ counts the number of pairs of ``massive'' magnons. In the absence of massive magnons the corresponding anomalous dimension vanishes, i.e. $\gamma_{k=0,N=0,n=0}=0$. The three first terms in this expression coincide with the result for the SU(2) sector found before (see equation (\ref{eigensu2})) with the values of $\epsilon_{0,1,2}$ being given in \eqref{eq:defe012} restricted to one-loop. In addition, we now have an extra term $\epsilon_3$ when $n>0$ which should arise from the interactions between different scalars. Its value is given at leading order in the large charge 't Hooft coupling by
\beq
\epsilon_{3}=8 g^2\,.
\eeq
These eigenvalues also come with some degeneracy. However the detailed pattern shown in Table~\ref{tabstatesbrutef} is rather complicated. We postpone their interpretation from the magnon picture to Section~\ref{sec:subleading}, where we verify that both the degeneracy and the energies $\epsilon_{0,1,2,3}$ can be fully explained from interactions among different magnons and determine the values of some of the ``energies'' $\epsilon_{0,1,2,3}$ as an exact function of the large charge 't Hooft coupling. (See sections \ref{sec:states} and  \ref{sec:eftcoefs}.)

	\subsection{SL(2) sector: bound states}\label{sl2data}
Finally, let us discuss the analysis of the weak coupling data for the SL(2) sector, which consists of gauge invariant operators made out of complex scalars $Z$ and light-cone derivatives acting on them. As we will see below, the analysis in this sector reveals yet another feature of the spectrum, i.e.~the existence of a new infinitely family of magnons, labelled by a positive integer. Owing to the similarity with magnon bound states in the planar $\mathcal{N}=4$ spin chain, we often refer to them ``bound states'' in this section. This time we restrict the analysis to the one-loop and  we will focus on the leading large $J$ anomalous dimension. The elementary fields in this sector will be denoted by the notation
\beq \label{excitationsl2}
Z_{n} := D^{n}_{+} Z
\eeq
where $ D^{n}_{+}$ is the $n$-th covariant derivative along a light-cone direction. The full non-planar dilatation operator at one-loop can be obtained for example from the dilation operator in the $\mathfrak{psu}(1,1|2)$ sector worked out in \cite{Zwiebel:2005er} by projecting out the fermions and one of the two scalars. The outcome is given by\footnote{This dilatation operator is in fact equivalent to the non-planar uplift of the Hamiltonian found in \cite{Beisert:2003jj}.}
\beq \label{sl2dila}
\hat D_1= \frac{g^2_{{\rm{YM}}}}{8 \pi^2} \sum_{m,n=0}^{\infty} \sum_{k=0}^{n-1} \frac{1}{k+1}
 \Tr :[Z_{m+k+1}, \check{Z}_{m}]::[Z_{n-1-k}, \check{Z}_n]:\,.
 \eeq

Note that the derivative of the first commutator also acts on the field of the second commutator. In order to study large charge operators in this sector, we consider the letters out of which we construct any operator
\beq
\mathfrak{a} := \Tr \,ZZ \qquad  \mathfrak{b}_{\bf{n}}:=  \Tr\,Z_{n_1} Z_{n_2} \dots\, \quad {\rm{with}}\quad {\bf{n}}=(n_1,n_2,\dots)\,.
\eeq
Similarly to the previous scalar sector, for a given total spin $S=\sum_i n_i$ we find a finite number of letters, as a single trace can be factorized into multiple traces of smaller length whenever there is a repeated sequence of fields in $\mathfrak{b}_{\bf{n}}$. We consider states of the form
\beq
| \mathfrak{a}^{\frac{J-\sum_i \ell_i}{2}} \prod_{i} \mathfrak{b}_{{\bf{n}}_i} \rangle\,,
\eeq
where $\ell_i$ is the length of the tuple ${\bf{n}}_i$. As before, in the large charge 't Hooft limit, the leading contribution arises when $\check{Z}$ acts on the background. We now analyze a couple of examples of low spin. For simplicity, below we assume that $J$ to be even. 
\paragraph{Spin two.} The simplest example corresponds to spin $S=2$ operators of arbitrary twist $\tau $ ($\tau=\Delta-S$), for which there are only three independent letters we can use to construct gauge invariant operators
\beq
\mathfrak{b}_{(0,1)}\,,\quad \mathfrak{b}_{(1,1)}\,,\quad \mathfrak{b}_{(0,2)}\,.
\eeq
The leading order result in the large $J$ expansion for the action of the dilatation operator on the large charge states is given by
\beq
\begin{aligned}
\hat D_1 \,|\mathfrak{a}^{\frac{J}{2}-1} \, \mathfrak{b}_{(1,1)} \rangle &= 16 g^2 \left( |\mathfrak{a}^{\frac{J}{2}-1} \, \mathfrak{b}_{(1,1)} \rangle -|\mathfrak{a}^{\frac{J}{2}-2} \, \mathfrak{b}_{(0,1)}^2 \rangle  \right) +\mathcal{O}(1/J) \\
\hat D_1 \,|\mathfrak{a}^{\frac{J}{2}-1} \, \mathfrak{b}_{(0,2)} \rangle &= -8 g^2 \left( |\mathfrak{a}^{\frac{J}{2}-1} \, \mathfrak{b}_{(1,1)} \rangle -|\mathfrak{a}^{\frac{J}{2}-2} \, \mathfrak{b}_{(0,1)}^2 \rangle  \right) +\mathcal{O}(1/J)\\
\hat D_1 \,|\mathfrak{a}^{\frac{J}{2}-2} \, \mathfrak{b}_{(0,1)}^2 \rangle &=0\,.
\end{aligned}
\eeq
This gives two eigenstates with $\gamma=0$ and only one non-zero eigenvalue with $\gamma = 16 g^2$.
\paragraph{Spin three.} At  spin three, we encounter a larger number of states made out of the following letters
\beq
\mathfrak{b}_{(0,1)}\,,\quad \mathfrak{b}_{(1,1)}\,,\quad \mathfrak{b}_{(0,2)}\,, \quad \mathfrak{b}_{(1,2)}\,, \quad \mathfrak{b}_{(0,3)}\,.
\eeq
The action on the states is simple to compute and we obtain two eigenstates with non-zero anomalous dimensions given by
$\gamma = 12 g^2$ and $\gamma= 16 g^2$.
\paragraph{General spin.} 

For general spin $S$, it is simple but tedious to find the eigenstates using the dilatation operator (\ref{sl2dila}). By  explicitly constructing them for several values of the spin, one finds that the spectrum and the degeneracy can be reproduced by combinations of magnons which are now labelled by a positive integer $n$
\beq
z_{n}^{0},\, z_{n}^{+},\,z_n^{-} \qquad (n\in \mathbb{Z}_{\geq 1})\,,
\eeq
and carry the following spin $S$ and the anomalous dimension $\gamma$:
\beq\label{eq:sl2energymagnon}
\begin{aligned}
&z_n^{0}:\quad S=n\,,\quad \gamma=0\comma\\
&z_n^{\pm}:\quad S=n\,,\quad \gamma=\frac{8g^2}{n}\period
\end{aligned}
\eeq
To select physical states, we impose the same conditions as in the SU(3) sector; namely we require the total numbers of $+$ and $-$ particles to be equal and impose the parity condition.

For instance, for spin $S=2$, we have three states
\beq
|z_{1}^{0}z_{1}^{0}\rangle \comma\quad |z_{2}^{0}\rangle \comma\quad |z_{1}^{-}z_{1}^{+}\rangle\period
\eeq
The first two states have $\gamma=0$ while the last state has $\gamma=16g^2$ being consistent with what we saw in the analysis above. Similarly, for spin $S=4$ we have the following states\footnote{Note that here we assumed $J$ to be even when performing the parity projection.} with non-zero $\gamma$
\beq\label{dataelemsl2}
\begin{aligned}
\gamma=8g^2:&\qquad|z_2^{-}z_{2}^{+}\rangle\comma \\
\gamma=\frac{32g^2}{3}:&\qquad \frac{|z_1^{-}z_3^{+}\rangle+|z_3^{-}z_1^{+}\rangle}{\sqrt{2}}\\
\gamma=12g^2:&\qquad\frac{|z_1^{0}z_1^{-}z_2^{+}\rangle+|z_1^{0}z_2^{-}z_1^{+}\rangle}{\sqrt{2}}\comma \\
\gamma=16g^2:&\qquad|z_1^{0}z_1^{0}z_1^{-}z_{1}^{+}\rangle\comma\quad |z_2^{0}z_1^{-}z_{1}^{+}\rangle\comma\\
\gamma=32g^2:&\qquad  |z_1^{-}z_1^{-}z_1^{+}z_1^{+}\rangle\period
\end{aligned}
\eeq 
As shown above, the eigenvalue $\gamma=16g^2$ has double degeneracy. In comparison with the previous scalar sectors, we find a larger number of non-trivial anomalous dimensions for a given spin. For example, with $n_{X}=4$ we have found only two non-trivial values whereas here there are five distinct dimensions for $S=4$. 

The energy of these magnons given by \eqref{eq:sl2energymagnon} coincides with the expansion of
\beq
\sqrt{n^2+16g^2}=n+\frac{8g^2}{n}+\cdots\period
\eeq
(Again the leading term corresponds to the tree-level conformal dimension.)
Later, we will discover that these distinct magnons sit in higher representations of the symmetry group and can be regarded as bound states of the fundamental ones. 
\subsection{Superconformal index and partition function at large charge}\label{sec:partition}
We now analyze the superconformal index and the partition function of free $\mathcal{N}=4$ SYM with the SU(2) gauge group and show that the results in the large charge sector agree with the index and the partition function of free magnons subject to the constraints discussed above. This consolidates the magnon interpretation advocated in the previous subsections. 

\paragraph{Toy model.}To understand the mechanism in a simple setup, let us first consider the partition function in the SU(2) sector, namely operators made out of $X$ and $Z$. Following the standard recipe (see e.g. \cite{Aharony:2003sx,Romelsberger:2005eg,Kinney:2005ej}), we find that the partition function is given by

\begin{align}
Z_{\rm SU(2)}(q_1,q_2)=&\int dU \exp \left[\sum_{n=1}^{\infty}\frac{f_{X,Z}(q_1^{n},q_2^{n})}{n}\chi_{\rm adj}(U^{n})\right]\comma\label{eq:toypartition}\\
=&\oint_{x=0} \frac{dx}{4\pi i x}(1-x^2)(1-x^{-2}) \exp \left[\sum_{n=1}^{\infty}\frac{f_{X,Z}(q_1^{n},q_2^{n})}{n}(1+x^{2n}+x^{-2n})\right]\comma\nn
\end{align}
with $f_{X,Z}$ being the single-letter index
\beq
f_{X,Z}(q_1,q_2)=q_1+q_2\period
\eeq
Here $dU$ is the Haar measure of SU(2), $q_1$ and $q_2$ are the fugacities for $Z$ and $X$ respectively, and $\chi_{\rm adj}$ is the SU(2) character for the adjoint representation. On the second line of \eqref{eq:toypartition}, we replaced the integral over SU(2) with the integral over the Cartan element $U={\rm diag}(x,1/x)$. 

To project onto the large charge sector, we perform the contour integral around the origin
\beq
Z_{J}(q_2)\equiv \oint \frac{dq_1}{2\pi i q_1^{J+1}}Z_{\rm SU(2)}(q_1,q_2)\comma
\eeq
and set $J\gg 1$. To evaluate this integral, we use the integral expression for $Z_{\rm SU(2)}(q_1,q_2)$ (the second line of \eqref{eq:toypartition}) and exchange the order of the $x$-integral and the $q_1$-integral. This leads to
\beq\label{eq:ZJIX}
\begin{aligned}
Z_J(q_2)&= \oint_{x=0} \frac{dx}{4\pi i x}(1-x^2)(1-x^{-2})\oint \frac{dq_1}{2\pi i q_1^{J+1}} \exp \left[\sum_{n}\frac{1}{n}f_{X,Z}(q_1^{n},q_2^{n})(1+x^{2n}+x^{-2n})\right]\\
&=\oint_{x=0} \frac{dx}{2\pi i x}\underbrace{\left(\frac{1}{2}-\frac{x^{-2-2J}+x^{4+2J}}{2(1+x^2)}\right)}_{=:\mathcal{I}_Z(x)}\underbrace{\left(\frac{1}{(1-q_2)(1-q_2/x^2)(1-q_2x^2)}\right)}_{=:\mathcal{I}_X(x,q_2)}\period
\end{aligned}
\eeq
The first factor $\mathcal{I}_Z$ is the contribution from $Z$ scalars while the second factor $\mathcal{I}_X$ is the contribution from $X$ scalars. 

In the limit $J\gg 1$, the term $x^{4+2J}$ in $\mathcal{I}_Z$ can be neglected as long as we perform the integral in the region $|x|\ll 1$. We thus have
\beq
Z_{J\gg 1}(q_2)=\frac{1}{2}\oint_{x=0} \frac{dx}{2\pi i x} \mathcal{I}_X(x,q_2)-\frac{1}{2}\oint_{x=0} \frac{dx}{2\pi i x}\frac{x^{-2-2J}}{1+x^2}\mathcal{I}_X(x,q_2)\period
\eeq
To evaluate the second term, we deform the contour to infinity.
The contribution from the contour at infinity vanishes due to $x^{-2-2J}$ and we are left with\footnote{$\mathcal{I}_X(x,q_2)$ has other poles as shown in \eqref{eq:ZJIX}. However, if we first expand $\mathcal{I}_X$ as a power series in $q_2$, each term in the expansion only contains poles at $x=0$ or $x=\infty$. Thus, for the purpose of determining the partition function as a power series in $q_2$, we do not need to take into account the contribution from poles in $\mathcal{I}_X$.} the contributions from residues at $x=\pm i$. Using $\mathcal{I}_X(x,q_2)=\mathcal{I}_X(1/x,q_2)$, we then arrive at the following expression:
\beq\label{eq:toyfinal}
Z_{J\gg 1}(q_2)=\frac{1}{2}\left(\oint_{x=0} \frac{dx}{2\pi i x} \mathcal{I}_X(x,q_2)\right) +\frac{(-1)^{J}}{2}\mathcal{I}_X(i,q_2)\period
\eeq
One can check explicitly that this expression correctly reproduces the counting of operators in the large charge sector.

We now show that the expression \eqref{eq:toyfinal} can be interpreted as a partition function of free magnons subject to the parity and the U(1) gauge constraints. In this SU(2) sector, we have three types of magnons $X^{0}$, $X^{+}$ and $X^{-}$. To impose the U(1) gauge invariance discussed above, we assign the gauge charge\fn{Assignment of the charges $\pm 2$ is purely a convention that we chose to make the comparison with \eqref{eq:toyfinal} easier. As long as we assign the opposite charges to $X^{\pm}$, the result will be the same.} $+2$ to $X^{+}$,  $-2$ to $X^{-}$ and $0$ to $X^{0}$. In addition each magnon carries a fugacity $q_2$. Thus the partition function of free magnons without parity constraint reads
\beq
\begin{aligned}
\tilde{Z}_0(q_2)=\oint_{x=0}\frac{dx}{2\pi i x}\exp \left[\sum_{n=1}^{\infty}\frac{q_2^{n}}{n}\left(1+x^{2n}+x^{-2n}\right)\right]=\oint_{x=0}\frac{dx}{2\pi i x}\mathcal{I}_{X}(x,q_2)\period
\end{aligned}
\eeq
This coincides (up to a factor of $1/2$) with the first term in \eqref{eq:toyfinal}. So the remaining task is to take into account the parity constraint. We follow the approaches developed in \cite{Zwiebel:2011wa,Henning:2017fpj,Komatsu:2019hgc} and write a parity-projected partition function as
\beq\label{eq:parityprojectedfreemagnon}
\tilde{Z}(q_2)=\frac{1}{2}\left(\tilde{Z}_{0}(q_2)+(-1)^{J}\tilde{Z}_{\sigma}(q_2)\right)\comma
\eeq 
where $\tilde{Z}_{\sigma}(q_2)$ is the parity-weighted partition function defined by a trace over the Hilbert space with the parity operator $\sigma$ inserted (see e.g.~section 3.3 of \cite{Komatsu:2019hgc}). When $J$ is even (odd), this projects to parity even (odd) magnon states. The actions of the U(1) gauge transformation $u$ and $\sigma$ on the basis of single magnons (i.e. $X^{0}$, $X^{+}$ and $X^{-}$) are given by
\beq
u=\left(\begin{array}{ccc}1&0&0\\0&x^2&0\\0&0&x^{-2}\end{array}\right)\comma\qquad  \sigma=\left(\begin{array}{ccc}-1&0&0\\0&0&-1\\0&-1&0\end{array}\right)\period
\eeq
Then, applying the general formula (see (3.13) of \cite{Komatsu:2019hgc}), we obtain
\beq
\begin{aligned}
\tilde{Z}_{\sigma}(q_2)&=\oint \frac{dx}{2\pi i x}\exp\left[\sum_{n=1}^{\infty}\frac{q_2^{n}}{n}{\rm tr}\left((\sigma u)^{n}\right)\right]=\oint\frac{dx}{2\pi i x}\exp\left[\sum_{n=1}^{\infty}\frac{q_2^{n}}{n}\left(1+2 (-1)^{n}\right)\right]\\
&=\oint\frac{dx}{2\pi i x} \mathcal{I}_X (i,q_2)\\
&=\mathcal{I}_X(i,q_2)\period
\end{aligned}
\eeq
We can then verify that the parity-projected magnon partition function \eqref{eq:parityprojectedfreemagnon} coincides precisely with the large charge sector of the partition function \eqref{eq:toyfinal}:
\beq
\tilde{Z}(q_2)=Z_{J\gg 1}(q_2)\period
\eeq
\paragraph{Generalization.} The argument above can be readily generalized to the full partition function of free $\mathcal{N}=4$ SYM. The building block for writing the full partition function is a single-letter partition function $f_{\rm single}$, which in general depends on several different fugacities. To take the large charge limit, we separate out the contribution from the $Z$ scalar as follows
\beq\label{eq:splitsingle}
f_{\rm single}(q_0,q_1,q_2,\ldots)=q_0+f_{\rm rest}(q_1,q_2,\ldots)\comma
\eeq
where the first term on RHS is the contribution from the $Z$ scalar while $f_{\rm rest}$ denotes the contributions from the rest of the single letters. 

Before proceeding, let us make a cautionary remark here. When writing \eqref{eq:splitsingle}, we made an assumption that $f_{\rm rest}$ does not depend on the fugacity $q_0$. This is not true in a standard convention since the letters $D^{k}Z$, which are part of $f_{\rm rest}$, are charged under the same ${\rm U}(1)_R$ symmetry as $Z$. To get \eqref{eq:splitsingle}, we need to introduce an extra fugacity $q_0$, which only counts the number of $Z$ fields (excluding $D^{k}Z$). This is in fact more suitable for analyzing small fluctuations around the large charge half-BPS operators since fixing the ${\rm U}(1)_R$ charge does not guarantee that the state is a small deformation of the half-BPS state.

Using \eqref{eq:splitsingle}, we express the full partition function as
\beq
\begin{aligned}
Z (q_0,q_1,\ldots)=&\oint_{x=0} \frac{dx}{4\pi i x}\frac{(1-x^2)(1-x^{-2})}{(1-q_0)(1-q_0/x^2)(1-q_0x^2)}\mathcal{I}_{\rm rest}(x,q_1,\ldots)\comma
\end{aligned}
\eeq
where $\mathcal{I}_{\rm rest}$ is the contribution from all the letters except $Z$:
\beq
\mathcal{I}_{\rm rest}(x,q_1,\ldots)=\exp \left[\sum_{n=1}^{\infty}\frac{f_{\rm rest}(q_1^{n},q_2^{n},\ldots)}{n}(1+x^{2n}+x^{-2n})\right]\period
\eeq
We then perform the projection to the large charge sector by
\beq
\begin{aligned}
Z_{J}(q_1,\ldots)&=\oint \frac{dq_0}{2\pi i q_0^{J+1}}Z (q_0,q_1,\ldots)\\
&=\oint_{x=0} \frac{dx}{2\pi i x}\left(\frac{1}{2}-\frac{x^{-2-2J}+x^{4+2J}}{2(1+x^2)}\right)\mathcal{I}_{\rm rest}(x,q_1,\ldots)\period
\end{aligned}
\eeq
Following the argument above, we obtain the partition function for $J\gg 1$,
\beq
Z_{J\gg 1}(q_1,\ldots)=\frac{1}{2}\left(\oint_{x=0} \frac{dx}{2\pi i x} \mathcal{I}_{\rm rest}(x,q_1,\ldots)\right) +\frac{(-1)^{J}}{2}\mathcal{I}_{\rm rest}(i,q_1,\ldots)\period
\eeq

On the other hand, the free magnon partition function is given by
\beq
\tilde{Z}(q_1,\ldots)=\frac{1}{2}\left(\tilde{Z}_0(q_1,\ldots)+(-1)^{J}\tilde{Z}_{\sigma}(q_2)\right)\comma
\eeq
with
\beq
\begin{aligned}
\tilde{Z}_0 (q_1,\ldots)&=\oint_{x=0}\frac{dx}{2\pi i x}\mathcal{I}_{\rm rest}(x,q_1,\ldots)\comma\\
\tilde{Z}_{\sigma}(q_1,\ldots)&=\oint \frac{dx}{2\pi i x}\exp\left[\sum_{n=1}^{\infty}\frac{f_{\rm rest}(q_1^{n},\ldots)}{n}{\rm tr}\left((\sigma u)^{n}\right)\right]=\mathcal{I}_{\rm rest}(i,q_2)\period
\end{aligned}
\eeq
We thus conclude that the partition function in the large charge sector is in precise agreement with the partition function of magnons subject to the gauge and the parity constraints. For a special choice of fugacities, the partition function preserves a fraction of supersymmetries and can be identified with the superconformal index \cite{Romelsberger:2005eg,Kinney:2005ej}. Therefore the argument above also establishes the agreement between the superconformal index in the large charge sector and the BPS particle index of free magnons subject to the constraints.

Let us make two comments before concluding this section. First the relation between the index at large charge and the index of free BPS magnons discussed here is reminiscent of the relation between the superconformal index and the BPS particle index found in \cite{Cordova:2015nma,Cordova:2016uwk}. It would be interesting to explore a potential connection. Second, another interesting future direction is to generalize the analysis here to higher-rank gauge groups and to less supersymmetric states (see Section~\ref{sec:generalization} for further comments). In higher-rank gauge theories, there is a `moduli' of large charge operators; namely there exist multiple of operators with the same $R$-charge. It would be interesting to see if the index at large charge exhibits the wall crossing phenomena as we change the moduli parameters.

\section{Symmetry and Operator Spectrum at Leading Large $J$\label{sec:leading}}
As we saw in the previous section, the spectrum in the large charge 't Hooft limit can be interpreted as a system of ``magnons'' obeying certain dispersion relations whose interaction is mediated by a coupling that scales as $\sim 1/J$. In this section, we show that the underlying symmetry of this system is a centrally-extended $\mf{psu}(2|2)^2$ symmetry and explain how to use this symmetry to fully determine the dispersion relation at finite $\lambda_J$. We also perform explicit semiclassical analysis around the large charge state and verify the results of the symmetry analysis.

\subsection{Symmetry and its central extension in the large charge 't Hooft limit}
Understanding the symmetry is of utmost importance in various branches of theoretical physics. In relativistic QFT, particles and their interactions are classified by the representation theory of the Poincar\'{e} group, which is a symmetry preserved by the Minkowski vacuum. In the (standard) large charge limit of CFT, a convenient way to construct the large-charge EFT is to use the coset construction of Callan-Coleman-Wess-Zumino based on the symmetry preserved and/or spontaneously-broken by the large charge state. Below we will discuss the symmetry of the large charge 't Hooft limit of $\mathcal{N}=4$ SYM with emphasis on the central extension.

\subsubsection{Residual superconformal symmetry at large charge} The large charge half-BPS states can be created by inserting $(\Tr ZZ)^{J/2} $ at the origin of $\mathbb{R}^{4}$ and $(\Tr \bar{Z}\bar{Z})^{J/2} $ at infinity. By the state-operator correspondence, this configuration maps to a half-BPS large charge state on $\mR_\tau\times {\rm S}^{3}$. The bosonic symmetries preserved by both of these operators are the $\mf{so}(4)$ R-symmetry that rotates the four real scalars, which are not in $Z$ or $\bar{Z}$, and the $\mf{so}(4)$ Lorentz (Euclidean rotation) symmetry around the origin. The generators of these symmetries in the chiral and anti-chiral spinor representations of $\mf{so}(4)$
are 
\beq
\begin{aligned}
\text{$\mf{so}(4)$ R-symmetry:} \quad &R^{a}{}_{b}\comma\quad \dot{R}^{\dot{a}}{}_{\dot{b}} \qquad (a,b, \dot{a},\dot{b}=1,2)\comma\\
\text{$\mf{so}(4)$ Lorentz symmetry:} \quad &L^{\alpha}{}_{\beta}\comma\quad \dot{L}^{\dot{\alpha}}{}_{\dot{\beta}} \qquad (\alpha,\beta, \dot{\alpha},\dot{\beta}=1,2)\,.
\label{twoSO4}
\end{aligned}
\eeq
 In addition, since these operators satisfy the BPS condition $\Delta=|J|$, they are invariant under the combination given by \beq\label{eq:centralcharged-j}
 C\equiv \frac{\hat D-\hat{J}}{2}\,,
 \eeq
  where $\hat D$ is the dilatation operator around the origin and $\hat{J}$ generates the ${\rm U}(1)_R$ symmetry under which $Z$ and $\bar{Z}$ have charge $J=1$ and $J=-1$ respectively (and $X,Y$ are uncharged). The configuration is invariant also under half of the supercharges which we denote as,
 \beq \label{eq:susygens}
 Q^{a}{}_{\alpha}\comma \quad S^{\alpha}{}_{a}\comma\qquad \dot{Q}^{\dot{a}}{}_{\dot{\alpha}}\comma \quad \dot{S}^{\dot{\alpha}}{}_{\dot{a}}\period
 \eeq
 where the chiral supercharges come from restricting the $\cN=4$ generators $(\cQ^A_\A,\cS^\A_A)$ to $A=3,4$, and the antichiral supercharges from the $\cN=4$ generators $(\dot\cQ_{A\dot \A},\dot\cS^{A \dot\A})$ with $A=1,2$.
 
The full symmetry  of the setup is given by the following half-BPS subalgebra of the $\cN=4$ superconformal algebra,\footnote{The bosonic subalgebra of $\mathfrak{psu}(2|2)\times\mathfrak{psu}(2|2)$ is 
\beq
\underbrace{\mathfrak{su}(2)_a\times \mathfrak{su}(2)_{\dot{a}}}_{R-\text{symmetry}} \times \underbrace{\mathfrak{su}(2)_{\alpha}\times \mathfrak{su}(2)_{\dot{\alpha}}}_{\text{Lorentz}}\,.\nn
\eeq
where the undotted algebras belong to the first $\mathfrak{psu}(2|2)$ while the dotted ones belong to the second $\mathfrak{psu}(2|2)$. 
} 
\ie 
\left( \mathfrak{psu}(2|2)\times\mathfrak{psu}(2|2)\right)\ltimes \mathbb{R}\,,
\label{centralsym}
\fe 
 with the last abelian factor being a central extension generated by $C$ in \eqref{eq:centralcharged-j}. The two $\mf{psu}(2|2)$ factors are generated by chiral (undotted) and anti-chiral (dotted) elements in \eqref{twoSO4} and \eqref{eq:susygens} respectively.

Explicitly, the nontrivial part of the (anti-)commutation relations  for the chiral $\mathfrak{psu}(2|2)$ reads
\beq\label{eq:anticompsu}
\begin{aligned}
_{}[R^{a}{}_{b},G^{c} ] &= - \delta^{c}_{b} G^{a} +\frac{1}{2} \delta^{a}_{b} G^{c}\comma\quad [R^{a}{}_{b},G_{c}] = \delta^{a}_{c} G_{b} -\frac{1}{2} \delta^{a}_{b} G_{c}\comma \\
 [L^{\alpha}{}_{\beta},G^{\gamma}] &= \delta^{\gamma}_{\beta} G^{\alpha} -\frac{1}{2} \delta^{\alpha}_{\beta} G^{\gamma}\,, \quad
[L^{\alpha}{}_{\beta},G_{\gamma}] = -\delta^{\alpha}_{\gamma} G_{\beta} +\frac{1}{2} \delta^{\alpha}_{\beta} G_{\gamma} \comma\\
[\hat D-\hat{J},Q^{a}{}_{\alpha}]&= [\hat D-\hat{J},S^{\alpha}{}_{a}]=0 \,, \\
\{S^{\alpha}{}_{a},Q^{b}{}_{\beta} \} &= \delta^{b}_{a} L^{\alpha}{}_{\beta}+\delta^{\alpha}_{\beta} R^{b}{}_{a}+ \delta^{b}_{a}\delta^{\alpha}_{\beta} \,(\hat D-\hat{J})/2\period
\end{aligned}
\eeq
The (anti-)commutation relations for the anti-chiral $\mathfrak{psu}(2|2)$ are identical. Here $G^{c}$ and $G_{c}$ represent any generators with a $R$-symmetry index while $G^{\gamma}$ and $G_{\gamma}$ represent any generators with a Lorentz index.

For later convenience, let us also write down the commutators between fermionic charges and the ${\rm U}(1)_R$ symmetry $\hat{J}$,
\beq
[\hat{J},Q^{a}{}_{\alpha}]=+\frac{1}{2}Q^{a}{}_{\alpha}\comma\qquad [\hat{J},S^{\alpha}{}_{a}]=-\frac{1}{2}S^{\alpha}{}_{a}\period
\eeq
 
 \subsubsection{The maximal central extension and its representations} \label{sec:centralext}
 The symmetry \eqref{centralsym} provides certain constraints on the dynamics of excitations. However, the constraints obtained in this way are not strong enough. Namely it does not determine the dependence on the coupling constant $\lambda_J$. This parallels the fact that the Poincar\'{e} symmetry in flat space only allows us to express the S-matrix  in terms of Mandelstam variables and  is ignorant about its dependence on the coupling constant. Nonetheless, as we will see below, the large charge half-BPS states are in fact invariant under a further central extension of \eqref{centralsym}, namely the {\it universal (maximal) central extension} \cite{univcentral},\footnote{The Lie superalgebra $\mf{psu}(2|2)$ admits a maximal three-dimensional central extension known as the universal central extension in \cite{univcentral} (see also \cite{Beisert:2005tm}). The fully centrally extended algebra is the contraction of the Lie superalgebra $D(2,1;\lambda)$ as $\lambda\to 0$ (which can be thought of a deformation of $\mf{osp}(4|2)$ at $\lambda=1$).}
  \ie
 \left( \mathfrak{psu}(2|2)\times\mathfrak{psu}(2|2)\right)\ltimes \mathbb{R}^3
 \label{univcentral}
 \fe 
 which will allow us to determine the full dependence on $\lambda_J$ at large $J$.
 
 \paragraph{Central extension.}The universal central extension \eqref{univcentral} of $\mathfrak{psu}(2|2)\times\mathfrak{psu}(2|2)$ was discussed first in the study of the spin-chain description of single-trace operators at large $N$ \cite{Beisert:2005tm}. The key insight there was to include field-dependent gauge transformations as part of the symmetry. Since the (non-abelian) gauge transformation depends on the gauge coupling $g_{\rm YM}$, this enabled non-perturbative determination of the dispersion relation of ``magnons'' (i.e.~excitations on the spin chain) and the matrix structure of magnon $S$-matrices. Below, we argue that the same symmetry is present in the large charge 't Hooft limit\footnote{For the origin of the centrally extended symmetry in the large $N$ limit, see e.g.~the review \cite{Komatsu:2017buu}.} and explain how it constrains the spectrum around the large charge state.
 
To understand the origin of the symmetry, we recall that the superconformal transformations of the fields in the $\mathcal{N}=4$ SYM
can be written compactly in the following way using the 10d SYM notation (see \cite{Berkovits:1993hx,Pestun:2007rz}),\footnote{Here the convention for the SYM action is such that ${1\over g_{\rm YM}^2}$ appears as an overall factor in the action. To obtain the supersymmetry transformation rules for the canonically normalized fields, we rescale $A_M \to  g_{\rm YM} A_M $ and $\Psi \to g_{\rm YM} \Psi$.}
\ie 
\D A_M=\ve \Gamma_M \Psi\,,\quad \D \Psi ={1\over 2} F_{MN} \Gamma^{MN}\ve+{1\over  2}\Gamma_{\m I} \Phi^I \nabla^\m \ve\,,
\label{N4SUSY}
\fe
where $\ve$ is a conformal Killing spinor (written as a 10d chiral spinor) that parametrize the 32 supercharges and $\Gamma_M$ denotes the 10d gamma matrices in the chiral basis. 
The 10d indices $M,N$ split into 4d spacetime indices $\m,\n$ and R-symmetry indices $I,J$. Correspondingly the 4d gauge fields and scalars are packaged together in $A_M=(A_\m,\Phi_I)$ and $F_{MN}=[D_M,D_N]$ with $D\equiv d+A$. The three complex adjoint scalars $X,Y,Z$ introduced previously are each made of the two out of the six real scalars $\Phi_i$ (see \eqref{Phi2Z}).

 From \eqref{N4SUSY}, one can show that the consecutive action of two (different) supercharges on a fermion leads to a term of the schematic form
 \beq
 \delta^{2}\Psi\sim [\Phi,\Psi]\,,
 \eeq
which represents a field-dependent gauge transformation and therefore carries information on the interaction term of the Lagrangian. Working out this relation more explicitly for the supercharges \eqref{eq:susygens} belonging to $\mathfrak{psu}(2|2)\times \mathfrak{psu}(2|2)$, we obtain the following central extension,
 \beq\label{eq:centralext}
 \{Q^{\alpha}_{a},Q^{\beta}_{b}\}\Psi = \{\dot{Q}^{\dot{\alpha}}_{\dot{a}},\dot{Q}^{\dot{\beta}}_{\dot{b}}\}\Psi =\epsilon_{ab}\epsilon^{\alpha\beta} [\mathcal{Z},\Psi]\comma
 \eeq
 with $\mathcal{Z}=\sqrt{2}Z$.
 In what follows, we denote the action of this field-dependent gauge transformation by\footnote{The action of the supercharges on a scalar field is also given by the same equation (\ref{eq:centralext}).} 
 $P\cdot \bullet \equiv [\mathcal{Z},\bullet]$.
 
Since $P$ is a gauge transformation, it acts trivially on gauge-invariant states. However, in the presence of large charge states, the SYM scalar fields acquire nontrivial profiles, whose presence breaks the gauge invariance. As a result, ``magnons'' around large charge states are nontrivially charged\fn{Of course, physical observables are always gauge invariant and are given in terms of gauge-invariant combinations of these excitations as we will see below.} with respect to $P$. As we see later in Section~\ref{sec:semicl}, the scalar expectation value induced by the large charge state on $\mR_\tau \times {\rm S}^3$ is given by
\beq\label{eq:classicalgZ}
\langle \mathcal{Z}\rangle=\sqrt{2}Z_{\rm cl} =\left(\begin{array}{cc}\frac{g_{\rm YM}\sqrt{J}}{2\pi} e^{i\varphi}&0\\0&-\frac{g_{\rm YM}\sqrt{J}}{2\pi}e^{i\varphi}\end{array}\right)\comma
\eeq
where $e^{i\varphi}$ is a (time-dependent) phase which does not affect the spectrum. Thus the upper and lower off-diagonal components of the SYM fields have charges $\pm g_{\rm YM}\sqrt{J}e^{i\varphi}/\pi$ under $P$ respectively while the diagonal components are uncharged\footnote{Recall the definition of $g$ as in $g=\sqrt{\lambda_J}/(4\pi)=g_{\rm YM}\sqrt{J}/(4\pi)$.}:
\begin{align}
&M=\left(\begin{array}{cc}m^{0}&m^{+}\\m^{-}&-m^{0}\end{array}\right)\comma\label{eq:su2matrices}\\
&P\cdot m^{\pm}=\pm 2g e^{i\varphi}\, m^{\pm} \comma\qquad P\cdot m^{0}=0\period\label{eq:actionofP}
\end{align}

Now, on $\mR_\tau\times {\rm S}^3$, the supersymmetry generators $Q$'s and the superconformal generators $S$'s are related by Hermitian conjugation, which  implies that the anti-commutator of the superconformal generators also get centrally extended in the following way:
\beq\label{eq:centralext2}
\{S_{\alpha}^{a},S_{\beta}^{b}\}\Psi = \{\dot{S}_{\dot{\alpha}}^{\dot{a}},\dot{S}_{\dot{\beta}}^{\dot{b}}\}\Psi =\epsilon^{ab}\epsilon_{\alpha\beta} [\mathcal{Z}^{-1},\Psi]\period
\eeq
The transformation above involves an inverse of the field $Z$, which might seem unusual at first sight. However, such a transformation is possible once we include quantum corrections. In fact, this relation was originally found in planar $\mathcal{N}=4$ SYM through the two-loop analysis of superconformal generators \cite{Beisert:2003ys}. Here we do not have such a direct field-theoretical derivation; instead we simply assume its presence, motivated by the closure of the algebra. In what follows, we denote this action by 
$K\cdot \bullet=[\mathcal{Z}^{-1},\bullet]$. The action of $K$ on fundamental fields can be deduced from the Hermiticity of the algebra. Namely by taking a Hermitian conjugate of \eqref{eq:actionofP}, we obtain
\beq\label{eq:actionofK}
K\cdot m^{\pm}=\pm 2g e^{-i\varphi}m^{\pm}\comma\qquad K\cdot m^{0}=0\period
\eeq

Together with the  superconformal subalgebra \eqref{centralsym} discussed earlier, these generators constitute the maximally-centrally-extended  $\mathfrak{psu}(2|2)^2$ symmetry \eqref{univcentral}, which has the following new nontrivial anti-commutation relations for fermionic generators in the chiral $\mf{psu}(2|2)$ factor as compared to \eqref{eq:anticompsu},
\beq
\begin{aligned}
\{ Q^{a}{}_{\alpha},Q^{b}{}_{\beta}\} &= \epsilon^{ab} \epsilon_{\alpha \beta} P\,,\quad \{S^{\alpha}{}_{a},S^{\beta}{}_{b}\} = \epsilon_{ab}\epsilon^{\alpha \beta} K\,,
\end{aligned}
\eeq
and similarly for the anti-chiral $\mf{psu}(2|2)$ factor.

\paragraph{Representations and short multiplets.} As is the case with the Poincar\'{e} symmetry in relativistic QFT, excitations around the large charge state are classified by the representations of the symmetry group. In the present case, the relevant symmetry is the maximally-centrally-extended $\mathfrak{psu}(2|2)^2$. In flat space, the central extension of supersymmetry allows for the BPS representations, which are short representations of the extended super-Poincar\'{e} algebra. Two important features of such representations are\footnote{See lecture notes from G.W. Moore on BPS states and wall-crossing in 4d $\cN=2$ QFTs available at \href{https://static.ias.edu/pitp/archive/2010files/Moore_LectureNotes.rev3.pdf}{here}.}
\begin{enumerate}
\item[{\bf 1.}] In order for a particle to be in a short representation, it needs to satisfy a special relation between mass and charge, namely the BPS condition $M=|Z|$. This allows one to determine the mass of the particle purely from its quantized charge.
\item[{\bf 2.}] The dynamics of BPS particles is severely constrained. For instance, the only way in which a particle stops being BPS is through a multiplet recombination process, in which several short representations combine into a long (generic) representation. This property allows us to define a variety of BPS indices, which are invariant under the continuous deformation of the theory such as a coupling constant.
\end{enumerate}
As we see in this section and the next,  these properties have natural counterparts for the maximally-centrally-extended $\mathfrak{psu}(2|2)^2$ symmetry.

The short representations of the maximally-centrally-extended $\mathfrak{psu}(2|2)$ are classified and explained in \cite{Beisert:2006qh}. They are labelled by two integers $m$ and $n$ and the three central charges satisfy the BPS condition $C^2-PK=(n+m+1)^2/4$. Among them, the ones that play an important role in $\mathcal{N}=4$ SYM are 
\beq
\begin{aligned}
\langle m,n\rangle =\langle m,0\rangle&:\qquad  \text{symmetric representations},\\
\langle m,n\rangle=\langle 0,n\rangle &:\qquad  \text{anti-symmetric representations}.
\end{aligned}
\eeq
The most special case $\langle m,n\rangle=\langle 0,0\rangle$, or more precisely $\langle 0,0\rangle\otimes \langle 0,0\rangle$ since we have two copies of $\mathfrak{psu}(2|2)$, corresponds to inserting fundamental fields (or letters) of $\mathcal{N}=4$ SYM into the large charge state.  To make manifest the product-algebra structure, we often express them as
 \beq
 \mathcal{X}^{A\dot{A}}=\chi^{A}\dot{\chi}^{\dot{A}}\comma
 \eeq
 where $\chi$ and $\dot{\chi}$ are the fundamental representations of the left and the right $\mathfrak{psu}(2|2)$,
 \beq\label{chichi}
 \chi^{A}=(\varphi^{1},\varphi^{2},\psi^{1},\psi^{2})\comma\qquad \dot{\chi}^{\dot{A}}=(\dot{\varphi}^{1},\dot{\varphi}^{2},\dot{\psi}^{1},\dot{\psi^{2}})\period
 \eeq
 The explicit relation between $\mathcal{X}^{A\dot{A}}$ and fundamental fields of $\mathcal{N}=4$ SYM is given by
 \beq\label{eq:fundamentalletter}
 \begin{aligned}
 \varphi^{1}\dot{\varphi}^{1}=X\comma\quad \varphi^{1}\dot{\varphi}^{2}=Y\comma\quad \varphi^{2}\dot{\varphi}^{1}=\bar{Y}\comma\quad \varphi^2\dot{\varphi}^2=-\bar{X}\comma\\
 \psi^{\alpha}\dot{\psi}^{\dot{\alpha}}=D^{\alpha\dot{\alpha}}Z\comma\qquad \psi^{\alpha}\dot{\varphi}^{\dot{a}}\comma \varphi^{a}\dot{\psi}^{\dot{\alpha}}:\text{fermion}\period
 \end{aligned}
 \eeq
 Here $D^{\alpha\dot{\alpha}}$ is a covariant derivative in the spinorial notation. Note also that here we are using an ``operator notation'' rather than a ``state notation'' to label the excitations; namely the ``excitation $X$'' means an insertion of the letter $X$ into the large charge operator, and corresponds to exciting an $s$-wave of $X$ on ${\rm S}^{3}$. Similarly, the ``excitation $D^{\alpha\dot{\alpha}}Z$'' means an insertion of $D^{\alpha\dot{\alpha}}Z$ into the large charge operator, which corresponds to exciting $Z$ with a unit angular momentum on ${\rm S}^3$.

In the planar limit, the other representations show up as the bound states\fn{The symmetric bound states correspond to bound states in the spin chain while the antisymmetric bound states correspond to bound states in the so-called mirror channel. For more explanation, see e.g. the review \cite{Arutyunov:2009ga}.} of the $\mathcal{N}=4$ SYM spin chain. In the current context, we need to know how the excitations around the large charge ``vacuum'' decompose into the irreps of the maximally-centrally-extended $\mathfrak{psu}(2|2)^2$. There are several ways to see this; either by decomposing the partition function computed in Section~\ref{sec:partition} or by expanding the Lagrangian around the large charge vacuum as we will do in Section~\ref{sec:semicl}. The result of such analyses is that the excitations decompose into anti-symmetric representations with a unit multiplicity:
\beq
(\text{excitations at large charge}):\quad  \bigoplus_{n=0}^{\infty}\,\, \langle 0,n\rangle \otimes \langle 0,n\rangle
\eeq
The states with $n=0$ correspond to fundamental letters discussed above, while states $n\geq 1$ correspond to  excitations with higher angular momenta on ${\rm S}^3$, such as the insertion of $D^{k}X$ to the large charge operator. All these representations are BPS representations  and satisfy the BPS condition
\beq\label{eq:BPScondition}
(\hat D-\hat{J})^2-4P K=(n+1)^2 \comma\qquad n\in \mathbb{N}\period
\eeq

\paragraph{Constraining the dynamics.} We are now in a position to use the symmetry and the BPS condition to constrain the dynamics of excitations. Let us first discuss the $\langle 0,0\rangle$ representation corresponding to fundamental fields. The action of bosonic generators on the $\langle 0,0\rangle$ follows directly from the superconformal subalgebra \eqref{centralsym},
\beq
\begin{aligned}
R^{a}{}_{b}\,|\varphi_{c}\rangle_{J} &= \delta^{a}_{c}|\varphi_{b}\rangle_{J} -\frac{1}{2}\delta^{a}_{b} |\varphi_{c}\rangle_{J}\,,\\ 
L^{\alpha}{}_{\beta}\,|\psi_{\gamma}\rangle_{J} &= \delta^{\alpha}_{\gamma}|\psi_{\beta}\rangle_{J} -\frac{1}{2}\delta^{\alpha}_{\beta} |\psi_{\gamma}\rangle_{J}\,,\\
\hat{J} |\psi_{\gamma}\rangle_{J} &= \left(J+\frac{1}{2} \right) |\psi_{\gamma}\rangle_{J}\\
\quad  \hat{J} |\varphi_{a}\rangle_{J} &=J  |\varphi_{a}\rangle_{J}
\end{aligned}
\eeq
Here the subscript in $|\bullet \rangle_{J}$ indicates the background ${\rm U}(1)_R$ charge coming from the BPS large charge operator. For the action of the fermionic generators, we write down the most general ansatz consistent with the bosonic symmetry,
\beq
\begin{aligned} \label{actions}
Q^{a}{}_{\alpha}|\varphi_{c}\rangle_{J} &={\bf a} \delta^{a}_{c}\, |\psi_{\alpha}\rangle_{J}\,,\\
Q^{a}{}_{\alpha}|\psi_{\gamma}\rangle_{J} &={\bf b}\, \epsilon^{a b}\epsilon_{\alpha \gamma} |\varphi_{b}\rangle_{J+1}\,, \\
S^{\alpha}{}_{a}|\varphi_{c}\rangle_{J} &={\bf c}\, \epsilon_{a c}\epsilon^{\alpha \gamma}\, |\psi_{\gamma}\rangle_{J-1}\,, \\
 S^{\alpha}{}_{a}|\psi^{m}_{\gamma}\rangle_{J} &={\bf d}\, \delta^{\alpha}_{\gamma} |\varphi_{a}\rangle_{J}\,. 
 \end{aligned}
\eeq
Since the central charges commute with other generators, their actions are the same for all the states in a given multiplet. We denote them as,
\beq
(\hat D-\hat{J})|\chi\rangle_{J}={\bf h}|\chi\rangle_{J}\comma\qquad P|\chi\rangle_{J}={\bf p}|\chi\rangle_{J+1}\comma\qquad K|\chi\rangle_{J}={\bf k}|\chi\rangle_{J-1}\,,
\eeq
where $\bf h,p,k$ are in principle non-trivial functions of the large charge 't Hooft coupling $\lambda_{J}$. To determine their  $\lambda_{J}$ dependence, we first impose the (anti-)commutation relations of the generators to relate them. The result of this straightforward exercise (see e.g. \cite{Beisert:2005tm}) is,\fn{The relation on the second line is the BPS condition. See \cite{Beisert:2005tm} for more details.}
\beq
\begin{aligned}
&{\bf h}={\bf a}{\bf d}+{\bf b}{\bf c}\comma\qquad {\bf p}={\bf a}{\bf b}\comma\qquad {\bf k}={\bf c}{\bf d}\comma\\
&({\bf a}{\bf d}-{\bf b}{\bf c})^2=1\period
\end{aligned}
\eeq
Inverting the above relations, we can then express  ${\bf h}$ in terms of ${\bf p}$ and ${\bf k}$ as follows,
\beq
{\bf h}=\sqrt{1+4\, {\bf p}\,{\bf k}}\period
\eeq
We now use the fact that the central charges $P$ and $K$ are field-dependent gauge transformations and their actions are given by \eqref{eq:actionofP} and \eqref{eq:actionofK}. This fixes the energy of the excitations to be
\beq\label{eq:energy1}
{\bf h}_{0}=1\comma\qquad {\bf h}_{\pm}=\sqrt{1+16g^2}\comma
\eeq
where the subscripts $0$ and $\pm$ signify the diagonal and off-diagonal components of the $\mathfrak{su}(2)$ matrix (see \eqref{eq:su2matrices}). This shows that the energy of the diagonal components is protected while the energy of the off-diagonal components depend nontrivially on the coupling constant. In what follows, we often add superscripts to excitations to distinguish the diagonal and off-diagonal components e.g.~$\chi^{0}$ and $\chi^{\pm}$.

Performing a similar analysis for the other representations following \cite{Beisert:2006qh}, we find their energies to be
\beq\label{eq:energy2}
\langle 0,n-1\rangle \otimes \langle 0,n-1\rangle: \qquad {\bf h}_{0}=n\comma\quad {\bf h}_{\pm}=\sqrt{n^2+16g^2}\period
\eeq
Here again, the subscripts are for distinguishing the diagonal and off-diagonal components of the adjoint fields. As mentioned in the introduction, \eqref{eq:energy1} and \eqref{eq:energy2} take the same form as the magnon dispersion relation. This is simply because both are consequences of the representations of the maximally-centrally-extended $\mathfrak{psu}(2|2)^2$.
\subsubsection{Gauge invariant operators and comparison with data}\label{sec:states}
So far we have focused on the spectrum of individual excitations around the large charge vacuum. To discuss the spectrum of operators, we need to make sure that the resulting state is gauge-invariant, since each excitation itself is not, as is clear from the fact that it transforms nontrivially under the action of the central charges $P$ and $K$, which are field-dependent gauge transformations.

To build gauge-invariant operators, we have to add particles in such a way that the total central charges $P$ and $K$ vanish:
\beq
P|\Psi\rangle_J= K|\Psi \rangle_J=0\period
\eeq
 This is the condition imposed also for the spectrum in the planer limit (in which case it is called the zero-momentum condition see e.g. \cite{Beisert:2005tm}). Roughly speaking, this condition comes from the diagonal ${\rm U}(1)$ gauge invariance which is unbroken in the presence of the large charge semiclassical state. This is however not sufficient for the invariance under the full ${\rm SU}(2)$ gauge group. One way to achieve this is to impose the invariance under the transformation\footnote{Combined with the diagonal ${\rm U}(1)$ gauge transformation mentioned above, the $\sigma_2$-transformation (which implements the $\mZ_2$ Weyl group reflection) generates any ${\rm SU}(2)$ transformation and it is therefore enough to impose these two conditions.} generated by $\sigma_2$, which acts on a field in the adjoint representation as
 \beq
 \sigma_2 \chi \sigma_2 =-\chi^{T}\quad  (=:\mathfrak{P}\cdot \chi)\comma
 \label{parity}
 \eeq
 where $\chi^{T}$ is the transposition of the SU(2) matrix $\chi$. 
 The advantage of using this transformation is that its action on the excitations around the large-charge vacuum is simple; it simply multiplies $(-1)$ to all the excitations and the background charges $Z$, and swaps the excitations $\chi^{+}$ with $\chi^{-}$. This is precisely the ``parity'' transformation discussed in Section~\ref{su3parity}.
  
As a result, the rules for constructing gauge-invariant operators can be summarized as follows:
\begin{enumerate}
\item The total U(1) charge has to be zero which implies that the number of upper triangular modes $\chi^{+}$ and lower triangular modes $\chi^{-}$ needs to match.
\item For even $J$, the state has to be symmetric under the parity transformation of all the fields, $\chi^{\pm}\mapsto -\chi^{\mp}$ and $\chi^{0}\mapsto -\chi^{0}$. 
\item For odd $J$, the state has to be anti-symmetric under the parity transformation.
\end{enumerate}

\begin{table}
\centering
\renewcommand{\arraystretch}{1.32}
\scriptsize
$\begin{array}{|c|c|c|c|c|c|c|}\hline
\rm{sector}& {\#\, \text{of excitations}} & J& \rm{state}&E\\ \hline
\multirow{7}{*}{SU(2)} &\multirow{2}{*}{2} & \multirow{2}{*}{\text{even}}& | \phi_1^{0} \phi_1^{0} \rangle_{J}   &2 \\
&&  & |\phi_1^{+} \phi_1^{-} \rangle_{J}   & 2\sqrt{1+16g^2} \\
\cline{2-5}
&  \multirow{2}{*}{3}& \multirow{2}{*}{\text{odd}} & |\phi_1^{0}\phi_1^{0}\phi_1^{0} \rangle_{J}   & 3 \\
 &&&|\phi_1^{+}\phi_1^{0}\phi_1^{-} \rangle_{J}   & 1+2\sqrt{1+16g^2} \\
\cline{2-5}
& \multirow{3}{*}{4}&   \multirow{3}{*}{\text{even}} & |\phi_1^{0}\phi_1^{0}\phi_1^{0}\phi_1^{0} \rangle_{J}   & 4 \\
 &&  &|\phi_1^{+}\phi_1^{0}\phi_1^{0}\phi_1^{-} \rangle_{J}   & 2+2\sqrt{1+16g^2} \\
&&   & |\phi_1^{+}\phi_1^{+}\phi_1^{-}\phi_1^{-} \rangle_{J}   & 4\sqrt{1+16g^2} \\
\hline 
\multirow{31}{*}{SU(3)} &\multirow{4}{*}{2} &   \multirow{2}{*}{\text{even}}&| \phi_1^{0} \phi_2^{0} \rangle_{J}   &2 \\
& &  &\frac{1}{\sqrt{2}}\left( |\phi_1^{+} \phi_2^{-} \rangle_{J} + |\phi_2^{+} \phi_1^{-} \rangle_{J}\right)  & 2\sqrt{1+16g^2} \\
\cline{3-5}  
&&\multirow{1}{*}{\text{odd}}& \frac{1}{\sqrt{2}}\left( |\phi_1^{+} \phi_2^{-} \rangle_{J} - |\phi_2^{+} \phi_1^{-} \rangle_{J}\right)  & 2\sqrt{1+16g^2} \\
\cline{2-5}
&  \multirow{10}{*}{3}&  \multirow{2}{*}{\text{even}} & \frac{1}{\sqrt{2}}\left( |\phi_1^{+}\phi_1^{0}\phi_2^{-} \rangle_{J} -|\phi_1^{-}\phi_1^{0}\phi_2^{+} \rangle_{J}  \right)  & 1+2\sqrt{1+16g^2} \\
 &&  & \frac{1}{\sqrt{2}}\left( |\phi_1^{+}\phi_2^{0}\phi_2^{-} \rangle_{J} -|\phi_1^{-}\phi_2^{0}\phi_2^{+} \rangle_{J}  \right)   & 1+2\sqrt{1+16g^2} \\
\cline{3-5}
&  &  \multirow{6}{*}{\text{odd}} & |\phi_1^{0}\phi_1^{0}\phi_2^{0} \rangle_{J}   & 3 \\
&&& |\phi_1^{0}\phi_2^{0}\phi_2^{0} \rangle_{J}   & 3\\
 &&  & \frac{1}{\sqrt{2}}\left( |\phi_1^{+}\phi_1^{0}\phi_2^{-} \rangle_{J} +|\phi_1^{-}\phi_1^{0}\phi_2^{+} \rangle_{J}  \right)   & 1+2\sqrt{1+16g^2} \\
  &&  & \frac{1}{\sqrt{2}}\left( |\phi_1^{+}\phi_2^{0}\phi_2^{-} \rangle_{J} +|\phi_1^{-}\phi_2^{0}\phi_2^{+} \rangle_{J}  \right)   & 1+2\sqrt{1+16g^2} \\
    &&  &  |\phi_1^{+}\phi_1^{-}\phi_2^{0} \rangle_{J} & 1+2\sqrt{1+16g^2} \\
    &&  &  |\phi_1^{0}\phi_2^{-}\phi_2^{+} \rangle_{J} & 1+2\sqrt{1+16g^2} \\
\cline{2-5}
& \multirow{26}{*}{4}& \multirow{14}{*}{\text{even}} & |\phi_1^{0}\phi_1^{0}\phi_2^{0}\phi_2^{0} \rangle_{J}   & 4 \\
&& & |\phi_1^{0}\phi_1^{0}\phi_1^{0}\phi_2^{0} \rangle_{J}   & 4 \\
&&& |\phi_1^{0}\phi_2^{0}\phi_2^{0}\phi_2^{0} \rangle_{J}   & 4 \\
&&& |\phi_1^{0}\phi_1^{0}\phi_2^{+}\phi_2^{-} \rangle_{J}   &  2+2\sqrt{1+16g^2} \\
&&& |\phi_1^{0}\phi_2^{0}\phi_2^{+}\phi_2^{-} \rangle_{J}   &  2+2\sqrt{1+16g^2} \\
&&& |\phi_1^{0}\phi_1^{+}\phi_1^{-}\phi_2^{0} \rangle_{J}   &  2+2\sqrt{1+16g^2} \\
&&& |\phi_1^{+}\phi_1^{-}\phi_2^{0}\phi_2^{0} \rangle_{J}   &  2+2\sqrt{1+16g^2} \\
&&  & \frac{1}{\sqrt{2}} \left( |\phi_1^{+}\phi_1^{0}\phi_2^{-}\phi_2^{0} \rangle_{J} + |\phi_1^{-}\phi_1^{0}\phi_2^{+}\phi_2^{0} \rangle_{J}  \right) & 2+2\sqrt{1+16g^2} \\
&&  & \frac{1}{\sqrt{2}} \left( |\phi_1^{+}\phi_1^{0}\phi_1^{0}\phi_2^{-} \rangle_{J} + |\phi_1^{-}\phi_1^{0}\phi_1^{0}\phi_2^{+} \rangle_{J}   \right) & 2+2\sqrt{1+16g^2} \\
&&  & \frac{1}{\sqrt{2}} \left( |\phi_1^{+}\phi_2^{0}\phi_2^{0}\phi_2^{-} \rangle_{J} + |\phi_1^{-}\phi_2^{0}\phi_2^{0}\phi_2^{+} \rangle_{J}   \right) & 2+2\sqrt{1+16g^2} \\
 &&  &\frac{1}{\sqrt{2}} \left( |\phi_1^{+}\phi_1^{+}\phi_2^{-}\phi_2^{-} \rangle_{J}+|\phi_1^{-}\phi_1^{-}\phi_2^{+}\phi_2^{+} \rangle_{J} \right)   & 4\sqrt{1+16g^2} \\
&&  & |\phi_1^{+}\phi_1^{-}\phi_2^{+}\phi_2^{-} \rangle_{J}   & 4\sqrt{1+16g^2} \\
&&  & \frac{1}{\sqrt{2}} \left( |\phi_1^{+}\phi_1^{+}\phi_1^{-}\phi_2^{-} \rangle_{J} + |\phi_1^{-}\phi_1^{-}\phi_1^{+}\phi_2^{+} \rangle_{J}   \right) & 4\sqrt{1+16g^2}  \\
&&  & \frac{1}{\sqrt{2}} \left( |\phi_1^{+}\phi_2^{+}\phi_2^{-}\phi_2^{-} \rangle_{J} + |\phi_1^{-}\phi_2^{-}\phi_2^{+}\phi_2^{+} \rangle_{J}   \right) & 4\sqrt{1+16g^2}  \\
\cline{3-5}
&& \multirow{8}{*}{\text{odd}} &  \frac{1}{\sqrt{2}} \left( |\phi_1^{+}\phi_1^{0}\phi_2^{-}\phi_2^{0} \rangle_{J} - |\phi_1^{-}\phi_1^{0}\phi_2^{+}\phi_2^{0} \rangle_{J}  \right) & 2+2\sqrt{1+16g^2} \\
&&  & \frac{1}{\sqrt{2}} \left( |\phi_1^{+}\phi_1^{0}\phi_1^{0}\phi_2^{-} \rangle_{J} - |\phi_1^{-}\phi_1^{0}\phi_1^{0}\phi_2^{+} \rangle_{J}   \right) & 2+2\sqrt{1+16g^2} \\
&&  & \frac{1}{\sqrt{2}} \left( |\phi_1^{+}\phi_2^{0}\phi_2^{0}\phi_2^{-} \rangle_{J} - |\phi_1^{-}\phi_2^{0}\phi_2^{0}\phi_2^{+} \rangle_{J}   \right) & 2+2\sqrt{1+16g^2} \\
 &&  &\frac{1}{\sqrt{2}} \left( |\phi_1^{+}\phi_1^{+}\phi_2^{-}\phi_2^{-} \rangle_{J}-|\phi_1^{-}\phi_1^{-}\phi_2^{+}\phi_2^{+} \rangle_{J} \right)   & 4\sqrt{1+16g^2} \\
  &&  &\frac{1}{\sqrt{2}} \left( |\phi_1^{+}\phi_1^{+}\phi_1^{-}\phi_2^{-} \rangle_{J}-|\phi_1^{-}\phi_1^{-}\phi_1^{+}\phi_2^{+} \rangle_{J} \right)   & 4\sqrt{1+16g^2} \\
   &&  &\frac{1}{\sqrt{2}} \left( |\phi_1^{+}\phi_2^{+}\phi_2^{-}\phi_2^{-} \rangle_{J}-|\phi_1^{-}\phi_2^{-}\phi_2^{+}\phi_2^{+} \rangle_{J} \right)   & 4\sqrt{1+16g^2} \\
\hline
\end{array}$
\caption{This table presents the spectrum of eigenstates for the dilatation operator, showcasing the dependence on the large charge 't Hooft coupling at leading order in the large charge expansion. The analysis focuses on the sector involving up to two complex scalars.}
\label{tabstates}
\end{table}
Using these rules, we recover the spectrum of some simple operators studied in the previous section. For example, let us consider the simplest states in the SU(2) sector with two excitations. By the rules above, $J$ needs to be even and the only states one can construct  are the following with the corresponding energies 
\beq
| \phi_1^{0} \phi_1^{0} \rangle_{J}\quad \text{with}\quad E=2\,,\quad |\phi_1^{+} \phi_1^{-} \rangle_{J}\quad \text{with}\quad E= 2\sqrt{1+16g^2}\,.
\eeq
We display in Table~\ref{tabstates}, a list of some states up to four excitations in the SU(2) and SU(3) scalar sectors. 
We recover the spectrum of the last section by expanding the results of this section for small $\lambda_{J}$ (equivalently small $g$).

A similar analysis can be performed also for the SL(2) sector once we include the higher representations ($\langle 0,n\rangle$). As expected, the results are in perfect agreement with the perturbative analysis in the previous section.
\subsubsection{Relation to Poincar\'{e} supersymmetry}\label{subsubsec:Poincare}
Before ending this subsection, let us comment on the relation to the Poincar\'{e} supersymmetry. So far we have been discussing the large charge 't Hooft limit, in which the double-scaled coupling $\lambda_J$ is fixed. If we instead take the standard large charge limit in which the coupling $g_{\rm YM}$ is fixed, the dynamics is described by the effective action on the Coulomb branch \cite{Hellerman:2017veg,Hellerman:2017sur,Bourget:2018obm,Hellerman:2018xpi,Beccaria:2018xxl,Hellerman:2020sqj,Sharon:2020mjs,Hellerman:2021yqz,Hellerman:2021duh}. Therefore in this latter limit, we expect that the relevant symmetry is the centrally-extended Poincar\'{e} supersymmetry. 

To see the relation between the two explicitly, we carefully take the limit of anti-commutation relations of the maximally-centrally-extended $\mathfrak{psu}(2|2)$:
\beq
\begin{aligned}
&\{S^{\alpha}{}_{a},Q^{b}{}_{\beta} \} = \delta^{b}_{a} L^{\alpha}{}_{\beta}+\delta^{\alpha}_{\beta} R^{b}{}_{a}+ \delta^{b}_{a}\delta^{\alpha}_{\beta} \,\frac{(\hat D-\hat{J})}{2}\comma\\
&\{ Q^{a}{}_{\alpha},Q^{b}{}_{\beta}\} = \epsilon^{ab} \epsilon_{\alpha \beta} P\,,\quad \{S^{\alpha}{}_{a},S^{\beta}{}_{b}\} = \epsilon_{ab}\epsilon^{\alpha \beta} K\period
\end{aligned}
\eeq
In the standard large charge limit, the eigenvalues of $\hat D-\hat{J}$, $P$ and $K$ all go to infinity as $\propto \sqrt{J}$. Therefore, it is more natural to rescale the generators in the following way:
\beq
\begin{aligned}
Q^{a}{}_{\alpha}=J^{1/4}{\bf Q}^{a}{}_{\alpha}\comma\quad S^{\alpha}{}_{a}=J^{1/4} \bar{\bf Q}^{a}{}_{\alpha}\comma\quad \frac{\hat D-\hat{J}}{2}=\sqrt{J}\,{\bf P}^{0}\comma\quad P=\sqrt{J}{\bf Z}\comma\quad K=\sqrt{J}\,\bar{\bf Z}\period
\end{aligned}
\eeq
After this redefinition, the anti-commutators become
\beq
\begin{aligned}
&\{{\bf Q}^{a}{}_{\alpha} \,, \bar{\bf Q}^{b}{}_{\dot{\beta}}\}=\delta^{a,b} \delta_{\alpha, \dot{\beta}}{\bf P}_0\comma\\
&\{{\bf Q}^{a}{}_{\alpha}\,,{\bf Q}^{b}{}_{\beta}\}=\epsilon^{ab}\epsilon_{\alpha\beta}{\bf Z}\comma\qquad\{\bar{\bf Q}^{a}{}_{\dot{\alpha}}\,,\bar{\bf Q}^{b}{}_{\dot{\beta}}\}=\epsilon^{ab}\epsilon_{\dot{\alpha}\dot{\beta}}\bar{\bf Z}\comma
\end{aligned}
\eeq
which can be identified with the anti-commutation relations\footnote{Here we only have ${\bf P}_0$ on the right hand side of the anti-commutator of ${\bf Q}$ and $\bar{\bf Q}$. We can see the other components of translation generators by rescaling $L^{\alpha}{}_{\beta}$ and $L^{\dot{\alpha}}{}_{\dot{\beta}}$ so that the three of the six generators survive in the limit. This is a standard Inonu-Wigner contraction of $\mf{so}(4)\simeq \mf{su}(2)\times \mf{su}(2)$ to $\mf{iso}(3)$.} for the centrally-extended Poincar\'{e} supersymmetry. In addition, the BPS condition for the maximally-centrally-extended $\mathfrak{psu}(2|2)^2$ \eqref{eq:BPScondition} becomes
\beq
(\hat D-\hat{J})^2-4PK=(n+1)^2\quad \mapsto \quad {\bf P}_0 =|{\bf Z}|\comma
\eeq
which coincides with the BPS condition of the extended Poincar\'{e} supersymmetry. Note that the representation index $n$ disappears from the BPS condition upon taking the large charge limit. This is a reflection of the fact that the entire tower of $\langle 0,n\rangle \otimes \langle 0,n\rangle$'s combine into a BPS single-particle representation on the Coulomb branch. We will see this more explicitly in Section~\ref{sec:coulomb}.
\subsection{Recovering the leading large $J$ spectrum from semiclassics}\label{sec:semicl}
In this section, we will recover the previous results from a simple semiclassical analysis. In particular, we will derive the relation (\ref{eq:energy1}). 

It proves useful to study the theory on the Euclidean cylinder ${\rm S}^3\times \mathbb{R}_{\tau}$ by a Weyl rescaling of $\mathbb{R}^{4}$. We profit from the state/operator map, and consider the insertion of large charged states created by the action on the vacuum of $\Tr Z^{J}$ and its conjugate at $\tau = -\infty$ and $\tau=+\infty$ respectively. As we have explained, these states break a particular R-symmetry subgroup ${\rm U}(1)_{R}\subset SO(6)_{R}$ and 
are protected by supersymmetry so that $\Delta = |J|$. The full superconformal subalgebra preserved by this setup is \eqref{centralsym}. The longitudinal direction of the cylinder provides an additional isometry corresponding to time translations and the energies of the states are equal to the conformal dimensions of the operators in flat space. Therefore the combination $H-\hat{J}$ (where $H$ is the time-translation generator) is preserved by these states, a situation that finds parallel in a superfluid state carrying a finite homogeneous charge density on the 3-sphere, namely $\rho \sim J/L^{3} $  \cite{Monin:2016jmo}, with the crucial difference that the energy density in this case is vanishing by supersymmetry. The SYM path integral is now performed with the boundary conditions set by these states and in particular the trajectories reflect the same symmetry breaking pattern discussed in the previous section. In the large charge limit, one expects a particular semiclassical trajectory to dominate the path integral. To see it explicitly, we notice that performing the path integral with insertions is equivalent to compute the partition function with the effective action
 \beq\label{largechargeSeff}
S_{\rm{eff}} = \frac{2}{g^{2}_{\rm{YM}}}\left[ \int d\tau d\Omega_{{\rm S}^3}\, \mathcal{L}_{\mathcal{N}=4} - \frac{\lambda_J}{2} \log \left( \Tr Z^{2} \right)(\tau=-\infty, \Omega_1)- \frac{\lambda_J}{2} \log \left( \Tr \bar{Z}^{2} \right) (\tau=+\infty, \Omega_2)\right]\,,
 \eeq
where we bring the insertions into the original action, thus modifying it, and we have used the metric on the cylinder as $ds^2=d\tau^2+L^2 d\Omega_{{\rm S}^3}^2$. We see that the whole effective action is of order $\frac{1}{g_{\rm{YM}}^{2}}\gg 1$ when $\lambda_{J}$ is fixed, which justifies a semiclassical approach. The effect of  the insertions is to source a nontrivial profile for the fields $Z$ and $\bar{Z}$ determined by the saddle point. In order to find the saddle, it is convenient to expand these fields in terms of spherical harmonics on ${\rm S}^3$,\footnote{The spherical harmonics are normalized by
\beq
\frac{1}{L^3}\int d^3x \sqrt{g} \,Y^{{\bf{k}}}_{{\rm{scalar}}} Y^{{\bf{m}}}_{{\rm{scalar}}}  =\delta^{{\bf{k}}{\bf{m}}}\,, 
\eeq with $g_{\mu\nu}$ being the metric on a $3$-sphere with radius $L$ and obey $\nabla^2 Y^{{\bf{k}}}_{{\rm{scalar}}}  = -\frac{k(k+2)}{L^2}Y^{{\bf{k}}}_{{\rm{scalar}}} $\,.}
\beq \label{spherical}
Z(t,\Omega) = \frac{z(t)}{\sqrt{2 \pi^2}} + \sum_{\bf{k}} z_{\bf{k}}(t) Y^{\bf{k}}_{{\rm{scalar}}}(\Omega)\,,
\eeq
where we have introduced a collective index ${\bf{k}}=(k,I)$ which comprises the spin $k$ and polarization $I$ and the sum in $\bf{k}$ entails a double sum running over $k=1,\dots, \infty$ and $I=1,\dots, (k+1)^2$.
Since we are focusing for now on operators that preserve the full isometries of ${\rm S}^3$, we consider only the lowest spherical harmonic (or $s$-wave) $z(t)$ and later we will add higher spherical modes which correspond to excited states from operator insertions with derivatives.
At leading order in the large $J$ limit, only the quadratic part of the action is relevant and when restricted to the $s$-wave, it reads 
\beq
S_{\rm{scalar}}= \frac{ 2L^3}{g^2_{\rm{YM}}} \int d\tau \left[ \Tr\left(\dot{\bar{z}}\dot{z} +\frac{ \bar{z} z}{L^2}\right)-
\frac{\lambda_J }{2L^3}\left( \log \left(\Tr z^2 \right)\delta(\tau-\tau_1) + \log \left( \Tr \bar{z}^2 \right) \delta(\tau-\tau_2)\right) \right] \,,
\eeq
with the limit $\tau_1 \rightarrow -\infty$ and  $\tau_2 \rightarrow +\infty$. The second quadratic term inside the trace arises from the conformal masses due to the curvature of the sphere. As in the previous section, we partially fix the gauge by considering diagonal fields, $z= {\rm{diag}}(z_0,-z_0)$ and $\bar{z}= {\rm{diag}}(\bar{z}_0,-\bar{z}_0)$. We then have the stationary equations
\beq
\begin{aligned}
&\ddot{\bar{z}}_0 -\frac{1}{L^2} \bar{z}_0 +\frac{\lambda_J}{L^3} \frac{1}{z_0} \delta(\tau-\tau_2) =0\,, \\
&\ddot{z}_0 -\frac{1}{L^2} z_0 +\frac{\lambda_J}{L^3} \frac{1}{\bar{z}_0} \delta(\tau-\tau_1) =0\,,
\end{aligned}
\eeq
which are solved by\footnote{In deriving the solution, we use that $\frac{\partial}{\partial x}{\rm{sign}}(x) = 2\delta(x)$.}
\beq \label{superfluid}
z_0= \kappa\, e^{-\frac{1}{L}|\tau-\tau_1|}\,,\quad  \bar{z}_0= \bar{\kappa}\, e^{-\frac{1}{L}|\tau-\tau_2|}\,,
\eeq
with the constraint $\kappa \bar{\kappa} = \frac{\lambda_J}{2L^2}\,e^{\frac{1}{L}|\tau_1-\tau_2|}$. The fact that only the product $\kappa \bar{\kappa}$ is fixed reflects the invariance of the action under the rescaling $z\rightarrow \alpha z,\, \bar{z}\rightarrow \alpha^{-1} \bar{z}$. In Appendix~\ref{backlc}, we derive the stationary solutions $z_0,\bar z_0$ directly in the flat space $\mR^4$ which are related to the above by a Weyl transformation (see around \eqref{saddlesol}).

\paragraph{Scalar excitations.}We now consider  excited states obtained by inserting other scalars on the background created by $\Tr Z^{J}$. In the discussion that follows, we will add a single additional scalar field $X$, but there is no obstacle to adding more. Technically, the steps are analogous to the Higgs mechanism in which the field $X$ acquires a mass term from the interactions with the nontrivial background $Z$. The interaction between these two fields in the original $\mathcal{N}=4$ SYM comes from the scalar commutator terms of the  Lagrangian.  We can write the scalar field $X$ and its conjugate as a matrix as in (\ref{eq:su2matrices}), expand each mode in spherical harmonics as in (\ref{spherical}) 
\beq \label{sphericalx}
X(t,\Omega) = \frac{x(t)}{\sqrt{2 \pi^2}} + \sum_{\bf{k}} x_{\bf{k}}(t) Y^{\bf{k}}_{{\rm{scalar}}}(\Omega)\,.
\eeq
Restricted to the lowest spherical mode, the quadratic part in $X$ of the action becomes
\beq
S_{\rm{quad}} = \frac{ 2L^3}{g^2_{\rm{YM}}} \int d\tau \left[ 2\, \dot{\bar{x}}^0\dot{x}^0+\dot{\bar{x}}^{+}\dot{x}^{+}+\dot{\bar{x}}^{-}\dot{x}^{-} +\frac{2}{L^2}\bar{x}^0 x^0+\frac{1}{L^2} \left(1+16 g^2\right)\left({\bar{x}}^{+}{x}^{+}+{\bar{x}}^{-}{x}^{-} \right) \right]\,.
\eeq
We identify two type of modes with distinct masses. We have \textit{massless} or \textit{diagonal} modes $x^0$ and its conjugate which do not get perturbative corrections to their mass given by $(m_{0})^{2}=1/L^2$. In addition, we also have \textit{massive} or \textit{off-diagonal} modes $x^{\pm}$ and their conjugate whose mass is given by $(m_{\pm})^{2}=(1+16g^2)/{L^2}$. Translating into  conformal dimensions, we see that each massless and massive mode contributes  
\beq
\quad \Delta_{0} =1\,,\quad \Delta_{\pm} =\sqrt{1+16 g^2}\,,
\eeq
to the total dimension of the operator at leading order in $1/J$. This is consistent with the findings of the previous section, see (\ref{eq:energy1}).

\paragraph{Spinning excitations.}
In order to study operators with spin we keep higher spherical modes of the complex scalar in (\ref{sphericalx}). The modes with index ${\bf{k}}$ correspond to states transforming in the representation $(k/2,k/2)$ of $\mf{so}(4)={\mf{su}}(2)\times {\mf{su}}(2)$
and in flat space, they amount to operators obtained by adding derivatives, which we can schematically write as
\beq
\mathcal{O}_{k,J}\sim \Tr (D^{k}X)Z^J \,.
\eeq
The corresponding polarization is specified  by the index $I$ in (\ref{spherical}) that can take  $(k+1)^2$ values. In order to write down the action for the higher spherical modes, we decompose them into \textit{diagonal} modes denoted by $x_{\bf{k}}^0$ and \textit{off-diagonal} ones denoted by $x_{\bf{k}}^{\pm}$, and in terms of these, the quadratic part of the action reads
\beq
\begin{aligned}
S_{\rm{quad}} = \frac{ 2L^3}{g^2_{\rm{YM}}}\sum_{\bf{k}} \int d\tau & \biggl[ 2\, \dot{\bar{x}}_{{\bf{k}}}^{0}\dot{x}_{{\bf{k}}}^0+\dot{\bar{x}}_{{\bf{k}}}^{+}\dot{x}_{{\bf{k}}}^{+}+\dot{\bar{x}}_{{\bf{k}}}^{-}\dot{x}_{{\bf{k}}}^{-} +2\, (m_{0,k})^2\bar{x}_{{\bf{k}}}^0 x^0_{{\bf{k}}}+(m_{+,k})^2{\bar{x}}_{{\bf{k}}}^{+}{x}_{{\bf{k}}}^{+}+(m_{-,k})^2{\bar{x}}_{{\bf{k}}}^{-}{x}_{{\bf{k}}}^{-}  \biggr]\,,
\end{aligned}
\eeq
where the masses are given by
\beq \label{spinmass}
(m_{0,k})^2 =\frac{(k+1)^2}{L^2}\,,\quad (m_{\pm,k})^2 =\frac{(k+1)^2+16 g^2}{L^2} \,.
\eeq
These masses translate into the conformal dimensions of operators made out of the corresponding modes by  (\ref{enconf}).
To draw a comparison with the leading order spectrum found in Section~\ref{sl2data}, we start by recalling that the excited states in that sector are formed by adding light-cone derivatives to the vacuum fields $Z$, as shown in (\ref{excitationsl2}).  Adding a derivative to a vacuum field $Z$ is a simple excitation, similar to adding a complex scalar $X$, and carries the same value of the central charge $C$, see (\ref{eq:centralcharged-j}). Thus simply by adding more derivatives, we can conclude that
\beq
C \, |\dots (D^{k+1} Z) \dots \rangle_{J-1} =C\, |\dots D^{k} X \dots \rangle_{J}
\eeq
We can then decompose the field $D^{k} Z$ into its diagonal part which we denote by ${z}_{{\bf{k}}}^{0}$ and off-diagonal ${z}_{{\bf{k}}}^{\pm}$ and they contribute to the anomalous dimension with
\beq \label{spinexciten}
\quad \Delta_{0} =k\,,\quad \Delta_{\pm} =\sqrt{k^2+16 g^2}\,,
\eeq
respectively, in agreement with (\ref{eq:energy2}). The comparison with the perturbative results now follows in a straightforward manner. We add excitations such that the resulting states are neutral under the U(1) gauge charge which requires the off-diagonal modes ${z}_{{\bf{k}}}^{\pm}$ to be added in pairs each one carrying an energy (\ref{spinexciten}) and at leading order, the correction in energy is
\beq
\Delta^{1{\rm{-loop}}}_{\pm} = \frac{8 g^2}{k}\,,
\eeq
precisely matching (\ref{dataelemsl2}). The degeneracy observed in that section  follows  from the distinct ways of adding modes at the cost same energy cost. For example, let us consider the spin 4 states. We can obtain states by adding modes up to spin $k=4$ while keeping the state neutral, see Table~\ref{spin4states}. The data obtained here is in line with the perturbative analysis of Section~\ref{sl2data}.
\begin{table}[t]
\centering
\renewcommand{\arraystretch}{1.30}
$\begin{array}{|c|c|c|}  \hline
\text{state}& \Delta-J & \text{weak coupling} \\ \hline
|z_1^{-}z_1^{-}z_1^{+} z_1^{+} \rangle_{J-4} & 4\times \sqrt{1+16g^2} & 4+32 g^2+\dots  
 \\ \hline
 |z_2^{-}z_2^{+} \rangle_{J-2} & 2\times \sqrt{2^2+16g^2} & 4+8 g^2+\dots  \\
\hline
|z_{\{1}^{-}z_{3\}}^{+} \rangle_{J-2} & \sqrt{1+16g^2}+\sqrt{3^2+16g^2} & 4+\frac{32 g^2}{3}+\dots  \\
\hline
|z_1^{0}z_1^{0}z_1^{-} z_1^{+} \rangle_{J-4} & 2+2\sqrt{1+16g^2} & 4+16 g^2+\dots  \\
\hline
|z_2^{0}z_1^{-} z_1^{+} \rangle_{J-3} & 2+2\sqrt{1+16g^2} & 4+16 g^2+\dots  \\
\hline
|z_1^{0}z_{[1}^{-} z_{2]}^{+} \rangle_{J-3} & 1+\sqrt{1+16g^2}+\sqrt{2^2+16g^2} & 4+12 g^2+\dots \\
\hline
\end{array}$ \caption{Eigenstates and respective spectrum for spin 4 states. Note that each excitation is labelled by the pair ${\bf{k}}=(k,I)$ but since we are projecting the derivative into a light-cone direction we keep the polarisation index $I$ fixed for all excitations and omit it here. The symmetrized and anti-symmetrized combinations showing up above reflect that we keep only parity even states as explained before for the case of pure scalar states. } \label{spin4states}
\end{table}

\section{Operator Spectrum at Order $1/J$\label{sec:subleading}}
The coupling constant of the effective theory around the non-trivial vacuum generated by the large charge operator scales with $1/J$. The reason is obvious: with a canonically normalized effective Lagrangian, the interacting terms are of order $1/\sqrt{J}$ (Yukawa terms and gauge interactions) or $1/J $ (purely scalar terms and  gauge interactions). This produces an expansion in $ 1/J$ for the anomalous dimensions which is a nontrivial function of $\lambda_{J}$ at each order. In this section, we will be interested  in determining the first subleading term in the expansion in $1/J$ (which we may call the \textit{one-loop} correction) exactly in the large charge 't Hooft coupling. We start by evoking again the maximally-centrally-extended symmetry \eqref{univcentral} and study how it constrains the dilatation operator and afterwards we complement the analysis by some perturbative computations from semiclassics.

	\subsection{Constraints from centrally-extended symmetry}
	It is natural to expect that the maximally-centrally-extended symmetry \eqref{univcentral} which fully constrained the one-magnon energies (up to the coupling constant redefinition) is also powerful in determining how they interact. In fact, such symmetry was put to good use in the large $N$ limit of $\mathcal{N}=4$ SYM in that it fixed the two-magnon dynamics in the spin chain. For example, in \cite{Beisert:2003ys} the symmetry completely constrained the one-loop spin-chain Hamiltonian  of a large compact subsector that included $\mathfrak{su}(2|2)$ (even without using the central extension) and in \cite{Beisert:2005tm} the two-magnon scattering matrix for asymptotically large spin chains was also completely determined up to a global phase. 
	
In our current problem, we aim to fix the dilatation operator using the same type of arguments. The superconformal generators (\ref{eq:susygens}) and (\ref{eq:centralcharged-j}) receive corrections of order $1/J$, and demanding closure of the algebra (\ref{eq:anticompsu}) at this order should constrain them. In this section, we focus on a single copy of $\mathfrak{psu}(2|2) \subset \mathfrak{psu}(2|2)^2$. However, extending the results to the entire algebra should pose no difficulty.
		
In order to reduce the space of parameters that the generators can depend on and therefore make the computations easier, we take advantage of the following simple considerations.
Since the bosonic subalgebra of $\mathfrak{psu}(2|2) $ does not get deformed at loop level, it is convenient to write down the generators in a manifest $\mathfrak{su}(2)_{a}\times \mathfrak{su}(2)_{\alpha}$ invariant way. 
Moreover, their form should also be compatible with the interacting part of the Lagrangian of the effective theory.
The interactions involve either three fields at order $1/\sqrt{J}$ (in the case of the Yukawa and gauge interactions) or four fields  at order $1/J$ (for the pure scalar terms and gauge interactions) from which we  expect a generator $\hat G$ to have an expansion of the form
\beq
\hat G = \sum_{k=0}^{\infty} \frac{1}{J^{k/2}} \hat G_{k/2}(g^2)\,.
\label{Gexp}
\eeq
At each order $k$, the perturbative correction $\hat G_{k/2}$ can have at most $k+2$ external legs. 
This means that up to order $1/J$ each generator can be built out of at most four external ``legs''. In a basis of states for which the classical dimensions are preserved by the interactions, this implies that we can have at most pairwise interactions. This is of course consistent with the magnon picture found previously in the weak coupling analysis of Section~\ref{sec:weak}.

With these constraints it is straightforward to write the most general ansatz for the generators at one-loop. We begin with the dilatation operator $\hat D$. Clearly at order $1/\sqrt{J}$ there is no possible term we can write compatible with $\mathfrak{su}(2)_{a}\times \mathfrak{su}(2)_{\alpha}$ symmetry and preserving the classical dimension. The lowest perturbative correction is therefore at order $1/J$ and we have for one-particle,
\beq
\begin{aligned}
\hat D_{1}\, |\phi_{a}^{m_1} \rangle_{J} &= A_{1}\,  |\phi_{a}^{m_1} \rangle_{J}\,, \\
\hat D_{1}\, |\psi_{\alpha}^{m_1} \rangle_{J} &= B_{1}\,  |\psi_{\alpha}^{m_1} \rangle_{J} \,,
\end{aligned}
\eeq
where the coefficients $A_{1}\equiv A_{m_1}(g^2)$ and $B_{1}\equiv B_{m_1}(g^2) $ are generic functions of the large charge 't Hooft coupling $g^2={\lambda_J\over 16\pi^2}$ but we will omit this dependence to simplify the notation.   
On a two-particle state, the transitions should preserve the ${\rm U}(1)$ gauge charge (besides the global $J$ charge) and together with the requirement that the classical dimension is not modified by the interaction this forces the gauge indices to be at most permuted. We can then write
\beq
\begin{aligned}\label{ansatzD2}
\hat D_{1}\, |\phi_{a}^{m_1} \phi_{b}^{m_2} \rangle_{J} = &\left(A_{1}+A_{2}+C_{J,12} \right)\,  |\phi_{a}^{ m_1}\phi_{b}^{m_2  } \rangle_{J}+D_{J,12}\,  |\phi_{a}^{m_2}\phi_{b}^{m_1} \rangle_{J} + E_{J,12}\,\epsilon_{ab}\,\epsilon^{\alpha \beta} |\psi^{m_1}_{\alpha}\psi^{m_2}_{\beta} \rangle_{J-1}\\
\hat D_{1}\, |\phi_{a}^{m_1}\psi_{\alpha}^{m_2} \rangle_{J} =& \left(  A_{1}+B_{2}+F_{J,12} \right)\,  |\phi_{a}^{m_1}\psi_{\alpha}^{m_2} \rangle_{J}+ G_{J,12}\,  |\phi_{a}^{m_2}\psi_{\alpha}^{m_1} \rangle_{J} \\
\hat D_{1}\, |\psi_{\alpha}^{m_1}\psi_{\beta}^{m_2} \rangle_{J} =& \left(B_{1}+B_{2}+H_{J,12}  \right)\,  |\psi_{\alpha}^{m_1}\psi_{\beta}^{m_2} \rangle_{J}+ I_{J,12}\,  |\psi_{\alpha}^{m_2}\psi_{\beta}^{m_1} \rangle_{J} +  J_{J,12}\,\epsilon_{\alpha \beta}\, \epsilon^{ab} |\phi^{m_1}_{a}\phi^{m_2}_{b} \rangle_{J+1}\,.
\end{aligned}
\eeq
where again we have used a shorthand notation for the two-particle coefficients, for example $C_{J,12} \equiv C_{J,m_1m_2}(g^2)$.
Our analysis at one-loop has revealed that the anomalous dimensions of multi-magnon states depended on $J$ in a discrete way. As seen in formula (\ref{su3eigens}), the dimensions of the operators for a given number of magnons vary for different parities of $J$. To account for this difference, we have added an extra index $J$ to the two-body interaction coefficients and a generic coefficient $\Sigma_{J,12}$ satisfies a parity condition
\beq \label{paritycoef}
\Sigma_{J,12} = \Sigma_{J\pm2,12}
\eeq
for any $J$. It is important to note that condition (\ref{paritycoef}) is an assumption based on lower loop perturbative data. It would be important to further verify its validity through a more robust perturbative analysis and to gain a deeper understanding of its origin.

Let us consider the superconformal generators $Q^{a}{}_{\alpha}$ and ${S}^{\alpha}{}_{a}$. Once again there is no term one can write at order $1/\sqrt{J}$\footnote{Note that even though in the original $\mathcal{N}=4$ SYM we have terms of order $g_{{\rm{YM}}}$ for these generators, for example $Q|\psi\rangle \sim g_{\rm YM}|Z \phi \rangle$, such $\mathcal{O}(g_{{\rm{YM}}})$ transitions involve $Z$ fields which are enhanced by a factor of $\sqrt{J}$ in the large charge limit.} and we have for the one-particle loop correction
\beq
\begin{aligned}
Q^{a}{}_{\alpha}\, |\phi_{b}^{m_1} \rangle &= C_{1}\, \delta^{a}_{b} |\psi_{\alpha}^{m_1} \rangle_{J} \\
Q^{a}{}_{\alpha}\, |\psi_{\beta}^{m_1} \rangle &= D_{1}\, \epsilon^{a b} \epsilon_{\alpha \beta} |\phi_{b}^{m_1} \rangle_{J+1} \\
S^{\alpha}{}_{a}\, |\phi_{b}^{m_1} \rangle &= \bar{D}_{1}\, \epsilon_{a b}\epsilon^{\alpha \beta} |\psi_{\beta}^{m_1} \rangle_{J-1} \\
S^{\alpha}{}_{a}\, |\psi_{\beta}^{m_1} \rangle &= \bar{C}_{1}\, \delta^{\alpha}_{\beta} |\phi_{a}^{m_1} \rangle_{J}\,,
\end{aligned}
\eeq
where $\bar{C}$ and $\bar{D}$ are in principle independent coefficients from $C$ and $D$ and not necessarily their complex conjugates.
The action involving a pair of magnons can be written as 
\beq
\begin{aligned}
Q^{a}{}_{\alpha}\, |\phi_{b}^{m_1}\phi_{c}^{m_2} \rangle_{J} &=   \delta^{a}_{b} \left(  K_{J,12}\, |\phi_{c}^{m_2}\psi_{\alpha}^{m_1} \rangle_{J}+ L_{J,12} \, |\phi_{c}^{m_1}\psi_{\alpha}^{m_2} \rangle_{J}\right) \\
&+\delta^{a}_{c} \left(M_{J,12}\, |\phi_{b}^{m_1}\psi_{\alpha}^{m_2} \rangle_{J} + \left(M_{J,21}+L_{J,21}-K_{J,12} \right)\,|\phi_{b}^{m_2}\psi_{\alpha}^{m_1} \rangle_{J}\right)\\
&+ \epsilon_{b c}\, \epsilon^{a d} \left(N_{J,12} \, | \phi_{d}^{m_2} \psi_{\alpha}^{m_1}\rangle_{J}+\left( M_{J,12}-K_{J,21}- N_{J,21}  \right)| \phi_{d}^{m_1} \psi_{\alpha}^{m_2}\rangle_{J} \right)\\
&+C_1 \, \delta^{a}_{b} \, |\phi_{c}^{m_2}\psi_{\alpha}^{m_1} \rangle_{J}+ C_{2} \, \delta^{a}_{c}\, |\phi_{b}^{m_1}\psi_{\alpha}^{m_2} \rangle_{J}
 \\
Q^{a}{}_{\alpha}\, |\phi_{b}^{m_1}\psi_{\beta}^{m_2} \rangle_{J} &=  \delta^{a}_{b}\left(O_{J,12}    \, |\psi_{\alpha}^{m_1}\psi_{\beta}^{m_2} \rangle_{J}+P_{J,12}\, |\psi_{\alpha}^{m_2}\psi_{\beta}^{m_1} \rangle_{J} \right) \\
&+\epsilon_{\alpha \beta}\, \epsilon^{a d} \left( Q_{J,12} | \phi_{b}^{m_1} \phi_{d}^{m_2}\rangle_{J+1}+R_{J,12} | \phi_{b}^{m_2} \phi_{d}^{m_1}\rangle_{J+1} \right)  \\
&+C_{1}\, \delta^{a}_{b} \, |\psi_{\alpha}^{m_1}\psi_{\beta}^{m_2} \rangle_{J}+D_{2}\,\epsilon_{\alpha \beta}\, \epsilon^{a d}  | \phi_{b}^{m_1} \phi_{d}^{m_2}\rangle_{J+1}
\\
Q^{a}{}_{\alpha}\, |\psi_{\beta}^{m_1}\psi_{\gamma}^{m_2} \rangle_{J} &=  
\epsilon^{ab}  \epsilon_{\alpha \beta} \left( S_{J,12}\,|\phi_{b}^{m_1} \psi_{\gamma}^{m_2} \rangle_{J+1}+T_{J,12} \,|\phi_{b}^{m_2} \psi_{\gamma}^{m_1} \rangle_{J+1} \right)\\
&+\epsilon^{ab} \epsilon_{\alpha \gamma} \left( U_{J,12} \,|\phi_{b}^{m_2}\psi_{\beta}^{m_1}\rangle_{J+1}-\left(S_{J,12}+T_{J,21}+U_{J,21}\right)\,|\phi_{b}^{m_1}\psi_{\beta}^{m_2}\rangle_{J+1} \right)\\
&+\epsilon^{ab}  \epsilon_{\beta \gamma} \left( V_{J,12} \, |\phi_{b}^{m_1}\psi_{\alpha}^{m_2} \rangle_{J+1}+\left(V_{J,21}-S_{J,21}-U_{J,12}\right)\,|\phi_{b}^{m_2} \psi_{\alpha}^{m_1}\rangle_{J+1}\right)\\
&+ D_1\, \epsilon^{ab}\epsilon_{\alpha \beta} |\phi_{b}^{m_1} \psi^{m_2}_{\gamma}\rangle_{J+1} -D_2 \,\epsilon^{ab}\epsilon_{\alpha \gamma} \, |\phi_{b}^{m_2} \psi^{m_1}_{\beta}\rangle_{J+1} \,.
\end{aligned}
\eeq
The coefficients $K_{J,12}$ to $V_{J,12}$ are unknown functions of the 't Hooft coupling.
In this ansatz, we have implemented the $\mathfrak{su}(2)_{a} \times \mathfrak{su}(2)_{\alpha}$ invariance which justifies certain odd looking combinations of the unknown coefficients. The last line in the RHS of each equation arises from the one-particle loop correction. We can write down a similar expression for the action of $S^{\alpha}{}_{a}$ on a two-particle state,
\beq
\begin{aligned}
S^{\alpha}{}_{a}\,  |\phi_{b}^{m_1}\phi_{c}^{m_2} \rangle_{J}  &=\epsilon_{a b} \epsilon^{\alpha \beta}\left( \bar{S}_{J,12} \,|\phi_{c}^{m_2}\psi_{\beta}^{m_1}\rangle_{J-1}+\bar{T}_{J,12} |\phi_{c}^{m_1}\psi_{\beta}^{m_2}\rangle_{J-1}\right)
\\
&+\epsilon_{a c} \epsilon^{\alpha \beta} \left( \bar{U}_{J,12}\,|\phi_{b}^{m_1}\psi_{\beta}^{m_2}\rangle_{J-1}+\left(-\bar{S}_{J,12}+\bar{T}_{J,21}+\bar{U}_{J,21}\right) \, |\phi^{m_2}_{b} \psi_{\beta}^{m_1} \rangle_{J-1} \right)\\
&+\epsilon_{b c} \epsilon^{\alpha \beta}\left(\bar{V}_{J,12}\, |\phi_{b}^{m_2}\psi_{\beta}^{m_1}\rangle_{J-1}  + \left(\bar{S}_{J,21}-\bar{U}_{J,12}-\bar{V}_{J,21}\right)|\phi_{a}^{m_1}\psi_{\beta}^{m_2}\rangle_{J-1} \right) 
\\
S^{\alpha}{}_{a}\, |\phi_{b}^{m_1}\psi_{\beta}^{m_2} \rangle_{J} &=  \delta^{\alpha}_{\beta} \left( \bar{O}_{J,12}  \, |\phi_{b}^{m_1} \phi_{a}^{m_2} \rangle_{J}+\bar{P}_{J,12}\,  |\phi_{b}^{m_2}\phi_{a}^{m_1} \rangle_{J}\right)  \\
&+\epsilon^{\alpha \gamma}\, \epsilon_{a b} \left( \bar{Q}_{J,12} | \psi_{\gamma}^{m_1} \psi_{\beta}^{m_2}\rangle_{J-1}+\bar{R}_{J,12} | \psi_{\gamma}^{m_2} \psi_{\beta}^{m_1}\rangle_{J-1} \right)\\
&+\bar{C}_{1} \delta^{\alpha}_{\beta} \, |\phi_{b}^{m_1} \phi_{a}^{m_2} \rangle_{J}+\bar{D}_{2} \,\epsilon^{\alpha \gamma}\, \epsilon_{a b} | \psi_{\gamma}^{m_1} \psi_{\beta}^{m_2}\rangle_{J-1}
\\
S^{\alpha}{}_{a}\, |\psi_{\beta}^{m_1}\psi_{\gamma}^{m_2} \rangle_{J} &=
\delta^{\alpha}_{\beta} \left( \bar{K}_{J,12}\, |\phi_{a}^{m_1}\psi_{\gamma}^{m_2}\rangle_{J}+\bar{L}_{J,12} | \phi_{a}^{m_2} \psi_{\gamma}^{m_1}\rangle_{J} \right)\\
&+\delta^{\alpha}_{\gamma} \left( \bar{M}_{J,12} \, |\phi_{a}^{m_2} \psi_{\beta}^{m_1} \rangle_{J} - \left( \bar{M}_{J,21}+\bar{K}_{J,12} + \bar{L}_{J,21}\right)|\phi_{a}^{m_1} \psi_{\beta}^{m_2} \rangle_{J} \right)\\
&+\epsilon_{\beta \gamma}\epsilon^{\alpha \delta} \left(\bar{N}_{J,12}\, |\phi_{a}^{m_1}\psi_{\delta}^{m_2}\rangle_{J} +\left(\bar{M}_{J,12}+\bar{N}_{J,21}+\bar{K}_{J,21} \right) |\phi_{a}^{m_2}\psi_{\delta}^{m_1}\rangle_{J} \right) 
\\
&+\bar{C}_1 \delta^{\alpha}_{\beta} |\phi_{a}^{m_1}\psi_{\gamma}^{m_2}\rangle_{J}-\bar{C}_2 \delta^{\alpha}_{\gamma} |\phi_{a}^{m_2}\psi_{\beta}^{m_1}\rangle_{J}
\end{aligned}
\eeq
A further constraint arises from the requirement that the theory is exactly symmetric under charge conjugation, as discussed in the previous section. The action of charge conjugation on a two-particle state is defined as 
\beq
\mf{P}|\chi_1^{m_1} \chi_2^{m_2} \rangle_{J} = (-1)^{J} |\chi_1^{-m_1} \chi_2^{-m_2}\rangle_{J}\,, \quad \chi_{1,2} = \phi_{a},\psi_{\alpha}\,,
\eeq
and we require $[\mf{P}, \hat G_{k}]=0$. This reduces the number of unknowns by relating coefficients with opposite U(1) gauge charges.

The algebraic relations impose strong constraints on the coefficients of the one-loop dilatation operator. For the one-particle energies, it reduces to a single independent coefficient function $A_{-}(g^2)$ due to the following conditions
\beq
A_{0}=0\,,\quad A_{+}=A_{-}\,, \quad B_{m} = A_{m}\quad \text{with}\quad m=0,\pm\,.
\eeq
For two-particle states involving at least one massless particle, we also obtain a single unfixed coefficient function
\beq
\begin{aligned}
&\hat D_{1}\, |\chi_{1}^{0}\, \chi_{2}^{0} \rangle_{J} = 0 \,, \quad \chi_{1,2} =\phi_{a},\psi_{\alpha}\,,\\
&\hat D_{1}\, |\chi_{1}^{0}\, \chi_{2}^{\pm} \rangle_{J} =(A_{-}+F_{J,+0})\, |\chi_{1}^{0}\, \chi_{2}^{\pm1} \rangle_{J} \,, \quad \chi_{1,2} =\phi_{a},\psi_{\alpha}
\end{aligned}
\eeq
where the coefficient $F_{J,+0}(g^2)$ remains as the only unknown leftover. On a pair of massive particles, we obtain instead
\beq
\begin{aligned} \label{hamiltonian1loop}
\hat D_{1}\, |\phi_{a}^{\pm} \phi_{b}^{\mp} \rangle_{J} & = \left(2A_{-}+F_{J,-+}\right)\,  |\phi_{a}^{\pm} \phi_{b}^{\mp} \rangle_{J}+G_{J,-+}\,  |\phi_{a}^{\mp} \phi_{b}^{\pm} \rangle_{J}\\
\hat D_{1}\, |\psi_{\alpha}^{\pm} \psi_{\beta}^{\mp} \rangle_{J}& = \left(2A_{-}+F_{J,-+}\right)\,  |\psi_{\alpha}^{\pm} \psi_{\beta}^{\mp} \rangle_{J}+G_{J,-+} \,  |\psi_{\alpha}^{\mp} \psi_{\beta}^{\pm} \rangle_{J}\\
\hat D_{1}\, |\phi_{a}^{\pm} \psi_{\beta}^{\mp} \rangle_{J} &= \left(2A_{-}+F_{J,-+} \right)\, | \phi_{a}^{\pm} \psi_{\beta}^{\mp} \rangle_{J} +G_{J,-+}\, |\phi_{a}^{\mp} \psi_{\beta}^{\pm} \rangle_{J}  
\\
\hat D_{1}\, |\chi_1^{\pm} \chi_2^{\pm} \rangle_{J} &= \left(2A_{-}+\hat{F}_{J,--} \right)\, |\chi_1^{\pm} \chi_2^{\pm} \rangle_{J}  
\end{aligned}
\eeq
where the coefficient $\hat{F}_{J,--}$ is given  in terms of the unknowns defined in the ansatz (\ref{ansatzD2}) by the combination 
\beq
\hat{F}_{J,--}= F_{J,--}+G_{J,--}\,.
\eeq 
Our analysis of algebraic constraints revealed that the transition between a pair of scalars and fermions, governed by the coefficients $E_{J,12}$ and $J_{J,12}$, vanishes. This can be attributed to the assumption made in condition (\ref{paritycoef}). While our findings are consistent with the data in Section~\ref{sec:weak}, it is important to note that this particular type of transition has not been probed by our perturbative analysis and further higher loop calculations are necessary to confirm the validity of this result.

\paragraph{Representation theory.} We want to understand these results  from the point of view of the representation theory of the centrally extended $\mathfrak{psu}(2|2)$. The dilatation operator we just constructed is one generator of the extended algebra and corresponds to one of its central charges that was previously called $C$. We have expanded each generator in the parameter $1/J$ which serves as a ``quantum'' parameter in this context. The first quantum correction to the dilatation operator that we have denoted by $\hat D_1$ is in fact invariant under the tree level algebra, that is
\beq
[\hat D_{1}-\hat{J},G_{0}] = 0\,,
\eeq
for any tree level generator $\hat G_{0}$ of $\mathfrak{psu}(2|2)$. Hence $\hat D_1-\hat{J}$ is an invariant operator under the (tree) $\mathfrak{psu}(2|2)$ algebra acting on a tensor product of two fundamental multiplets. On general grounds, we expect the number of degrees of freedom of such operator to be  in one-to-one correspondence to the number of irreducible components in the tensor product. 

We recall that each fundamental multiplet is characterised by the triplet of central charges $\vec{C}=(C,P,K)$ obeying the shortening condition, see the discussion in Section~\ref{sec:centralext}. As before, we denote the fundamental representation by 
$\langle 0,0\rangle_{m}$ where we included the subindex $m=\pm,0$ that stands for the U(1) charge. We therefore have three distinct cases
\beq
\langle 0,0\rangle_{+}  \otimes \langle0,0 \rangle_{+}\,,\quad \langle 0,0 \rangle_{+} \otimes \langle 0,0\rangle_{-}\,,\quad \langle 0,0 \rangle_{+} \otimes \langle 0,0 \rangle_{0}
\eeq
with all other possibilities related to these by parity. 

In the first case 
$\langle 0,0\rangle_{+}  \otimes \langle0,0 \rangle_{+}$, we have that the resulting central charge of the tensor product is $\vec{C}'= (2 C_{+},2 P_{+},2K_{+})$ and it obeys again a shortening condition
\beq
\vec{C}'^2 = 1\,.
\eeq
This means that there are in principle two eight-dimensional short representations in the decomposition of the tensor product. The states in these two representations can be constructed explicitly as follows. Let us denote the fundamental representation as 
\beq
\chi_{A} :=\left(
\begin{array}{c}
\phi_a\\
\psi_{\alpha}\\
\end{array}
\right)\,,
\eeq
where for the moment we have omitted the gauge charge index and $A$ is a super-index that runs over the bosonic and fermionic components $A=(a,\alpha)$. The first eight-dimensional irrep is built by symmetrising the super-indices in the following way
\beq
|{\rm{sym}}\rangle_{J}:=|\chi_{A}\chi_{B} \rangle_{J}+(-1)^{|A||B|}|\chi_{B}\chi_{A} \rangle_{J}
\eeq
with $|\bullet|$ denoting the corresponding grading. The second is obtained instead by anti-symmetrising the super-indices as
\beq
|{\rm{asym}}\rangle_{J}:=|\chi_{A}\chi_{B} \rangle_{J}-(-1)^{|A||B|}|\chi_{B}\chi_{A} \rangle_{J}\,.
\eeq
Since the particles are commuting or anti-commuting depending on the component, this second representation is  trivial. Therefore, there is only one non-trivial irreducible component in the decomposition of the tensor product of 
$\langle 0,0\rangle_{+}  \otimes \langle0,0 \rangle_{+}$ and therefore we expect only one degree of freedom. This degree of freedom corresponds precisely to the coefficient function $\hat{F}_{J,--} $.

We now consider the sector of states obtained from the tensor product 
$\langle 0,0\rangle_{+}  \otimes \langle0,0 \rangle_{-}$.
In this case, the total central charge, 
\beq
\vec{C}' = ( C_{+}+ C_{-}, P_{+}+P_{-},K_{+}+K_{-}) = (2C_{+},0,0)\,,
\eeq  
does not fulfill the shortening condition, resulting in a unique, sixteen-dimensional long multiplet. However, as previously noted, when dealing with massive particles of opposite charges, we can construct two distinct sets of states with different quantum numbers under charge conjugation, depending on the parity of $J$. As a result, we expect to find two distinct copies in the tensor product  
$\langle 0,0\rangle_{+}  \otimes \langle0,0 \rangle_{-}$. This is why we anticipate two degrees of freedom in this tensor product, which explains the presence of the two undetermined coefficient functions $F_{J,-+}$ and $G_{J,-+}$.

We observe that the tensor product 
$\langle 0,0\rangle_{+}  \otimes \langle0,0 \rangle_{0}$ results in states with total central charge 
\beq
\vec{C}' = ( C_{+}+ C_{0}, P_{+}+P_{0},K_{+}+K_{0}) = (C_{+}+C_{0},P_{+},K_{+})\,,
\eeq
which again does not obey a shortening condition and results in a single long multiplet. In this case, there is only a single degree of freedom corresponding to the  unfixed function $F_{J,+0}$.

We have finally narrowed down the problem to five unknown coefficients
\beq
\{A_{-}, F_{J,-+},G_{J,-+},F_{J,+0},\hat{F}_{J,--} \}\,.
\eeq These coefficients can be determined through a one-loop computation in the effective field theory. In the following discussion, we will perform a partial analysis and determine some of them.
	\subsection{Constraints from semiclassics at one-loop} \label{sec:eftcoefs}
	We now turn to the computation of some coefficients of the previous section and compare them with the perturbative data from the first section. 
	
	The effective action which we will use is obtained by expanding the $\mathcal{N}=4$ SYM Lagrangian around the classical profile $Z_{{\rm{cl}}}$  of the elementary complex scalars $Z$ and $\bar{Z}$ sourced by the large charge operators,
	\beq
Z= Z_{\rm{cl}}+\delta Z\,,\quad Z_{\rm{cl}}= \begin{pmatrix}
	z_{{\rm{cl}}}^0 & 0\\
	0 & -z_{{\rm{cl}}}^0
\end{pmatrix}\,,
\eeq
where $\delta Z$ denotes the fluctuation around the saddle point
and similarly for $\bar{Z}$. In Section~\ref{sec:semicl}, we analyzed the leading semiclassics on the cylinder $\mR_\tau\times {\rm S}^3$. Here we find it convenient to work directly on the flat space $\mR^4$. 
The diagonal entry $z_{{\rm{cl}}}^0$ is a nontrivial function of the large charge 't Hooft coupling and fixed by the saddle point equation. This computation is given in the Appendix~\ref{backlc} and the answer is\footnote{They are related to \eqref{superfluid} by a Weyl transformation.}
\beq
z^{0}_{\rm{cl}}(x) = \frac{e^{i\phi} \, |x_1-x_2|}{2\pi |x-x_2|^2}\,\sqrt{\frac{\lambda_{J}}{2}}\,,\quad \bar{z}^{0}_{\rm{cl}}(x) = \frac{e^{-i\phi} \, |x_1-x_2|}{2\pi |x-x_1|^2}\,\sqrt{\frac{\lambda_{J}}{2}}\,.
\eeq
where $x_{1,2}$ are the location of the operator insertions and $\phi$ is an undetermined phase which reflects a symmetry of the saddle point equation and it will not play any role in what follows.
	
We aim to study the one-loop correction to the two-point function of large charge operators with impurities. This calculation poses significant challenges compared to the analogous computation in the original $\mathcal{N}=4$ SYM. Firstly, we must contend with massive fields whose mass has a non-trivial kinematic dependence given by 
\beq\label{mass}
m^{2}(x) = 8 |z^0_{\rm cl}|^2 =\frac{ |x_1-x_2|^2}{ \pi^2 |x-x_1|^2|x-x_2|^2} \,\lambda_{J}\,.
\eeq 
This makes even the computation of propagators more complicated. Secondly, there are additional vertices that couple massive and massless particles induced by the background field, for example, new cubic couplings for scalars and additional scalar-gluon interactions. Thirdly, the mass terms are not diagonal in the flavor indices and there is a quadratic term coupling scalars and gauge fields. To make the computations feasible, we need to bring the quadratic part of the action into a canonical diagonal form. These are three important differences that make the one-loop computation significantly more challenging. 

\begin{figure}[t] 
\includegraphics[width=15cm]{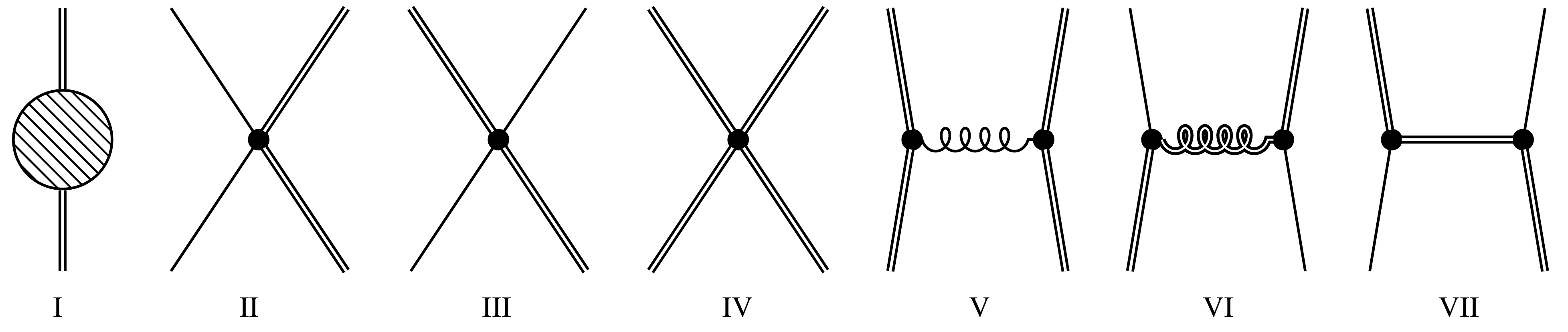}
\centering
\caption{The figure displays the complete set of diagrams involving external scalars. The single (double) line represents the propagation of massless (massive) scalar modes and the single (double) coiled line denotes the massless (massive) gluon propagator. The blob amounts to the sum of all self-energy diagrams.}\label{diagrams}
\end{figure}
We have determined the scalar massive propagator in the Appendix~\ref{backlc}, and fermion and gauge field propagators are computable by similar methods. We also strongly benefit from the symmetry analysis of the previous section, which reduces the number of graphs to be analyzed. For example, processes with external fermions are completely determined in terms of processes having only scalars as external fields. We display in Figure~\ref{diagrams}, all the diagrams involved with external scalars. For our purposes, we will only determine a few of those diagrams. The third point is the most technically involved and will be left for future analysis.	
	
\paragraph{From diagrams to coefficients.} In order to read off the undetermined coefficient functions of the previous section, we have to extract the logarithm divergence of each of the Feynman diagrams above. In this paper, we will restrict ourselves to the diagrams with four external massive legs, namely the ones represented in $\rm{IV}$ and $\rm{V}$ in Figure~\ref{diagrams}. Those will be enough to fix three out of the five unknowns of the previous section namely $ F_{J,-+},G_{J,-+}$ and $\hat{F}_{J,--}$. The constant $A_{-}$ is fixed from the self-energy diagram  $\rm{I}$ and the  interaction of a massive and massless particle mediated by $F_{J,+0}$ is determined by the remaining diagrams. These diagrams involve exchange of intermediate massive virtual particles which are technically harder to deal with. 

The diagrams $\rm{IV}$ and $\rm{V}$ are simple to compute from the effective action and we display the result here while leaving the details in the Appendix~\ref{perturbapp}:
  \begin{align*}
       & \scalebox{0.88}{$G_1:=$}\eqnDiag{\includegraphics[scale=0.9]{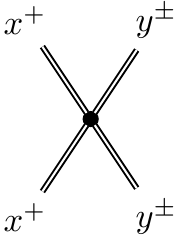}} \scalebox{0.88}{ $=-\frac{2\lambda_{J}}{J} \frac{1}{1+\frac{\lambda_{J}}{ \pi^2}} \,(2\pi)^4\,|x_1-x_2|^4 X_{1122}$}\quad\quad \scalebox{0.88}{$G_2:=$} \eqnDiag{\includegraphics[scale=0.9]{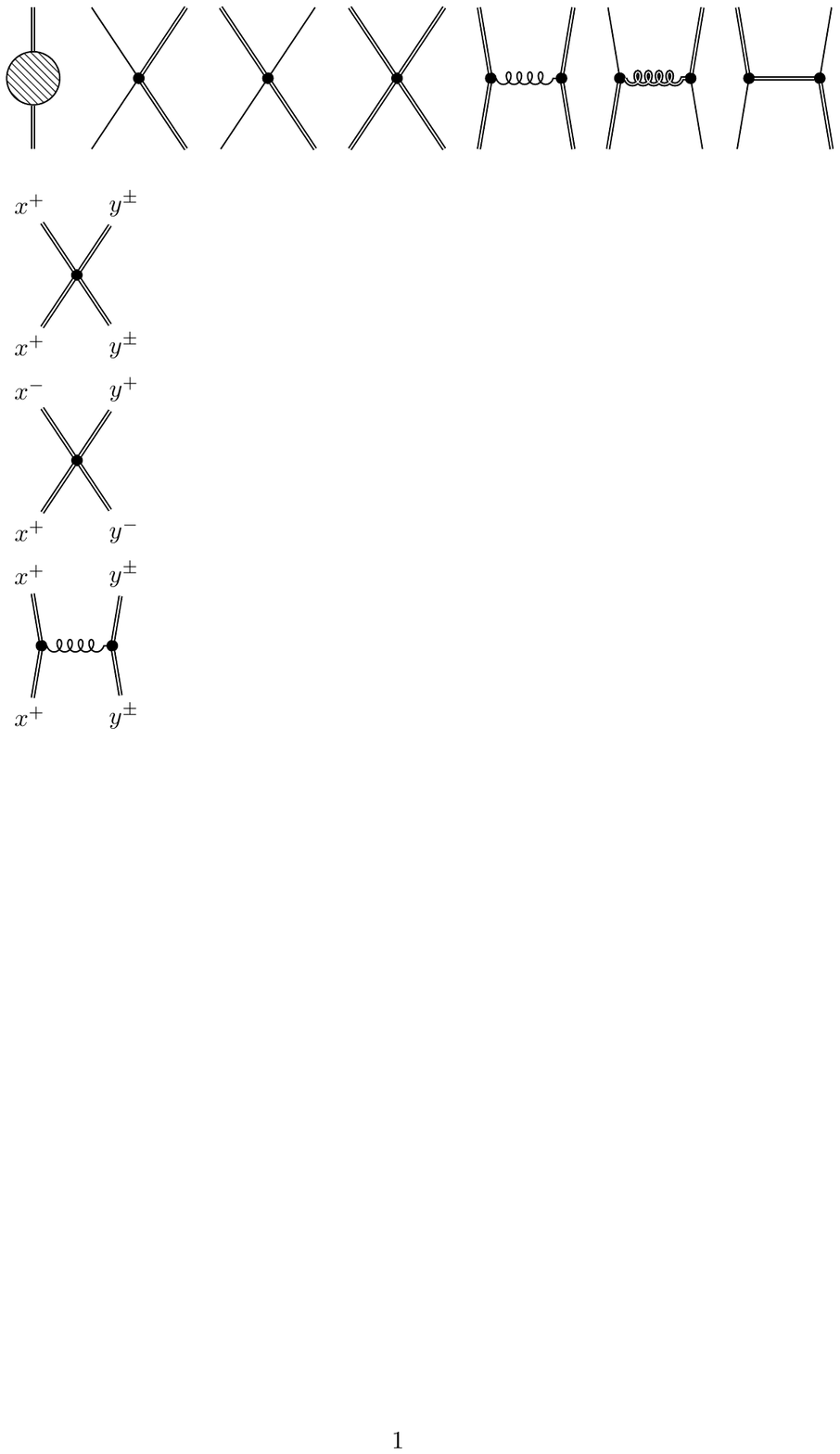}}\scalebox{0.88}{ $=\frac{4 \lambda_{J}}{J} \frac{1}{1+\frac{\lambda_{J}}{ \pi^2}} \,(2\pi)^4\,|x_1-x_2|^4 X_{1122}$}\\
&\scalebox{0.88}{$G^{\pm}_3:=$} \eqnDiag{\includegraphics[scale=0.9]{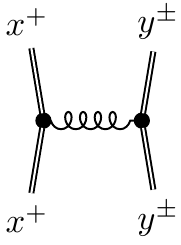}}\scalebox{0.88}{ $=\mp \frac{16 \lambda_{J}}{J}\frac{1}{1+\frac{\lambda_{J}}{\pi^2}} (2\pi)^4|x_1-x_2|^4\lim_{\substack{x_3\rightarrow x_1 \\ x_4 \rightarrow x_2}} \partial_{2} \cdot \partial_{4} H_{12,34}$}\,.
  \end{align*}
  The double line corresponds to the massive scalar propagator and the coiled line represents the propagator of the massless mode of the gluon field.  The functions $X_{1234}$ and $H_{12,34}$ are standard integrals with four external legs defined in the Appendix~\ref{perturbapp} and the repeated label in $X$ means taking a colliding limit of the respective points.  
  	
The coefficient functions are obtained by combining the above integrals as follows
\beq
\begin{alignedat}{3}
&F_{J,-+} = \left(G_{1}+G_{3}^{-}\right)\big\rvert_{\log} &&= -\frac{12 g^2}{1+16 g^2} &&\simeq -12 g^2 +192 g^4 +\dots\\
&G_{J,-+}= \left(G_{2}\right)\big\rvert_{\log} &&= -\frac{8g^2}{1+16g^2} && \simeq-8 g^2+ 128 g^4+\dots\\
&\hat{F}_{J,--} = \left(G_{1}+G_{3}^{+}\right)\big\rvert_{\log} &&=\frac{20 g^2}{1+16 g^2} &&\simeq 20 g^2-320 g^4+\dots
\end{alignedat} \label{eftpred}
\eeq
where we are extracting the coefficient of the logarithmic divergence of the respective integral (and also factorizing out the tree-level respective contribution).

\paragraph{Comparison with data.}We can partially fix the coefficients of the dilatation operator obtained in (\ref{hamiltonian1loop}) by using the perturbative data summarized in Table~\ref{tabstatesbrutef}. The result is 
\beq
\begin{alignedat}{3}
&F_{J,-+} +2 A_{-} &&\simeq24 g^2+\dots \\
&F_{J,+0}  &&\simeq-8 g^2+\dots \\
&\hat{F}_{J,--} -2 A_{-} &&\simeq-16 g^2+\dots \\
\end{alignedat}
\eeq
which is fully consistent with the effective field theory computation (\ref{eftpred}).

	\section{Higher-point Functions with Large Charges}\label{sec:higher} 

	In this section, we will analyze correlation functions involving two large charge operators (referred to as ``heavy'') and additional operators with small charges. In these examples, the $n+2$ correlation function of  two operators $\cO_J$ of large ${\rm U}(1)_R$  charge $J$  and $n$ small charge operators $\cO_{i_m}$ with $m=1,2,\dots,n$  are obtained in the large charge 't Hooft limit as the $n$-point function of $\cO_{i_m}$ computed with the effective action \eqref{largechargeSeff},
	\beq
	\begin{aligned}\label{genlargechargecorrelator}
	\langle \mathcal{O}_{J}(0) \mathcal{O}_{i_1}(x_1) \dots \mathcal{O}_{i_n}(x_n) \mathcal{O}_{J}(\infty) \rangle &= \langle J
	|\mathcal{O}_{i_1}(x_1) \dots \mathcal{O}_{i_n}(x_n)|J\rangle \overset{\text{'t\,Hooft}}{=} \langle \mathcal{O}_{i_1}(x_1) \dots \mathcal{O}_{i_n}(x_n)\rangle_{\rm{eff}},
	\end{aligned}
	\eeq 
	where the heavy state $|J\rangle$ is defined in radial quantization as $\mathcal{O}_{J}(0)|0\rangle$. We will focus on two specific examples: the first is the three-point function of the Konishi operator with two heavy half-BPS operators, while the second is the four-point function of two light half-BPS operators and two heavy half-BPS operators.
	
	Before delving into the computation, let us summarize some basic facts about the half-BPS superconformal primaries in $\mathcal{N}=4$ SYM in order to set the notation. The single-trace half-BPS superconformal primary operator of $\Delta=J\in \mZ_{\geq 2}$ is defined in terms of the adjoint scalars $\Phi_I$ by
\ie 
S_J(x,Y)=  \tr (Y^I \Phi_I)^J
\fe
where $Y^I$ is a null $\mf{so}(6)_R$ polarization vector. The stress tensor multiplet is generated by the primary operator $S_2$. As discussed in Section~\ref{sec:weak}, when the gauge group is SU(2), $S_{J>2}$ can be expressed as products of $S_2$ using the trace relation. Therefore, below we use the following basis of half-BPS operators from the multi-trace operators, which will play the role of the large charge heavy operators in \eqref{genlargechargecorrelator} at large $J$,
\beq\label{eq:formlargeBPS}
\mathcal{O}_J (x,Y)\equiv N_J\left(S_2(x,Y)\right)^{J/2}\comma
\eeq
where $N_J$ is a normalization needed to make the two-point function unit-normalized.
       \subsection{Resummed ladder integrals and worldline instantons\label{subsec:resummedladder}} Before discussing the correlation functions, let us study the propagator of a massive particle around the large charge background which will play a key role in the discussions below.
       
       \paragraph{Massive propagator.}
       The propagator of a massive particle can be computed by expanding the $\mathcal{N}=4$ SYM Lagrangian around the large charge semiclassical solution. As shown in \cite{Giombi:2020enj} and reviewed in Appendix~\ref{backlc}, the result can be expressed as a sum of conformal ladder integrals
    \ie 
       G(x_1,x_2)&=\frac{g_{\rm YM}^2}{8\pi^2|x_{12}|^2}t(z,\bar{z})\comma\\
       t(z,\bar{z})&=\sum_{k} \left( -4 g^2 \right)^k   (1-z)(1-\bar{z})F^{(k)} (z,\bar{z})\period\label{eq:tzzbardef}
\fe
         Here $F^{(k)}$ is a $k$-loop conformal ladder integral (see (\ref{eq:ladder}))
       \beq\label{eq:conformalladder}
       F^{(k)}(z,\bar{z})=\frac{1}{z-\bar{z}}\left[\sum_{r=0}^{k}\frac{(-1)^{r}(2k-r)!}{k!(k-r)!r!}(\log (z\bar{z}))^{r}({\rm Li}_{2k-r}(z)-{\rm Li}_{2k-r}(\bar{z}))\right]\comma
       \eeq
       and $z$ and $\bar{z}$ are the conformal cross ratios\footnote{Note that here we adopted a non-standard definition of the cross ratios. The main reason is to simplify the expressions appearing in the heavy-light OPE.}
       \beq\label{crossratioszzb}
       z\bar{z}=  \frac{x_{14}^2x_{23}^2}{x_{13}^2x_{24}^2}\comma\qquad (1-z)(1-\bar{z})=\frac{x_{12}^2x_{34}^2}{x_{13}^2x_{24}^2}\comma
       \eeq
       and $x_{3,4}$ are positions of insertions of large charge operators.  The vertical lines in the figure of \eqref{eq:ladder} correspond to background scalar fields produced by the large charge operators while the horizontal line corresponds a probe scalar propagating in the presence of the large charge background. This is precisely the resummation of ladder integrals discussed by Broadhurst and Davydychev \cite{Broadhurst:2010ds} but direct physical applications were missing. It is worth emphasizing that the large charge 't Hooft limit provides a physical setup for this quantity. 
       
       A similar but different resummation of the conformal ladder integral shows up in the integrability approach to the four-point function \cite{Fleury:2016ykk} in the planar limit.  Taking inspiration from it, one can rewrite \eqref{eq:tzzbardef} into the following {\it integrability-like} representation,
       \beq
       \begin{aligned}\label{masspropmain}
       t(z,\bar{z})&=\frac{(1-z)(1-\bar{z})}{\sqrt{z\bar{z}}}\sum_{a=1}^{\infty}\int_{-\infty}^{\infty}\frac{du}{2\pi i}\frac{a}{u^2+\frac{a^2}{4}+4g^2}\frac{\sin (a\varphi)}{\sin (\varphi)}e^{-2i u\sigma}\comma
       \end{aligned}
       \eeq
       where $e^{i\varphi}=\sqrt{z/\bar{z}}$ and $e^{-\sigma}=\sqrt{z\bar{z}}$.
       An alternative representation of the massive propagator was found in \cite{Broadhurst:2010ds} by using the integral representation for the conformal ladder integrals:
       \beq\label{tintegralform}
      t(z,\bar{z})=\frac{(1-z)(1-\bar{z})}{2\sqrt{z\bar{z}}}\int_{\sigma}^{\infty} dt \frac{ \sinh t\, J_{0}\left( 4g\sqrt{ t^2-\sigma^2}\right)}{\left(\cos\varphi-\cosh t\right)^2}\period
       \eeq
       Amusingly, the expression resembles the one that appear in the octagon form factor \cite{Bargheer:2019kxb,Kostov:2019stn,Kostov:2019auq,Belitsky:2019fan,Bargheer:2019exp,Basso:2020xts,Belitsky:2020qrm,Belitsky:2020qir,Belitsky:2020qzm,Kostov:2021omc} (see e.g.~(1.31) of \cite{Kostov:2019auq}), which computes the correlation function of four large charge operators in the planar limit. 
       \paragraph{Strong coupling expansion and worldline instantons.} At strong coupling $\lambda_{J} \gg 1$, we will find below exponentially small corrections of order $\mathcal{O}(e^{-\#\sqrt{\lambda_J}})$ to the propagator, which can be interpreted as coming from worldline instantons from the virtual propagation of massive W-bosons.

In order to see this explicitly, we first perform the integral in \eqref{masspropmain} 
\beq\label{eq:tsumbound}
t(z,\bar{z})=\frac{(1-z)(1-\bar{z})}{\sqrt{z\bar{z}}}\sum_{a=1}^{\infty}\frac{ae^{-\sigma \sqrt{a^2+16g^2}}}{\sqrt{a^2+16g^2}}\frac{\sin (a\varphi)}{\sin(\varphi)}\,.
\eeq
As we will see later, this can be interpreted as a conformal block expansion in the heavy-light channel. In order to obtain a strong coupling expansion, we now convert this into the  {\it conformal Regge} representation \cite{Costa:2012cb}.
First we split $a\sin (a\varphi)$ into two terms $(ae^{ia\varphi}-ae^{-ia\varphi})/2i$ and convert the sum over $a$ into an integral using
\beq
\sum_{a}ae^{ia\varphi} \bullet \mapsto \oint_{\mathcal{C_{+}}}du\frac{ue^{i\varphi u}}{1-e^{2\pi i u}}\bullet \comma\qquad -\sum_{a}ae^{-ia\varphi u}\bullet \mapsto \oint_{\mathcal{C_{-}}}du\frac{ue^{i\varphi  u}}{1-e^{2\pi i u}}\bullet\comma
\eeq 
where $\mathcal{C}_{+}$ ($\mathcal{C}_{-}$) is a contour that encircles positive (negative) integers on the real axis. As a result, we obtain
\beq
t(z,\bar{z})=\frac{(1-z)(1-\bar{z})}{z-\bar{z}}\oint_{\mathcal{C}_{+}\cup\mathcal{C}_{-}}du\frac{ue^{i\varphi  u}}{1-e^{2\pi i u}} \frac{e^{-\sigma \sqrt{u^2+16g^2}}}{\sqrt{u^2+16g^2}}\period
\eeq
We then deform the contour so that it encircles the branch cuts $[4g i, i\infty]$ and $[-4g i,-i\infty]$. Each contour can be decomposed as 
\beq
\mathcal{C}_{\pm}=\pm[ i \infty - \epsilon, 4gi- \epsilon] \cup C^{\pm}_{\epsilon} \cup \pm[ 4gi+ \epsilon, i \infty+ \epsilon]
\eeq
where $C^{\pm}_{\epsilon}$ are infinitesimal semicircles of radius $\epsilon$ centered on $\pm 4g i$ which connect the two semi-infinite lines. As $\epsilon \rightarrow 0$ the contribution of the two semicircles vanishes and we can combine the remaining four semi-infinite line integrals composing $\mathcal{C}_{+}\cup\mathcal{C}_{-}$ into a single integral
\beq
\begin{aligned}
&t(z,\bar{z})=\frac{(1-z)(1-\bar{z})}{z-\bar{z}} \int_{4 gi}^{i\infty} du\, \frac{2i u  \sin( u(\pi-\varphi))}{\sin \pi u}\frac{\cosh(\sigma\sqrt{u^2+16g^2})}{\sqrt{u^2+16g^2}} \\
&= \frac{(1-z)(1-\bar{z})}{z-\bar{z}} \, (-2i\partial_{\varphi})  \int_{1}^{\infty} dx\, \sum_{n=0}^{\infty} \left( e^{-4 g x (2 \pi  n+\varphi )}+e^{-4 g x (2 \pi  (n+1)-\varphi )}\right)\frac{\cosh(4g\sigma\sqrt{1-x^2})}{\sqrt{1-x^2}}\comma
\end{aligned} 
\eeq
where in the second line we have changed variables with $u = 4 g i x$, and rewritten the integrand as a series expansion around the strong coupling regime (assuming $0<\Re(\varphi) <2\pi$).

Performing the integral, we arrive at the following expression for the massive propagator,
\beq\label{eq:massivestrong}
\begin{aligned}
t(z,\bar{z})&=\frac{(1-z)(1-\bar{z})}{\sqrt{z\bar{z}}} \,  \sum_{n=0}^{\infty}W(\varphi+2\pi n)+W(2\pi -\varphi+2\pi n) \comma
\end{aligned}
\eeq
where $W(x)$ is given in terms of the modified Bessel function as below,
\beq\label{eq:worldlineinst}
W(x)=\frac{4gx K_{1}(4g\sqrt{x^2+\sigma^2})}{\sin (x) \sqrt{x^2+\sigma^2}}\period
\eeq
To our knowledge, the expression \eqref{eq:massivestrong} has never been written down in the literature, and it makes manifest the structure of the propagator at strong coupling: each term in the sum in \eqref{eq:massivestrong} can be interpreted as the contributions from the worldline instantons\footnote{It should be possible to reproduce the expression \eqref{eq:worldlineinst} by directly performing the worldline path integral, but we leave it for future investigations.} (see \cite{Dondi:2021buw,Hellerman:2021duh,Antipin:2022dsm} for other setups in which the worldline instantons show up in the large charge expansion). The integer $n$ counts how many times the worldline wraps the great circle of ${\rm S}^3$ while the two terms in the summand correspond to different orientations of the worldlines. 

The explicit expansion at strong coupling can be obtained straightforwardly by expanding the Bessel functions,
\beq
\begin{aligned}\label{tstrong}
&t(z,\bar{z})\overset{g\gg1}{\simeq} \sqrt{8 \pi g}\,\frac{(1-z)(1-\bar{z})}{z-\bar{z}}
\\ 
&\times \sum_{n=0}^{\infty}  \Biggl(\frac{ (2 \pi  n+\varphi)  \,e^{-4 g \sqrt{\sigma ^2+(2 \pi  n+\varphi) ^2}}}{\left(\sigma ^2+(2 \pi  n+\varphi) ^2\right)^{3/4}}+\frac{(2 \pi  (n+1)-\varphi) \,e^{-4 g \sqrt{\sigma ^2+(2 \pi  (n+1)-\varphi)^2}}}{\left(\sigma ^2+(2 \pi  (n+1)-\varphi)^2\right)^{3/4}}\Biggr)\,,
\end{aligned}
\eeq
which manifests the existence of an infinite series of non-perturbative corrections.
 The exponential suppression of the resummed ladder integral at strong coupling was pointed out already by Broadhurst and Davydychev \cite{Broadhurst:2010ds}, but the analysis here provides a physical interpretation of such suppression in terms of the worldline instantons in the large charge background. 
These worldline instantons come from the virtual propagation of massive W-bosons of the SYM in the large charge background. Indeed, working on the cylinder $\mR_\tau\times {\rm S}^3$, the mass of the W-boson is 
 \ie 
 m_{\rm W}={4g\over L}\,,
 \label{mw}
 \fe
 which follows from \eqref{mass} after the Weyl transformation (see also \cite{Grassi:2019txd,Hellerman:2021duh}). In the conformal frame where the large charge states are inserted at $\tau=\pm \infty$, and one of the light operator at $\tau=0$ and the north pole on the ${\rm S^3}$, then the other light operator is located at $\tau=\sigma$ and at azimuth angle $\varphi$ on the  ${\rm S^3}$. There is an integer family of geodesics in the form of helices that join the two light operator insertions (see Figure~\ref{helix}). The length of such a geodesic of winding number $n$ in units of the radius $L$ of the $\rm S^3$ is either $\sqrt{\sigma^2+(2\pi n+\varphi)^2}$ or $\sqrt{\sigma^2+(2\pi (n+1)-\varphi)^2}$ depending on its orientation. Putting together with \eqref{mw}, this precisely explain the exponentially suppressed worldline instanton contributions in \eqref{tstrong}.

The resummation of other classes of diagrams in $\mathcal{N}=4$ SYM (the double pentaladders and integrals arising from the negative amplituhedron geometry) were discussed recently \cite{Caron-Huot:2018dsv,Arkani-Hamed:2021iya} and a similar exponential suppression was found at strong coupling. It would be interesting to see if these integrals compute some physical observables in the large charge limit.

\subsection{Three-point function of Konishi and two large charge operators}\label{subsec:Konishi}
We consider here the three-point function involving the Konishi operator $O^{\bf b}_{K}$,
\beq \label{kon}
O^{\bf b}_{K}=\frac{4 \pi ^2}{3 g^2_{{\rm{YM}}}} \tr \left( \Phi_{I}\Phi_{I} \right)=\frac{4 \pi ^2}{3 g^2_{{\rm{YM}}}}\left(2 \Phi^0_{I}\Phi^0_{I}+2\Phi^{+}_{I}\Phi^{-}_{I} \right)\,,
\eeq
 and two large charge operators of the form (\ref{eq:formlargeBPS}). Here we put the superscript ${\bf b}$ since the operator is bare (i.e.~subject to multiplicative renormalization). In the large charge limit, 
 the massless fields evaluated on the  classical background
 produce the tree level  structure constant  
   \beq
\frac{\langle O^{\bf b}_{K}(x_1) O_{J}(x_3) O_{J }(x_4)\rangle_{\rm{tree}}}{\langle O_{J}(x_3) O_{J}(x_4) \rangle} =  \frac{8 \pi ^2}{3 g^2_{{\rm{YM}}}} \int \frac{d\theta}{2\pi}\langle \Phi^0_{I}\rangle_{\rm{cl}} \langle \Phi^0_{I}\rangle_{\rm{cl}} = \frac{Jx_{34}^2}{3x_{13}^2x_{14}^2 } \,,
 \eeq
 where the classical solution\fn{The semiclassical configuration for $Y_3=\frac{1}{\sqrt{2}}(1,i,0,0,0,0)$ and $Y_4=\frac{1}{\sqrt{2}}(1,-i,0,0,0,0)$ is written down in the Appendix~\ref{backlc}. Performing the $R$-symmetry transformation, one can find the analog for general $Y_{3,4}$.} is given by
\beq\label{genbackground}
\langle \Phi_I^{0}\rangle_{\rm{cl}}=\frac{\sqrt{\lambda_J}|x_3-x_4|}{2\pi \sqrt{2Y_{34}}}\left(\frac{e^{i\theta}(Y_3)_{I}}{|x-x_3|^2}+\frac{e^{-i\theta}(Y_4)_{I}}{|x-x_4|^2}\right)\comma
\eeq
where $\theta$ is a `moduli' of the solutions that we need to integrate over (see e.g.~\cite{Monin:2016jmo,Yang:2021kot,Bajnok:2014sza}).

The massive fields on the other hand are  responsible for the non-trivial dependence on the 't Hooft coupling. At leading order in $J$, the non-trivial part of the structure constant arises  from the massive propagator connecting the two massive fields in the limit of coincident points.  The expression of the massive propagator is given in \eqref{eq:tzzbardef} and we will consider the case $k\geq1$, as $k=0$ corresponds to a self-contraction of the Konishi operator.
We take the OPE limit of the propagator, where the fields at the positions $x_{1}$ and $x_{2}$ are brought infinitesimally close to each other with $|x_{12}|=\epsilon$. This is analogous to a perturbative point-splitting regularization for the three-point function (see \cite{Caetano:2014gwa}). We can implement this limit in the cross-ratios by
\beq \label{cratioslimit}
z\to 1-\delta \,, \quad \bar{z}\to 1-\delta.
\eeq 
where $\delta^2 = \frac{\epsilon^2 x_{34}^2}{x_{13}^2 x_{14}^2}$. From the expression of the conformal ladder integrals (\ref{eq:conformalladder}) we obtain in this limit
\beq\label{eq:Fkexpanded}
\begin{aligned}
&F^{(1)}=2-2\log \delta \comma\\
&F^{(k\geq 2)}= \frac{4^{k}\Gamma (k+\frac{1}{2})}{\sqrt{\pi}\Gamma (k+1)}\zeta (2k-1)\period
\end{aligned}
\eeq
Let us first consider the cases with $k\geq2$ which remain finite as we consider (\ref{cratioslimit}).
Using the integral representation for the $\zeta$-function, one can resum the ladders into
\beq
\tilde{c}_K\equiv \sum_{k=2}^{\infty}(-4g^2)^{k}F^{(k)}(z,\bar{z})=2g\int_0^{\infty}dw \frac{4gw-J_{1}(8g w)}{\sinh^2(w)}\period 
\eeq
We then obtain
\beq
\frac{\langle O^{\bf b}_{K} O_{J} O_{J }\rangle}{\langle O_J O_J \rangle}=\frac{ Jx_{34}^2}{3x_{13}^2x_{14}^2 } \left(1- \frac{48g^2}{J}\left(1-\log \delta\right)+\frac{6\tilde{c}_{K}}{J} \right)\period
\eeq

In order to extract the structure constant, we need the Konishi operator normalization given by the corresponding two-point function. At the order we are working, we will only need the one-loop result because higher loop corrections  will only be relevant for higher terms in the $1/J$ expansion.
It has been observed for example in \cite{Bianchi:2000hn}, that the one-loop two-point function of the Konishi operator for any gauge group SU($N_c$) does not receive non-planar corrections. Using this observation and putting together the explicit results for the relevant diagrams listed in the Appendix B of \cite{Caetano:2014gwa}, we compute the two-point function in point-splitting regularization
\beq \label{twoptKon}
\frac{\langle O^{\bf b}_{K}(x_1) O^{\bf b}_{K}(x_2) \rangle}{\langle O^{\bf b}_{K}(x_1) O^{\bf b}_{K}(x_2)\rangle_{\rm{tree}}} =1- \frac{3 g^2_{\rm{YM}} N_c }{4\pi^2} \left(1- \log \left(\frac{\epsilon ^2}{x_{12}^2}\right)\right)\,.
\eeq
Thus, the renormalized three-point function reads
\beq
\frac{\langle O^{\bf ren}_{K} O_{J} O_{J }\rangle}{\langle O_J O_J \rangle}=\frac{ Jx_{34}^2}{3x_{13}^2x_{14}^2 } \left(1+ \frac{24g^2}{J}\log \frac{x_{34}^2}{x_{13}^2x_{24}^2}+6\frac{(\tilde{c}_{K}-4g^2)}{J} \right)
\eeq
From this, we can read off the one-loop anomalous dimension $\Delta_K=2+g_{\rm YM}^2\Delta_1+\cdots$ and the large charge expansion of the structure constant $C_K\,=JC_{K}^{(0)}+C_K^{(1)}+C_{K}^{(2)}/J+\cdots$,
\beq\label{eq:konishideltaC}
\begin{aligned}
&\Delta_1=\frac{48g^2}{g_{\rm YM}^2 J}=\frac{3}{2\pi^2}\comma\qquad C_K^{(0)}=\frac{1}{3}\comma\qquad C_K^{(1)}=-8g^2+4g\int_0^{\infty}dw \frac{4gw-J_{1}(8g w)}{\sinh^2(w)}\period
\end{aligned}
\eeq
The anomalous dimension agrees with the result in the literature \cite{Bianchi:2000hn} while the structure constant $C_K$ is a genuinely new prediction, which we will study in more detail below. See Figure~\ref{interpolplots} for a plot of this function. 

\paragraph{Weak- and strong-coupling expansions.} Let us study the weak and strong coupling expansions of $C_K^{(1)}$. The weak coupling expansion follows directly from \eqref{eq:Fkexpanded} and is given by
\beq
C_{K}^{(1)}=-8g^2+\sum_{k=2}^{\infty}\frac{\zeta (2k-1)\Gamma (k+\frac{1}{2})}{\sqrt{\pi}\Gamma(k+1)}(-16g^2)^{k}\period
\eeq
The series has a finite radius of convergence
\beq\label{weakradiusconv}
|\lambda_J|<\pi^2\period
\eeq
This parallels the fact that observables in the planar limit has a finite radius of convergence as a function of the 't Hooft coupling. A physical interpretation of the radius of convergence will be presented in Section~\ref{subsubsec:convergence}.

On the other hand, the strong coupling expansion can be derived by taking the limit, $z,\bar{z}\to 1-\delta$, of \eqref{eq:massivestrong} and subtracting the contributions from $F^{(0,1)}$. As a result, we obtain
\beq\label{eq:KonishiCnext}
C_K^{(1)}=\frac{\lambda_J}{2\pi^2}\left[2\gamma_{E}+\log\left(\frac{\lambda_J}{4\pi^2}\right)\right]-\frac{2\lambda_J}{\pi^2}\sum_{n=1}^{\infty}\left(K_{2}(2n\sqrt{\lambda_J})+K_{0}(2n\sqrt{\lambda_J})\right)\period
\eeq
The Bessel functions on the RHS are exponentially suppressed for large $\lambda_J$, and they capture the worldline instanton contributions. The structure of the result resembles the ones for the extremal correlator in $\mathcal{N}=2$ SCFT \cite{Grassi:2019txd,Hellerman:2021duh,Hellerman:2021yqz}. In that case, the coefficient of $\log \lambda_J$ is given by the $a$-anomaly \cite{Grassi:2019txd}. Here it is instead given by the anomalous dimension of the Konishi operator as we will show in Section~\ref{subsubsec:coulomb3pt}
	\subsection{Heavy-heavy-light-light four-point functions}\label{subsec:HHLL}
Four-point functions of BPS primaries $\mathcal{O}_J$ contain a wealth of information including the non-BPS operator spectrum that appear in the OPE channels and their corresponding OPE coefficients. 
In this subsection, we focus on the HHLL four-point function $\la \mathcal{O}_2 \mathcal{O}_2\mathcal{O}_J\mathcal{O}_J\ra$ in the large charge 't Hooft limit. 

\subsubsection{Four-point functions at large charge}
\begin{figure}[t] 
\includegraphics[width=13cm]{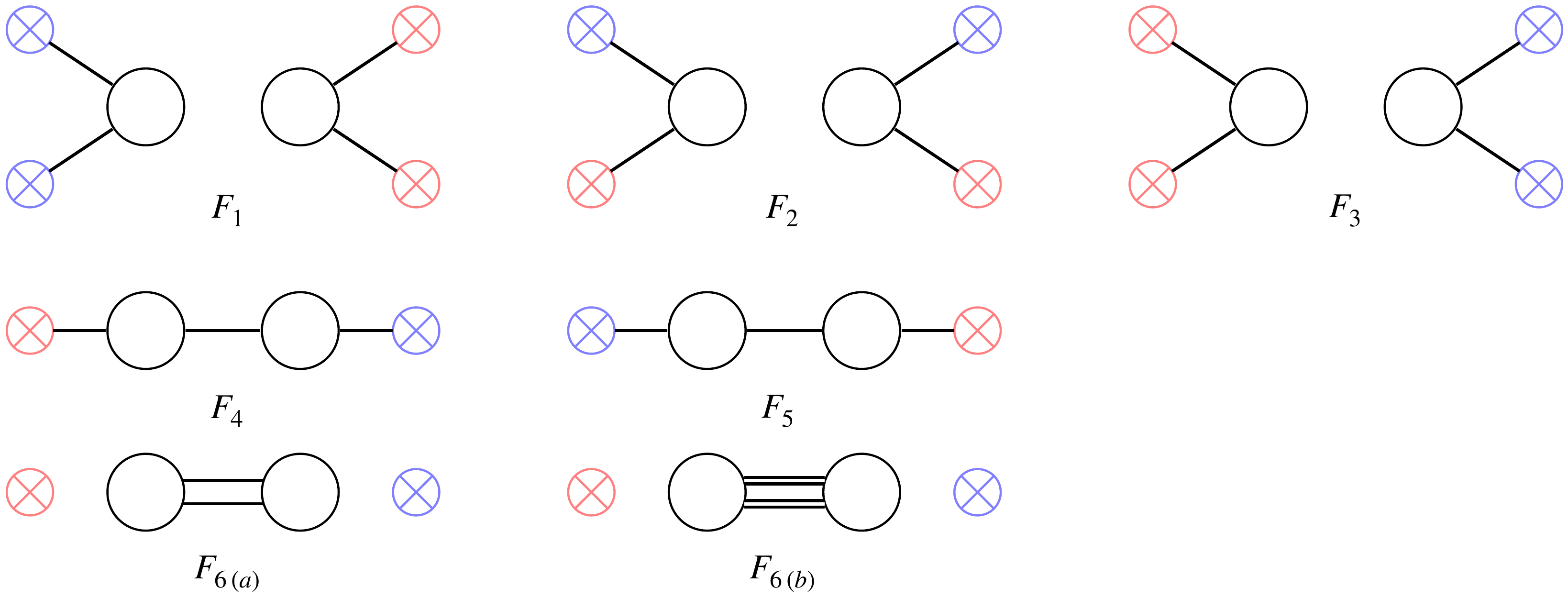}
\centering
\caption{The figure represents the diagrammatics of the four-point correlation function for the leading term in the large charge expansion. The light operators are depicted as black circles, while connections to crossed circles indicate the evaluation of fields on the classical solution. The red and blue crossed circles serve to differentiate between the two terms of the classical solution (see \eqref{genbackground}). The last row corresponds to $F_6$ and we distinguish the massless ($F_{6\, (a)}$) and massive propagators ($F_{6\, (b)}$) by single and double lines respectively.}\label{fourptgraphs}
\end{figure}

The HHLL four-point function $\la \mathcal{O}_2 \mathcal{O}_2\mathcal{O}_J\mathcal{O}_J\ra$ is a nontrivial function of the conformal cross ratios and a polynomial in the $R$-symmetry polarization \cite{Dolan:2001tt},
\beq\label{eq:defF123456}
\begin{aligned}
&\langle \mathcal{O}_2(x_1,Y_1)\mathcal{O}_2(x_2,Y_2)\mathcal{O}_J(x_3,Y_3) \mathcal{O}_J(x_4,Y_4)\rangle
=\\
&\frac{Y_{13}Y_{24}Y_{14} Y_{23}Y_{34}^{J-2}}{x_{13}^2x_{24}^2x_{14}^{2}x_{23}^{2}x_{34}^{2J-4}} \left[\frac{F_{1}(z,\bar{z})}{\alpha\bar{\alpha}}+F_{2}(z,\bar{z})+\alpha\bar{\alpha}F_{3}(z,\bar{z})\right.\\
&\left. +\frac{(1-\alpha)(1-\bar{\alpha})}{\alpha\bar{\alpha}}\left(F_4(z,\bar{z})+\alpha\bar{\alpha}F_{5}(z,\bar{z})\right)+(1-\alpha)^2(1-\bar{\alpha})^2\frac{F_{6}(z,\bar{z})}{\alpha\bar{\alpha}}\right]\comma
\end{aligned}
\eeq
where the conformal cross ratios are given by \eqref{crossratioszzb} 
and $\alpha$ and $\bar{\alpha}$ are the R-symmetry cross ratios
\beq
\alpha\bar{\alpha}\equiv \frac{Y_{14}Y_{23}}{Y_{13}Y_{24}}\comma\qquad (1-\alpha)(1-\bar{\alpha})\equiv \frac{Y_{12}Y_{34}}{Y_{13}Y_{24}}\,,
\eeq
with $Y_{ij}\equiv Y_i\cdot Y_j$.
As we see below, the scaling of $F_j$'s in the large charge 't Hooft limit is non-uniform,
\beq
F_{1,2,3} \sim O(J^2)\comma\qquad F_{4,5}\sim O(J) \comma\qquad F_{6}\sim O(1)\period
\eeq
In what follows, we compute the leading term\fn{Since the dependence on the $R$-symmetry polarization is different, one can easily distinguish the subleading corrections to ${F}_{1,2,3}$ from the leading contributions to ${F}_6$ although both can be of order $1$.} for each ${F}_j$ in the large charge 't Hooft limit by evaluating the diagrams depicted in Figure~\ref{fourptgraphs}.

\paragraph{Computation of $F_{1,2,3}$.} At the leading order in the large charge 't Hooft limit, the four-point function is given by expectation value of the light operators in the classical background,
\beq
\frac{\langle \mathcal{O}_2\mathcal{O}_2\mathcal{O}_J\mathcal{O}_J\rangle}{\langle \mathcal{O}_J \mathcal{O}_J\rangle} \overset{O(J^2)}{=} \int \frac{d\theta}{2\pi}\langle \mathcal{O}_2\rangle_{\text{cl}}\langle \mathcal{O}_2\rangle_{\text{cl}}\period
\eeq
Plugging the expression (\ref{genbackground}) into $\mathcal{O}_2$,
\beq\label{O2norm}
\mathcal{O}_2= \frac{1}{\sqrt{6}}\left(\frac{8\pi^2}{g_{\rm YM}^2}\right){\rm tr}\left(Y\cdot \Phi\right)^2\comma
\eeq
and integrating over $\theta$, we find
\beq\label{eq:f123basic}
\int \frac{d\theta}{2\pi}\langle \mathcal{O}_2\rangle_{\text{cl}}\langle \mathcal{O}_2\rangle_{\text{cl}}=\frac{J^2}{6}\frac{x_{34}^4}{(Y_{34})^2}\left[\frac{(Y_{13})^2(Y_{24})^2}{x_{13}^4x_{24}^4}+\frac{4Y_{13}Y_{14}Y_{23}Y_{24}}{x_{13}^{2}x_{14}^{2}x_{23}^{2}x_{24}^{2}}+\frac{(Y_{14})^2(Y_{23})^2}{x_{14}^4x_{23}^4}\right]\period
\eeq
Comparing this expression with \eqref{eq:defF123456} using $\langle \mathcal{O}_J\mathcal{O}_J\rangle=(Y_3\cdot Y_4)^{J}/x_{34}^{2J}$, we find
\beq\label{F123}
\frac{F_1}{\alpha\bar{\alpha}}+F_2+\alpha\bar{\alpha}F_3=\frac{J^2}{6}\left(\frac{z\bar{z}}{\alpha\bar{\alpha}}+4+\frac{\alpha\bar{\alpha}}{z\bar{z}}\right)\period
\eeq
\paragraph{Computation of $F_{4,5}$.} At the next order, the four-point function receives a correction from a diagram in which a massless particle is exchanged between the $\mathcal{O}_2$'s and the remaining scalar in each light operator is replaced with the semiclassical VEV (see Figure~\ref{fourptgraphs}). Computing this diagram, we find
\beq
\frac{\langle \mathcal{O}_2\mathcal{O}_2\mathcal{O}_J\mathcal{O}_J\rangle}{\langle \mathcal{O}_J \mathcal{O}_J\rangle} \overset{O(J)}{=}\frac{J}{3}\frac{x_{34}^2Y_{12}}{x_{12}^2Y_{34}}\left[\frac{Y_{13}Y_{24}}{x_{13}^2x_{24}^2}+\frac{Y_{14}Y_{23}}{x_{14}^2x_{23}^2}\right]\period
\eeq
This leads to
\beq\label{F_45}
\frac{(1-\alpha)(1-\bar{\alpha})}{\alpha\bar{\alpha}}\left(F_4(z,\bar{z})+\alpha\bar{\alpha}F_{5}(z,\bar{z})\right)=\frac{J}{3}\left(1+\frac{\alpha\bar{\alpha}}{z\bar{z}}\right)\frac{z\bar{z}(1-\alpha)(1-\bar{\alpha})}{\alpha\bar{\alpha}(1-z)(1-\bar{z})}\period
\eeq
\paragraph{Computation of $F_6$.} Finally at the next-to-next order, the four-point function receives a contribution from a diagram that includes exchanges of two massless or massive particles between the $\mathcal{O}_2$'s.\footnote{Recall that in the particle picture, $\mathcal{O}_2 = \frac{2}{\sqrt{6}}\left(\frac{8\pi^2}{g_{\rm YM}^2}\right) Y^{I}Y^{J} \left( \Phi_{I}^{0}\Phi_{J}^{0}+\Phi_{I}^{+}\Phi_{J}^{-}\right)$.} This diagram produces the following result
\beq \label{massexchange}
\frac{\langle \mathcal{O}_2\mathcal{O}_2\mathcal{O}_J\mathcal{O}_J\rangle}{\langle \mathcal{O}_J \mathcal{O}_J\rangle} \overset{O(J^0)}{=}\frac{Y_{12}^2}{3x_{12}^4}\left(1+2t(z,\bar{z})^2\right)\comma
\eeq
where the first term arises from the massless exchange and the second from the massive propagator (see also Figure~\ref{fourptgraphs}).
From this result, we read off the answer for $F_6$,
\beq \label{F6}
(1-\alpha)^2(1-\bar{\alpha})^2\frac{ F_6}{\alpha\bar{\alpha}} =\left(\frac{(1-\alpha)(1-\bar{\alpha})}{(1-z)(1-\bar{z})}\right)^2 \frac{z\bar{z}}{\alpha\bar{\alpha}} \frac{1+2t(z,\bar{z})^2}{3}\period
\eeq

\subsubsection{Heavy-light OPE at large charge}\label{subsec:heavylightOPE}
We will now show that the result of the four-point function (\ref{eq:defF123456}) and in particular the non-trivial piece (\ref{F6}) given in terms of ladder integrals naturally manifests the OPE structure. 

\paragraph{General OPE structure.}Let us begin by discussing the general structure of the OPE of two superconformal half-BPS primaries in the $[0,p,0]$ representation of $\mf{su}(4)_R$ in the large charge limit. We will be studying the heavy-light OPE channel ($s$-channel) of the four-point function for which we take the limit $z,\bar{z} \rightarrow 0$. Each term in (\ref{eq:defF123456}) admits an OPE expansion in terms of the standard four-dimensional conformal blocks 
that we can write as
\beq \label{opeexp}
F_{i}(z,\bar{z})=\sum_{\Delta, S} c_{\Delta,S} \,\mathcal{G}_{\Delta,S}(z,\bar{z})\,,
\eeq
where $c_{\Delta,S}$ denotes the product of the two three-point couplings between the two external operators $\mathcal{O}_{J}$ and $\mathcal{O}_{2}$ together with the exchanged primary $\mathcal{O}_{\Delta,S}$
\beq
c_{\Delta,S}=|C_{\mathcal{O}_{J}\mathcal{O}_{2}\,\mathcal{O}_{\Delta,S}}|^2\,,
\eeq
while the function $\mathcal{G}$ is given by $\mathcal{G}\equiv (z\bar{z})^{-\frac{J}{2}} G^{2-J,2-J}_{\Delta,S}$ with $G^{2-J,2-J}_{\Delta,S}$ being the conformal block for a primary operator of spin $S$ and dimension $\Delta$ given in \cite{Dolan:2000ut}.

As first noted in \cite{Jafferis:2017zna}, the large charge limit suppresses the contribution to the OPE from the conformal descendants  and what remains is the exchange of the primary operator of the conformal multiplet. Indeed, in the limit $J\rightarrow\infty$ the function $\mathcal{G}_{\Delta,S}$ simplifies to 
\beq  \label{confblocklc}
\mathcal{G}_{\Delta,S}(z,\bar{z}) \overset{J\to\infty}{=}  e^{-\sigma (\Delta-J)} \,\frac{\sin (S+1)\varphi}{\sin\varphi}
\eeq
where we have used the parametrization of the cross-ratios in terms of the $\sigma, \varphi$ defined as $e^{i\varphi} = \sqrt{z/\bar{z}}$ and $e^{-\sigma}=\sqrt{z\bar{z}}$ and $C^{(0)}_{S}$ denotes a Gegenbauer polynomial,
\beq  \label{blocklc}
C^{(0)}_{S}(\cos \varphi) = \frac{\sin(S+1)\varphi}{\sin\varphi}\,.
\eeq

It is useful to describe the primary operators  exchanged in the OPE from the magnon picture. The external operators are described in terms of the following magnon states,
\ie \label{external}
&\mathcal{O}_{J}  \leftrightarrow |0\rangle_{J}\,,
\\
&\mathcal{O}_2(Y) \leftrightarrow \frac{2}{ \sqrt{6}} \left(\frac{8\pi^2}{g_{\rm YM}^2}\right) \,Y^{I}Y^{K}\left( | \Phi^{0}_{I} \Phi_{K}^{0} \rangle_0+| \Phi^{+}_{I} \Phi_{K}^{-} \rangle_0\right)  \equiv \frac{2}{ \sqrt{6}}  \left(\frac{8\pi^2}{g_{\rm YM}^2}\right)\left( | x_1^{0} x_1^{0} \rangle_0 +| x_1^{+} x_1^{-} \rangle_0 \right) \,,
\fe
where we have used the normalization as in (\ref{O2norm}).
The subindices in the magnons reinforce that these are in the fundamental representation. 
For each R-symmetry structure in (\ref{eq:defF123456}), we find a different set of states appearing in the OPE as we now describe.

\paragraph{OPE expansion of $F_{1,2,3}$\,.}The contribution to $F_{1,2,3}$ arises from the part of $\mathcal{O}_2$ in \eqref{external} where both elementary fields are aligned with the vacuum polarizations. Therefore, the exchanged state is also a vacuum state at leading order in $J$. We can summarise the large charge OPE by
\beq
\begin{aligned}
|0\rangle_{J} \otimes 
 | 0 \rangle_2 &=\bigoplus_{n={-1,0,1}}| 0\rangle_{J+2n} \,. \label{vacvac}
\end{aligned}
\eeq
The three states in the right hand side have dimensions $\Delta= J+2,J,J-2$ and no spin and each of them  give rise to $F_{1,2,3}$ respectively. Combining (\ref{opeexp}) and (\ref{confblocklc}) with $S=0$ we get
\beq
\begin{aligned}
F_{1}&= (z\bar{z}) \, c_{J+2,0}\,, \qquad F_{2}&=   c_{J,0}\,, \qquad F_{3}&=  \frac{c_{J+2,0}}{z\bar{z}} \,.
\end{aligned}
\eeq
This matches the structure in (\ref{F123}) and the comparison determines the structure constants
\beq
c_{J+2,0} =\frac{J^2}{6}
\,,\quad c_{J,0} = 
\frac{2J^2}{3}\,,\quad 
c_{J-2,0} = \frac{J^2}{6}\,.
\eeq

\paragraph{OPE expansion of $F_{4,5}$\,.} 
From the R-symmetry structure of $F_{4,5}$, it follows that the polarization of one of the elementary fields in $\mathcal{O}_2$ is aligned with the vacuum while the other field is exchanged and contracted with the remaining short state. Therefore, we find two states in the OPE each one with an excitation carrying arbitrary spin and differing by their R-charges. This excitation is identified with a massless magnon in a bound-state representation whose energy does not receive corrections at leading order in the large charge limit. The OPE structure in terms of magnon states is then given by
\beq
|0\rangle_{J} \otimes | x_1^{0}  \rangle_1 = \bigoplus_{S=0}^{\infty}\bigoplus_{n=-1,1}| x_{1+S}^{0}\rangle_{J+n}  \label{vacone}
\eeq
 The dimensions of these operators are $\Delta =J+ S+2,J+S$ in one-to-one correspondence with $F_4$ and $F_5$ respectively. From the OPE expansion of these four-point function, we have
\beq
\begin{aligned} \label{opef45}
F_{4} &= z\bar{z} \sum_{S=0}^{\infty}c_{J+S+2,S}\, ( z\bar{z})^{\frac{S}{2}} \,\frac{\sin(S+1)\varphi}{\sin \varphi} 
\\
F_{5} &=\sum_{S=0}^{\infty}c_{J+S,S}\, ( z\bar{z})^{\frac{S}{2}} \,\frac{\sin(S+1)\varphi}{\sin \varphi } 
\end{aligned}
\eeq
In order to see this structure from the explicit formulas (\ref{F_45}), we make use of the integrability-like representation of the massless propagator which can be obtained by setting $g=0$ in \eqref{integlike}, leading to the following identity after performing the integral in $u$,
\beq \label{facttosum}
\frac{1}{(1-z)(1-\bar{z})} =\sum_{a=1}^{\infty}  \, e^{-\sigma (a-1)}\,\frac{ \sin a\varphi}{\sin \varphi}\,.
\eeq
The index $a$ labels the bound-state representation and it relates to the spin by $a=S+1$. By plugging this identity in (\ref{F_45}), we immediately recover the structure (\ref{opef45})
and by comparison we obtain that
\beq
c_{J+S+2,S} = \frac{J}{3}\,,\qquad c_{J+S,S} =\frac{J}{3}\,,
\eeq
and they are independent of the spin at this order.
\paragraph{OPE expansion of $F_{6}$\,.} In $F_{6}$ we find that the light operators are fully contracted. The exchanged operators contain two excitations which may be massless or massive depending on which term of  $\mathcal{O}_2$ we consider, see (\ref{external}), and they carry arbitrary spin. Hence, the OPE can be summarized by 
\beq
\begin{aligned}
|0\rangle_{J} \otimes | x_1^{\ell_1} x_1^{\ell_2} \rangle_0 &= \bigoplus_{S=0}^{\infty }\bigoplus_{ n=0}^{\infty}\bigoplus_{\substack{\texttt{a}=1}}^{ \left\lceil \frac{S+1}{2}\right\rceil }| x_{\texttt{a}+n}^{\ell_1}\,x_{S+2-\texttt{a}+n}^{\ell_2} 
\rangle_{J}\,,\label{vactwo}
\end{aligned}
\eeq
where $x_{p}^{\ell_1}$ and $x_{q}^{\ell_2}$ denote  magnons in the $p$-th and $q$-th bound-state representation and $n$ counts the number of pairs of contracted derivatives acting on both fields. The spin of the state appearing on the right hand side is $S$.
The description above is schematic and in order to obtain the correct physical states one should properly symmetrize/anti-symmetrize the U(1) indices of the two excitations depending on the parity of $J$  (to construct gauge-invariant operators) as described in earlier sections, thus the upper limit in the sum.

Our aim now is to rewrite the result (\ref{F6}) in the form (\ref{opeexp}) and read off the relevant three-point coefficients along the way. 
Let us start with the massless exchange that corresponds to the first term in (\ref{F6}). The operators with $\ell_1=\ell_2=0$ in (\ref{vactwo}) do not carry anomalous dimension at this order in the large charge limit and their dimensions are given by $\Delta=J+S+2n+2$. We can now proceed by using the relation (\ref{facttosum}) in (\ref{F6}), and rewrite the result using the identity
\beq \label{sinident}
\sin a\varphi \sin b \varphi = \sin \varphi \sum_{n=0}^{b-1} \sin(a+b-1-2n)\varphi
\eeq
to obtain
\beq
F^{\rm{massless}}_6 = \frac{z\bar{z}}{3} \sum_{a,b=1}^{\infty}\sum_{n=0}^{b-1} e^{-\sigma(a+b-2)}  \frac{\sin(a+b-1-2n)\varphi}{\sin \varphi}\,,
\eeq
where $F^{\rm{massless}}_6 $ denotes the first term of $F_6$ in (\ref{F6}).
To make the tower of operators of (\ref{vactwo}) more evident we simply redefine the labels according to $a=\texttt{a}+n$ and $b=S+2-\texttt{a}+n$, where it becomes clear that $a-1$ counts the total number of  derivatives acting on a field with $n$ of them being contracted to the other field. This equation turns into
\beq
F_6^{\rm{massless}} = \frac{(z\bar{z})^{-\frac{J}{2}}}{3} \sum_{S=0}^{\infty}\sum_{n=0}^{\infty}
(S+2n+1)
(z\bar{z})^{\frac{J+S+2n+2}{2}} \frac{\sin(S+1)\varphi}{\sin \varphi}\,.
\eeq
from where we read off the structure constant
\beq
c_{J+S+2n+2,S}=\frac{S+2n+1}{3}. 
\eeq
We proceed with the massive exchange given by the second term in (\ref{F6}) and make use of the integrability-like representation  (\ref{masspropmain}) to obtain
\beq \label{4ptope}
F_6^{\rm{massive}}  =\frac{2}{3} \sum_{a,b=1}^{\infty} C_{ab}\, e^{\sigma \left(\sqrt{a^2+16g^2}+\sqrt{b^2+16g^2} \right)}\frac{\sin(a\varphi) \sin(b\varphi)}{\sin^2(\varphi)}\,,
\eeq
with
\beq
C_{ab} = \frac{ \,a b}{\sqrt{a^2+16g^2}\sqrt{b^2+16g^2}}\,.
\eeq
The OPE structure becomes manifest once we use again the formula (\ref{sinident})
to rewrite (\ref{4ptope}) as 
\beq 
F_6^{\rm{massive}} =\frac{2\left(z\bar{z}\right)^{-\frac{J}{2}}}{3}  \sum_{S=0}^{\infty}\sum_{n=0}^{\infty}\sum_{\texttt{a}=1}^{\left\lceil \frac{S+1}{2}\right\rceil}C_{\texttt{a}+n\,S+2n+2-\texttt{a}}\, (z\bar{z})^{\frac{\Delta_{J,\texttt{a},n,S}}{2}}
\,\frac{\sin(S+1)\varphi}{\sin\varphi} \,,
\eeq
where we have redefined the indices $a$ and $b$ as before $a=\texttt{a}+n$ and and $b=S+2-\texttt{a}+n$, and 
\beq 
\Delta_{J,\texttt{a},n,S} =J+\sqrt{(\texttt{a}+n)^2+16g^2}+\sqrt{(S+n+2-\texttt{a})^2+16g^2}
\eeq 
are the dimensions of the exchanged operator. Finally, we read off the structure constant $c_{\Delta,S} $,
\beq
c_{\Delta,S} = \frac{ 4\,(\texttt{a}+n) (S+n+2-\texttt{a})}{3\sqrt{(\texttt{a}+n)^2+16g^2}\sqrt{(S+n+2-\texttt{a})^2+16g^2}}\,.
\eeq

\subsection{Relation to the Coulomb branch}\label{sec:coulomb}
It is known that the large charge limit of SCFTs is described by the effective action on the Coulomb branch, see e.g.~\cite{Hellerman:2017veg,Hellerman:2017sur}. From the point of view of the large charge 't Hooft limit discussed in this paper, the standard large charge limit corresponds to the strong coupling limit, in which $\lambda_{J}=g_{\rm YM}^2 J/2$ is sent to infinity. In this subsection we explain how the physics on the Coulomb branch is reproduced from the large charge 't Hooft limit at strong coupling.

For this purpose, we use the following formula relating correlation functions at large charge and correlation functions on the Coulomb branch (see \cite{toappear} for more details and the derivation):
\beq\label{eq:generalformulacoulomb}
\lim_{\substack{r,J\to \infty\\\frac{\sqrt{\lambda_J}}{\pi r}:\text{ fixed}}}\int \frac{d^3\vec{n}}{2\pi^2}\frac{\left<\mathcal{O}_J(r\vec{n},Y)\,\mathcal{O}_J(-r\vec{n},\bar{Y})\prod_j\mathcal{O}_j(x_j)\right>}{\left<\mathcal{O}_J(r\vec{n},Y)\,\mathcal{O}_J(-r\vec{n},\bar{Y})\right>}=\int \frac{d\theta}{2\pi}\left<\prod_{j}\mathcal{O}_j(x_j)\right>_{v_{I}(\theta)}
\eeq
Here the LHS is the correlation function with two large charge operators inserted at antipodal positions (with respect to the origin) while the RHS is the correlation function on the Coulomb branch where the scalar acquires the expectation value
\beq\label{eq:chargeVEV}
\langle\Phi_I\rangle =\left(\begin{array}{cc}v_I(\theta)&0\\0&-v_I(\theta)\end{array}\right)\comma\qquad \qquad v_I(\theta)=\frac{\sqrt{\lambda_J}}{\pi r}\frac{e^{i\theta}y_I +e^{-i\theta}\bar{y}_I}{\sqrt{2(y\cdot \bar{y})}} \comma
\eeq
 with $|v|=\frac{\sqrt{\lambda_J}}{\pi r}$. In what follows, we call this limit the {\it macroscopic limit} following the terminology in \cite{Jafferis:2017zna}.
 
 Note that the formula \eqref{eq:generalformulacoulomb} was derived for the standard large charge limit (i.e. fixed $g_{\rm YM}$ and large $J$). Strictly speaking, this is not exactly the same\footnote{In other words, there could potentially be an order-of-limits issue.} as the strong coupling limit of the large charge 't Hooft limit, which corresponds to first taking $J\to\infty$ with fixed $\lambda_J$ and then sending $\lambda_J\to\infty$.  Nevertheless, below we see that many of the features of the physics on the Coulomb branch can be reproduced from the large charge 't Hooft limit at strong coupling. 
 \subsubsection{Three-point function}\label{subsubsec:coulomb3pt}
 Applying the general formula \eqref{eq:generalformulacoulomb} to the three-point function of scalar primary $\mathcal{O}$, we obtain the relation 
 \beq\label{eq:comparethreepoint}
\lim_{\substack{r,J\to \infty\\\frac{\sqrt{\lambda_J}}{\pi r}:\text{ fixed}}} C_{JJ\mathcal{O}}\left(\frac{2}{r}\right)^{\Delta_{\mathcal{O}}}=\int \frac{d\theta}{2\pi}\langle \mathcal{O}(0)\rangle_{v_I (\theta)}\comma
 \eeq
 where $C_{JJ\mathcal{O}}$ is the structure constant of two BPS large charge operators $\mathcal{O}_J$ and a general scalar primary $\mathcal{O}$. The conformal symmetry fixes the dependence on $|v|$ on the RHS to be
 \beq
 \int\frac{d\theta}{2\pi}\langle \mathcal{O}(0)\rangle_{v_I (\theta)}=|v|^{\Delta_{\mathcal{O}}}\int \frac{d\theta}{2\pi} c_{\mathcal{O}}(\theta)\comma
 \eeq
 where $c_{\mathcal{O}}(\theta)$ is $O(1)$ quantity independent of $|v|$. Combining the two equations, we then obtain the prediction for the behavior of the three-point function at large charge 
 \beq\label{eq:predictionthree}
 C_{JJ\mathcal{O}} \overset{J\to\infty}{=} \left(\frac{\sqrt{\lambda_J}}{2\pi}\right)^{\Delta_{\mathcal{O}}}\int \frac{d\theta}{2\pi}c_{\mathcal{O}}(\theta)\period
 \eeq
 This in particular implies that the structure constant $C_{JJ\mathcal{O}}$ must scale as $\propto J^{\Delta_{\mathcal{O}}/2}$ at large $J$.
 To compare this with the strong coupling limit of the result in Section~\ref{subsec:Konishi}, we expand $\Delta_{\mathcal{O}}$ and $c_{\mathcal{O}}(\theta)$ in powers of $g_{\rm YM}$,
 \begin{align}
 \Delta_{\mathcal{O}}=\Delta_0+g_{\rm YM}^2 \Delta_1 +\cdots\comma\qquad \int \frac{d\theta}{2\pi}c_{\mathcal{O}}(\theta)=\frac{1}{(g_{\rm YM})^{\Delta_0}}\left(c_0+g_{\rm YM}^2 c_1+\cdots\right)\comma
 \end{align}
 where we have used the fact that the tree-level one-point function of scalar primary on the Coulomb branch scales as $1/g_{\rm YM}^{\Delta_0}$ (see \cite{toappear}). 
We can then rewrite \eqref{eq:predictionthree} in terms of $\lambda_J$ and $J$ as
 \beq
 C_{JJ\mathcal{O}}\overset{J\to\infty}{=}\left(\frac{\sqrt{J/2}}{2\pi}\right)^{\Delta_0}\left[c_0+ \frac{2\lambda_J}{J}\left(c_0\Delta_1\log\left(\frac{\sqrt{\lambda_J}}{2\pi}\right)+c_1\right)+\cdots\right]
 \eeq
 This is consistent with the structure of the result in Section~\ref{subsec:Konishi} and in particular explains the origin of the term proportional to $\log \lambda_J$ that we found in \eqref{eq:KonishiCnext} for the Konishi operator. 
 
 Indeed, by comparing the two expressions, we can read off $c_{0,1}$ and $\Delta_{0,1}$ for Konishi operator as follows,
 \beq
 \begin{aligned}
 \Delta_0= 2 \comma\qquad \Delta_1=\frac{3}{2\pi^2}\comma\qquad 
 c_0=\frac{8\pi^2}{3}\comma\qquad c_1=4\gamma_{\rm E}\period
 \end{aligned}
 \eeq
 The results for the scaling dimensions $\Delta_{0,1}$ are in perfect agreement with the direct perturbative computation \cite{Bianchi:2000hn} (see also \eqref{eq:konishideltaC}), thereby providing evidence for the formula \eqref{eq:comparethreepoint}.
 \subsubsection{Four-point function: dilaton exchange and form factor expansion}
 Let us next consider the limit of the four-point function \eqref{eq:defF123456}. The general formula \eqref{eq:generalformulacoulomb} relates the four-point function at large charge to the two-point function on the Coulomb branch. At long distance, the two-point function on the Coulomb branch can be expanded as a sum over intermediate states:
 \beq\label{eq:fourpoint}
 \langle \mathcal{O}_1(x_1)\mathcal{O}_2(x_2)\rangle_{v}=\langle \mathcal{O}_1\rangle_{v}\langle \mathcal{O}_2\rangle_{v}+\frac{{}_v\langle \Omega |\mathcal{O}_1|\pi \rangle\langle \pi |\mathcal{O}_2|\Omega \rangle_v}{|x_1-x_2|^2}+\int dm^2 \rho (m^2)G_{m}(|x_1-x_2|)\period
 \eeq
 Here the first term (the disconnected term) corresponds to the vacuum intermediate state while the second and the third terms are contributions from a dilaton exchange and massive particle exchanges respectively. $G_{m}$ is a massive propagator given by the Bessel function,
 \beq\label{eq:massiveflat}
 G_{m} (x)=\frac{m}{4\pi^2 x}K_{1}(m x)\,.
 \eeq
 In what follows, we explain how each term in \eqref{eq:fourpoint} can be reproduced from the HHLL four-point function studied in Section~\ref{subsec:HHLL} in the large charge 't Hooft limit at strong coupling.
 \paragraph{Disconnected term.} Let us first analyze the limit of $F_{1,2,3}$. Using  \eqref{eq:f123basic}, we find that the LHS of the general formula \eqref{eq:generalformulacoulomb} gives
 \beq\label{eq:disclargecharge}
 F_{1,2,3}\to \frac{2J^2}{3r^4}\frac{(Y_{13})^2(Y_{24})^2+4Y_{13}Y_{14}Y_{23}Y_{24}+(Y_{14})^2(Y_{23})^2}{(Y_{34})^2}\period
 \eeq 
 On the other hand, evaluating the disconnected term on the Coulomb branch with
 \beq
 v_{I}(\theta)=\frac{\sqrt{\lambda_J}}{2\pi r}\frac{e^{i\theta}(Y_3)_{I}+e^{-i\theta}(Y_4)_I}{\sqrt{Y_{34}}}\comma
 \eeq
 and averaging over $\theta$, we obtain \eqref{eq:disclargecharge} precisely. Therefore we conclude that the four-point function at large charge correctly reproduces the disconnected term in the Coulomb branch.

 Let us analyze this also using the heavy-light OPE (see Section~\ref{subsec:heavylightOPE}). In the kinematics in \eqref{eq:generalformulacoulomb}, the intermediate operator with dimension $\Delta$ and spin $S$ in the heavy-light OPE contributes to the four-point function as
 \beq
\frac{\langle \mathcal{O}_2\mathcal{O}_2\mathcal{O}_J\mathcal{O}_J\rangle}{\langle \mathcal{O}_J\mathcal{O}_J\rangle}=\frac{2^4}{r^4}\sum_{\tilde{\mathcal{O}}}\left|C_{J\tilde{\mathcal{O}}2}\right|^{2}(z\bar{z})^{\frac{\Delta-J}{2}}\frac{\sin (S+1)\varphi}{\sin \varphi}\comma
 \eeq
 with $e^{i\varphi}=\sqrt{z/\bar{z}}$. In the macroscopic limit $r\to \infty$, the cross ratios can be approximated as
 \beq\label{eq:crosslimit}
 z=1-\frac{2x_{12}e^{i\tilde{\varphi}}}{r}\comma\qquad \bar{z}=1-\frac{2x_{12}e^{-i\tilde{\varphi}}}{r}\comma
 \eeq
 with
 \beq
 \cos \tilde{\varphi}\equiv \vec{n}\cdot \frac{\vec{x}}{|x|}\period
 \eeq
As discussed in Section~\ref{subsec:heavylightOPE}, there are only a finite number of intermediate operators with $\Delta-J =O(1)$ contributing to $F_{1,2,3}$. Then we can simply replace $z$ and $\bar{z}$ with the limiting value, $1$:
\beq
\frac{\langle \mathcal{O}_2\mathcal{O}_2\mathcal{O}_J\mathcal{O}_J\rangle}{\langle \mathcal{O}_J\mathcal{O}_J\rangle}\overset{O(J^2)}{=}\sum_{\tilde{\mathcal{O}}}\left(\frac{4\left|C_{J\tilde{\mathcal{O}}2}\right|}{r^2}\right)^2 (S+1)\period
\eeq
Furthermore, the results in Section~\ref{subsec:heavylightOPE} shows that the intermediate operators are all scalars and the structure constant $C_{J\tilde{\mathcal{O}}2}$ asymptotes to $C_{JJ2}$ in the large $J$ limit:
\beq
C_{JJ2},\,\, C_{J\tilde{\mathcal{O}}2}\propto J\comma\qquad \qquad \frac{C_{J\tilde{\mathcal{O}}2}}{C_{JJ2}}\to 1\period
\eeq
Together with the results for the three-point function at large charge \eqref{eq:comparethreepoint}, these properties guarantee that the large charge limit of the four-point function reproduces the disconnected contribution to the two-point function in the Coulomb branch.
\paragraph{Dilaton exchange.} We next discuss $F_{4,5}$ and show that they give rise to the dilaton exchange on the Coulomb branch. Again using the general formula, we find that the large charge limit of $F_{4,5}$ gives
\beq
F_{4,5}\to \frac{2J}{3r^2}\frac{Y_{12}}{x_{12}^2 }\frac{Y_{13}Y_{24}+Y_{14}Y_{23}}{Y_{34}}\period
\eeq
Comparing this with the expansion \eqref{eq:fourpoint}, we find that
\beq
\begin{aligned}
{}_v\langle \Omega |\mathcal{O}_1|\pi \rangle\langle \pi |\mathcal{O}_2|\Omega \rangle_v&=\frac{2JY_{12}}{3r^2}\frac{Y_{13}Y_{24}+Y_{14}Y_{23}}{Y_{34}}\\
&=\frac{8\pi^2Y_{12}}{3g_{\rm YM}^2}\int \frac{d\theta}{2\pi}\, (v(\theta)\cdot Y_1)(v(\theta)\cdot Y_2)\comma
\end{aligned}
\eeq
where, in the second equality, we used the relation between the Coulomb branch VEV and the charge \eqref{eq:chargeVEV}. This can be further rewritten in terms of the expectation value of $\mathcal{O}_{1,2}$ on the Coulomb branch,
\beq
\langle \mathcal{O}_{1,2}\rangle_{v}=\frac{1}{\sqrt{6}}\frac{16\pi^2}{g_{\rm YM}^2}(v\cdot Y_{1,2})^2\comma
\eeq
in the following way,
\beq\label{eq:dilatonward1}
\frac{8\pi^2Y_{12}}{3g_{\rm YM}^2} (v(\theta)\cdot Y_1)(v(\theta)\cdot Y_2)=\frac{\left(\Delta_1\Delta_2+\sum_{I}\hat{R}_I^{(1)}\hat{R}_I^{(2)}\right)\langle \mathcal{O}_1\rangle_{v}\langle \mathcal{O}_2\rangle_{v}}{f_{\pi}^2v^2}\comma
\eeq
where $f_{\pi}=8\pi /g_{\rm YM}$ and $\hat{R}_{I}^{(1,2)}$ is given by
\begin{align}
\hat{R}_{I}^{(j)}&=\frac{(v\cdot Y_{j})\partial_{Y^{I}_j}-Y^{I}_j(v\cdot \partial_{Y})}{\sqrt{(v\cdot v)}}\comma
\end{align}
As will be explained in more detail in \cite{toappear}, the RHS of \eqref{eq:dilatonward1} is precisely of the form predicted by the supersymmetric dilaton Ward identity, and $f_{\pi}$ corresponds to the dilaton decay constant.\footnote{See \cite{Karananas:2017zrg} for a derivation of the dilaton Ward identity in non-supersymmetric theories.} Thus, the large charge limit correctly reproduces the physics on the Coulomb branch.

Let us interpret the result from the OPE point of view. Substituting the expression for the cross ratio in the macroscopic limit \eqref{eq:crosslimit} into
\beq
\frac{1}{(1-z)(1-\bar{z})}=\sum_{S=0}^{\infty}e^{-\sigma S}\frac{\sin (S+1)\varphi}{\sin\varphi}\comma
\eeq
we find that the dilaton exchange contribution $1/|x_{12}|^2$ come from a tower of primary operators with increasing spin whose conformal dimension is proportional to the spin $\Delta\sim S$. Note that apart from the $S=0$ contribution in the sum, all the operators exchanged in this channel are in a long multiplet, and there is a priori no reason why their conformal dimensions are integer-spaced. Nevertheless, this property is crucial for reproducing the dilaton exchange as we saw above and we can think of it as a highly nontrivial constraint on the CFT data at large charge for theories with a spontaneously broken conformal symmetry. We will discuss this more in Section~\ref{subsubsec:CBconformal} below.
\paragraph{Exchange of massive particles.} Finally we analyze $F_6$ and discuss how it reproduces the form factor expansion; namely the expansion in terms of exchanges of massive particles. For this purpose, it is enough to verify that the massive magnon propagator $t(z,\bar{z})$ becomes the massive propagator in flat space $G_{m^2}(x)$ upon taking the limit (since a product of two massive or massless propagators in flat space can be written in terms of the form factor expansion).

This can be checked rather easily using the strong coupling expansion of $t(z,\bar{z})$ \eqref{eq:massivestrong}. In the macroscopic limit $r\to \infty$, $\sigma$ and $\varphi$ can be approximated by
\beq
\sigma=-\frac{1}{2}\log z\bar{z}\to \frac{2x_{12}\cos\varphi}{r}\comma\qquad \varphi=\frac{1}{2i}\log \frac{z}{\bar{z}}\to \frac{2x_{12}\sin\varphi}{r}\period
\eeq
Substituting this into \eqref{eq:massivestrong}, we find that only the first term with $n=0$ survives in the limit. As a result, the contribution from a massive propagator is given by
\beq
\begin{aligned}
G(x_1,x_2)=\frac{g_{\rm YM}^2}{8\pi^2|x_{12}|^2}t(z,\bar{z})&\to  \frac{g_{\rm YM}^2}{8\pi^2|x_{12}|}\frac{8g}{r}K_{1}(8g |x_{12}|/r)=\frac{g_{\rm YM}^2}{2}G_{m=2|v|}(|x_{12}|)\comma
\end{aligned}
\eeq
where, in the second equality, we used the relation $|v|=\sqrt{\lambda_J}/(\pi r)=4g/r $ and $G_{m}(x)$ is a massive propagator in flat space \eqref{eq:massiveflat}. The result agrees precisely with the massive propagator on the Coulomb branch with $\langle \phi\rangle ={\rm diag}(v,-v)$. 

Again, let us interpret this result from the heavy-light OPE.  For this purpose, let us consider the following OPE sum, 
\beq\label{eq:Salphasum}
S(\alpha)\equiv\frac{1}{|x_{12}|^2}\frac{(1-z)(1-\bar{z})}{\sqrt{z\bar{z}}}\sum_{a=1}^{\infty}\frac{ae^{-\sigma \sqrt{a^2+\alpha^2g^2}}}{\sqrt{a^2+\alpha^2g^2}}\frac{\sin (a\varphi)}{\sin(\varphi)}\comma
\eeq
which corresponds to the contribution from a Regge trajectory of operators with 
\beq\label{eq:Reggemassive}
\begin{aligned}
\Delta (S)-J=\sqrt{(S+1)^2+\alpha^2g^2}-1\comma\qquad c_{\Delta(S), S}=\frac{S+1}{\sqrt{(S+1)^2+\alpha^2g^2}}\period
\end{aligned}
\eeq
To accurately evaluate this sum at strong coupling, we approximate the sum by an integral as follows
\beq
S(\alpha)\overset{g\to\infty}{\sim}\frac{g\alpha}{|x_{12}|^2}\frac{(1-z)(1-\bar{z})}{\sqrt{z\bar{z}}}\int_{0}^{\infty}dx \frac{xe^{-g\alpha\sigma\sqrt{1+x^2}}}{\sqrt{1+x^2}}\frac{\sin(a\varphi)}{\sin(\varphi)}\period
\eeq
This integral can be performed analytically, and taking the macroscopic limit, we obtain
\beq
S(\alpha)\overset{g\to\infty}{\sim} \frac{1}{|x_{12}|^2}\frac{2g\alpha}{r}K_{1}(2\alpha g|x_{12}|/r)=4\pi^2 G_{m=\frac{\alpha v}{2}}(|x_{12}|)\comma
\eeq
which coincides with a propagator for a particle with mass $m=\frac{\alpha v}{2}$. There are two lessons that we can learn from this analysis:
\begin{enumerate}
\item Operators in the Regge trajectory  \eqref{eq:Reggemassive} together reconstruct a single massive propagator.
\item If we truncate the sum over spin (i.e.~the sum over $a$), we will not be able to reproduce the correct massive propagator. This implies that the asymptotic behavior of the Regge trajectory \eqref{eq:Reggemassive} is important for reproducing the physics on the Coulomb branch.
\end{enumerate}
\subsubsection{Coulomb branch from conformal blocks: summary and conjectures \label{subsubsec:CBconformal}}
We now summarize the lessons we learned in the analyses above and propose conjectures on how the physics on the moduli space is reproduced from the large charge limit of CFTs with vacuum moduli.

To keep the discussion general, below we consider a macroscopic limit of general CFTs with $\Delta (J) \sim J$ discussed in section 5.2 of \cite{Jafferis:2017zna}. The macroscopic limit in \cite{Jafferis:2017zna}  is defined by the following limit of the correlation functions:
\beq
\begin{aligned}
&G(z_i,\bar{z}_i)\equiv \left< \mathcal{O}_H(0)\left(\prod_{k=1}^{n-1}\mathcal{O}_k(z_k,\bar{z}_k)\right)\mathcal{O}_n(1) \mathcal{O}_H^{\dagger}(\infty)\right>\comma\\
&G_{\rm macro}(w_i,\bar{w}_i)=\lim_{J\to\infty}\Delta_H^{-\beta\sum_{i}\Delta_i}G\left(1-\frac{w_i}{\Delta_H^{\beta}},1-\frac{\bar{w}_i}{\Delta_H^{\beta}}\right)\period
\end{aligned}
\eeq
Here $\mathcal{O}_H$ is the heavy (i.e.~large charge) operator with $\Delta_H\propto J$ and $z_k$'s are coordinates on the two-dimensional plane. For the four-point function, $z$ can be identified with the cross ratio we used in \eqref{crossratioszzb}. The exponent $\beta$ was left arbitrary in \cite{Jafferis:2017zna}, but in all the known examples including $\mathcal{N}=4$ SYM, it satisfies\footnote{It is likely that $\beta=1/(d-2)$ is a universal relation since it follows from the requirement that the macroscopic limit of the three-point function with a current $\langle \mathcal{O}_H J_{\mu}\mathcal{O}^{\dagger}_H\rangle$ is finite.} 
\beq
\beta=\frac{1}{d-2}\period
\eeq with $d$ being the dimension of the spacetime. Applying this to the three-point function and requiring it to be finite, one can show that the three-point function at large charge should scale as
\beq
C_{HH\mathcal{O}}\overset{J\to\infty}{\propto} \Delta_H^{\beta\Delta_{\mathcal{O}}}\period
\eeq
In what follows, we discuss constraints on the four-point functions for $d=4$. Before proceeding, let us emphasize that the results presented below should not be taken as rigorous proofs, rather they should be taken as well-motivated conjectures based on our analysis for $\mathcal{N}=4$ SYM. We leave it for future to provide rigorous proofs using the conformal bootstrap (assuming the existence of the macroscopic limit). 
\paragraph{Disconnected term.} As we saw in Section~\ref{subsec:heavylightOPE}, the disconnected contribution to the two-point function on the vacuum moduli arises from finitely many low-lying operators in the heavy-light OPE.  More concretely, we conjecture that the existence of the disconnected term implies the following sum rule for finitely many heavy operators $H^{\prime}$ in the OPE sum,
\beq\label{eq:sumruledisconnected}
1=\lim_{J\to\infty} \frac{1}{C_{HH\mathcal{O}_1}C_{HH\mathcal{O}_2}}\sum_{\Delta_{H^{\prime}}=\Delta_H+O(1/J)}C_{HH^{\prime}\mathcal{O}_1}C_{HH^{\prime}\mathcal{O}_2}(S_{H^{\prime}}+1)\period
\eeq
Here $S_{H^{\prime}}$ is the spin of the operator $H^{\prime}$. 
\paragraph{Dilaton exchange.} As we saw in the analysis of $\mathcal{N}=4$ SYM, the sum
\beq
\frac{1}{(1-z)(1-\bar{z})} =\sum_{a=1}^{\infty}  \, e^{-\sigma (a-1)}\,\frac{ \sin a\varphi}{\sin \varphi}\,,
\eeq
reproduces the spacetime dependence of massless dilaton exchange. Then the dilaton Ward identity\fn{Here we are not imposing supersymmetry and therefore considering the non-supersymmetric dilaton Ward identity discussed in \cite{Karananas:2017zrg}.} \cite{Karananas:2017zrg} in the vacuum moduli implies the existence of an asymptotically linear Regge trajectory $H^{\prime}(S)$,
\beq
\begin{aligned}
&\Delta_{H^{\prime}(S)}\overset{1\ll S}{\sim}\Delta_{H}+S\comma
\qquad \lim_{J\to\infty} \Delta_H^{\beta}\frac{C_{HH^{\prime}(S)\mathcal{O}}}{C_{HH\mathcal{O}}}=\frac{\Delta_{\mathcal{O}}}{f_{\pi}}\period
\end{aligned}
\eeq
Here $f_{\pi}$ is the dilaton decay constant and the factor $\Delta_H^{\beta}$ in the second equality comes from $(1-z)(1-\bar{z})=w\bar{w}/\Delta_H^{2\beta}$.
\paragraph{Massive exchanges.} Let us finally discuss the exchange of massive states in the form factor expansion of the two-point function in the vacuum moduli. Generalizing the analysis for $\mathcal{N}=4$ SYM \eqref{eq:Salphasum}, we conclude that the existence of a massive state with mass $m$ implies the existence of a Regge trajectory satisfying
\beq
\Delta_{H^{\prime}(S)}\sim \Delta_H+\sqrt{S^2+m^2\Delta_H^{2\beta}}\comma\qquad \lim_{J\to\infty}\Delta_H^{\beta}\frac{C_{HH^{\prime}(S)\mathcal{O}}}{C_{HH\mathcal{O}}}\sim c_0 \left(\frac{S}{\sqrt{S^2+m^2\Delta_H^{2\beta}}}\right)^{\frac{1}{2}}\period
\eeq
Here we conjecture that the relation holds for $S\sim \mathcal{O}(\Delta_H^{\beta})$, and the constant $c_0$ is related to the form factor in the vacuum moduli.

\subsection{Matrix model for integrated correlators}\label{subsec:MM}
We now study the HHLL correlation functions in the SYM with general gauge group ${\rm SU}(N)$  in the large charge 't Hooft limit using the supersymmetric localization \cite{Pestun:2007rz}. We will provide a strong consistency check of our results in the previous sections for the case of ${\rm SU}(2)$ SYM as well as hints for  the generalization of our large charge analysis to higher ranks. 

The supersymmetric localization is a powerful non-perturbative technique which is applicable when the relevant observable preserves certain supercharges \cite{Pestun:2007rz}. 
However for generic  $\mf{so}(6)_R$ polarizations and local BPS operators of definite spacetime  positions, the four- and higher-point functions do not preserve any residual supersymmetry. This can be remedied by choosing special kinematic variables and/or integrating over the kinematic variables with an appropriate measure. An example of the former is the extremal correlator for Coulomb branch chiral primaries in general 4d $\cN=2$ SCFTs \cite{Gerchkovitz:2016gxx}, which in the $\cN=4$ SYM amounts to a $(n+1)$-point correlator $\la S_{J_1}(x_1,Y)\dots S_{J_n}(x_n,Y)S_{\sum_{i=1}^n J_i}(\infty,\bar Y)\ra$ with aligned $\mf{so}(6)_R$ polarization $Y=(1,i,0,0,0,0)$. In general 4d $\cN=2$ SCFTs, the extremal correlators preserve half of the Poincar\'e supercharges and are independent of the positions $x_i$ of the operators, but can depend nontrivially on the conformal manifold (marginal couplings). Nonetheless, such dependence on the complexified coupling $\tau$ is forbidden in the $\cN=4$ SYM due to the enhanced symmetry and thus $\cN=4$ extremal correlators are rather trivial. 
The other possibility amounts to certain integrated correlators for the BPS operators which turn out to be much richer. Previous works on these integrated correlators in the SYM can be found in \cite{Binder:2019jwn,Chester:2019jas, Chester:2020dja,Chester:2020vyz} for ${\rm SU}(N)$ gauge group in the large $N$ limit, later at finite $N$ in \cite{Dorigoni:2021bvj,Dorigoni:2021guq,Collier:2022emf,Hatsuda:2022enx,Paul:2022piq,Brown:2023cpz}, with generalization to other classical gauge groups in \cite{Dorigoni:2022cua}, and more recently in \cite{Hatsuda:2022enx,Paul:2022piq,Dorigoni:2022cua,Brown:2023cpz,Fiol:2023cml} and especially  \cite{Paul:2023rka,Brown:2023why}  which will be relevant for our discussion below. In particular,  a detailed study of the integrated correlators of the ${\rm SU}(N)$ SYM in the large charge limit using the harmonic analysis on SL(2,$\mathbb{Z}$) and the recursion equations in both $N$ and the charge $J$ was performed in \cite{Paul:2023rka,Brown:2023why}. In what follows, we develop a complementary approach which recasts the integrated correlators in the large charge limit into an ``emergent'' matrix model of size $J/2$. Similar matrix models (whose sizes are proportional to the charge) showed up previously in the study of correlation functions on the supersymmetric Wilson loop in planar $\mathcal{N}=4$ SYM \cite{Giombi:2018qox,Giombi:2018hsx,Giombi:2020amn,Giombi:2021zfb,Giombi:2022anm}, and the extremal correlators in rank-1 $\mathcal{N}=2$ SCFTs  \cite{Grassi:2019txd}.\footnote{The result was subsequently generalized to superconformal QCD with higher-rank gauge groups in \cite{Beccaria:2020azj} under the assumption that certain mixings among operators are absent. Although the results were tested against the first few orders in perturbation theory, it is not fully understood why such mixings are absent. By contrast, such mixings are known to be absent in $\mathcal{N}=4$ SYM thanks to the structure of the $tt^{\ast}$-equation and this is why we succeeded in deriving the results for ${\rm SU}(N)$ gauge theories.} Here we present the results applicable to general ${\rm SU}(N)$ gauge groups. As we will see below, the matrix model approach is particularly useful for analyzing the large charge 't Hooft limit and allows us to efficiently compute the correlator in this limit. Furthermore, in the case of ${\rm SU}(2)$ SYM, we will explicitly confirm the un-integrated large charge correlators derived in Section~\ref{subsec:HHLL}.

\subsubsection{${\rm SU}(2)$ gauge theory}
We first focus on $\mathcal{N}=4$ SYM with the ${\rm SU}(2)$ gauge group and study the four-point function $\la \cO_2\cO_2\cO_J\cO_J\ra$ where $\cO_2$ is the primary in the stress tensor multiplet and $\cO_J$ is its multi-trace cousin (see \eqref{eq:formlargeBPS}).

\paragraph{Definition.}Let us first write down the general form of the four-point function that solves the $\cN=4$ superconformal Ward identity \cite{Dolan:2001tt,Belitsky:2014zha},
\ie \la &\mathcal{O}_2(x_1,Y_1)\mathcal{O}_2(x_2,Y_2)\mathcal{O}_J(x_3,Y_3) \mathcal{O}_J(x_4,Y_4)\ra
={Y_{34}^{J-2}\over x_{12}^4 x_{34}^{2J}}
\left[
\cT_{J,\rm free}(U,V,Y_{ij})
+\cT_{J,\rm loop}(U,V,Y_{ij})
\right ]
\label{22pp}
\fe
where $\mathcal{T}_{J,{\rm tree}}$ coincides with the free $\mathcal{N}=4$ SYM answer and $\mathcal{T}_{J,{\rm loop}}$ is given by 
\beq
\cT_{J,\rm loop}(U,V,Y_{ij}) =\Theta(U,V,Y_{ij})\cT_J(U,V)\,,
\eeq
with $U=\frac{x_{12}^2x_{34}^2}{x_{13}^2x_{24}^2}$ and $V=\frac{x_{14}^2x_{23}^2}{x_{13}^2x_{24}^2}$.
The function $\Theta(U,V,Y_{ij})$  is fixed by superconformal symmetry  \cite{Dolan:2001tt,Belitsky:2014zha}. Therefore the nontrivial information in the four-point function is completely captured by a single function $\cT_J(U,V)$. The four-point functions of the descendant operators satisfies similar relations and are determined by the same function $\cT_J(U,V)$ \cite{Belitsky:2014zha} (see also \cite{Binder:2019jwn}).

The integrated four-point function introduced in \cite{Binder:2019jwn} 
is a linear combination of the four-point functions \eqref{22pp} of the BPS primaries and their descendants integrated over ${\rm S}^4$.
In particular, it can be thought of as the four-point function of two types of integrated operator insertions on ${\rm S}^4$ both involving $\cO_2$ and its descendants. Such integrated insertions probe infinitesimal deformations of the SYM on ${\rm S}^4$ while preserve an $\cN=2$ subalgebra. In the $\cN=2$ language, the first type of integrated insertion amounts to an F-term deformation which is the
exactly marginal deformation of the complex coupling $\tau={4\pi i\over g_{\rm YM}^2}+{\theta\over 2\pi}$ (and $\bar\tau$), while the second type is a mass deformation with real mass parameter $m$.\footnote{The two deformations involve different R-symmetry components of $\cO_2$ and different supersymmetry descendant operators.}
The integrated correlator takes the following simplified form \cite{Chester:2020dja},
\ie \label{eq:integratedexplicitly}
G_J(\tau)\equiv -{2\over \pi}\int_0^\infty  dr\int_0^{\pi} d\varphi \left. {r^3 \sin^2\varphi \over U^2}\cT_J(U,V) \right|_{U=1+r^2-2r \cos\varphi,V=r^2}
\fe
where we emphasize $\cT_J(U,V)$ depends nontrivially on the complexified gauge coupling $\tau$.  
 
\paragraph{Relation to localization.}
By supersymmetric localization, the ${\rm S}^4$ partition function of the ${\rm SU}(2)$ SYM with $\mathcal{N}=2$ preserving mass deformation is equivalent to the following integral \cite{Pestun:2007rz}, 
\ie 
Z(\tau,m)=\int da \,(2a^2) e^{-4\pi {\rm Im}\tau a^2}Z_{\text{1-loop}}(m,a,\tau) |Z_{\rm inst}(m,a,\tau)|^2\,.
\label{su2mm}
\fe
with
\beq
Z_{\text{1-loop}}(m,a,\tau)={ H(2a)^2\over H(m) H(2a+m)H(2a-m)}\period
\eeq
Here the function $H(x)$ captures the one-loop determinants of the fields and is defined in terms of the Barnes $G$-function by
\ie 
H(x)=e^{-(1+\C_E)x^2} G(1+ix)G(1-ix)\,.
\fe
The function $Z_{\rm inst}(m,a,\tau)$ depends nonperturbatively on $g_{\rm YM}$ via $e^{-{8\pi^2\over g_{\rm YM}^2}}$ and contains the instanton contributions (the anti-instanton contributions are captured by $\overline {Z_{\rm inst}(m,a,\tau)}$) \cite{Nekrasov:2002qd,Nekrasov:2003rj}. For $m=0$, this becomes the familiar Gaussian Hermitian matrix model \cite{Pestun:2007rz,Okuda:2010ke}. One important property of $Z_{\text{1-loop}}$ and $Z_{\rm inst}$ is that they are even functions of $m$ and satisfy
\beq\label{eq:evenZ1}
\left.\partial_m Z_{\text{1-loop}}\right|_{m=0}=\left.\partial_m Z_{{\rm inst}}\right|_{m=0}=0\period
\eeq

Since the integrated insertions are naturally associated to derivatives of the $\cN=2$ preserving parameters $\tau,\bar\tau$ and $m$ in the SYM action on ${\rm S}^4$, we expect that the integrated four-point function can be computed from
\ie 
 \left.{\partial_\tau^{J\over 2} \partial_{\bar\tau}^{J'\over 2} \partial_m^2 Z(\tau,m) \over Z(\tau,m)}
\right|_{m=0}
\label{Fpqfromd}
\fe
for even non-negative integers $J,J'$. Importantly here we do not restrict $J=J'$. The reason is to take into account operator mixing on ${\rm S}^4$, which arises owing to curvature and mass deformation (see e.g.\cite{Gerchkovitz:2016gxx}),
\ie  
\cO_J=\mathcal{O}_J^{\prime}+\sum_{K=1}^{J\over 2} u_{J,J-2K} \cO'_{J-2K}\period
\fe
Here $\cO'_J$ is the natural basis of operators on ${\rm S}^4$ that are generated by the action of $\partial_{\tau}^{J/2}$ or $\partial_{\bar{\tau}}^{J/2}$, and corresponds to an insertion $a^J$ in the integral \eqref{su2mm}. On the other hand, $\cO_J$ is the basis of operators on $\mR^4$ whose two-point functions are orthogonal, $\langle \cO_J, \cO_{J^{\prime}}\rangle\propto \delta_{J,J^{\prime}}$.
Thus using a polynomial 
\beq
Q_{J}(a)=a^{J}+\sum_{K=1}^{\frac{J}{2}}u_{J,J-2K}a^{J-2K}\period
\eeq
we can compute $G_{J}$ by the following matrix integrals
\beq\label{eq:GJQJ}
G_{J}=\left.\frac{\partial_{m}^{2}\left(\int d\tilde{\mu}(m)\, Q_{J}(a)Q_{J}(a)\right)
}{\int d\tilde{\mu}(m)\, Q_J(a)Q_J(a)}\right|_{m=0}
\eeq
where the denominator gives the two-point function of $\mathcal{O}_J$'s and is needed to make $\mathcal{O}_J$ unit-normalized. The mass-deformed measure $d\tilde{\mu}(m)$ is given by
\beq
\begin{aligned}
d\tilde{\mu}(m)\equiv da \,(2a^2) e^{-4\pi {\rm Im}\tau a^2}Z_{\text{1-loop}}(m,a,\tau)|Z_{\rm inst}(m,a,\tau)|^{2}\period
\end{aligned}
\eeq
Thanks to the orthogonality of the two-point functions (in the absence of mass deformation), $Q_J$'s satisfy the condition,
\beq
\begin{aligned}
&\int d\mu\,Q_{J}(a) Q_{J^{\prime}}(a) \propto \delta_{J,J^{\prime}}\comma\qquad \qquad d\mu\equiv  d\tilde{\mu}(0)=da\, (2a^2)e^{-4\pi {\rm Im}\tau a^2}\,,
\end{aligned}
\eeq
which uniquely fixes the coefficients $u_{J,J-2K}$ as functions of $\tau$ and $\bar{\tau}$. Applying the Gram-Schmidt procedure following \cite{Gerchkovitz:2016gxx,Giombi:2018qox}, we can write down $Q_J$ explicitly as follows:
\begin{align}
&Q_{J}(a)=\frac{1}{D_J}\left|\begin{array}{cccc}\mathcal{I}_0&\mathcal{I}_2&\cdots&\mathcal{I}_J\\\mathcal{I}_2&\mathcal{I}_4&\cdots&\mathcal{I}_{J+2}\\\vdots&\vdots&\ddots&\vdots\\\mathcal{I}_{J-2}&\mathcal{I}_{J}&\cdots&\mathcal{I}_{2J-2}\\1&a^2&\cdots&a^{2J}\end{array}\right|\comma\\
&D_J=\det \left(\mathcal{I}_{2(j+k-2)}\right)_{1\leq j,k\leq \frac{J}{2}}\comma\qquad \mathcal{I}_{j}\equiv \int d\mu \,a^{j}\,.
\end{align}
Two important identities that follow from the Gram-Schmidt procedure is the expression for the normalization of the two-point function,
\beq\label{eq:orthogonal}
\int d\mu \, Q_{J}(a) Q_{J^{\prime}}(a)=\frac{D_{J+2}}{D_J}\delta_{J,J^{\prime}}\comma
\eeq
and an ``emergent'' matrix-model representation for $D_J$ \cite{Giombi:2018qox,Grassi:2019txd},
\beq
D_{J}=\frac{1}{(J/2)!}\int_{0}^{\infty}\prod_{i=1}^{\frac{J}{2}}\underbrace{\left(dx_i \sqrt{x_i}e^{-4\pi {\rm Im}\tau x_i}\right)}_{d\mu}\prod_{i<j}(x_i-x_j)^2\period
\eeq
As shown above, the factor in the middle is the measure $d\mu$ written in terms of the variable $x=a^2$.

\paragraph{Matrix model for integrated correlators.} We now derive the matrix model representation for the integrated correlator \eqref{eq:GJQJ} following the approach in \cite{Giombi:2021zfb}. We first consider orthogonal polynomials under the mass-deformed measure $d\tilde{\mu}(m)$,
\begin{align}\label{eq:tildeQorth}
&\int d\tilde{\mu}(m)\, \tilde{Q}_{J}(a)\tilde{Q}_{J^{\prime}}(a)= \frac{\tilde{D}_{J+2}}{\tilde{D}_J}\delta_{J,J^{\prime}}\comma\qquad \tilde{Q}_J(a)=a^{J}+\sum_{K=1}^{\frac{J}{2}}\tilde{u}_{J,J-2K}a^{J-2K}\comma
\end{align}
where $\tilde{u}_{J,K}$ now depends both on $\tau$ and $m$, and $\tilde{D}_J$ admits a matrix-model type representation
\beq\label{eq:matrixmodeldeformed}
\begin{aligned}
&\tilde{D}_J=\frac{1}{(J/2)!}\int_{0}^{\infty}\prod_{i=1}^{\frac{J}{2}}\left(dx_i\sqrt{x_i}e^{\mathcal{F}(\sqrt{x_i})}e^{-4\pi {\rm Im}\tau x_i}\right)\prod_{i<j}(x_i-x_j)^2\comma\\
&\mathcal{F}(a)\equiv \log Z_{\text{1-loop}}(m,a,\tau)+\log |Z_{\rm inst}(m,a,\tau)|^2\period
\end{aligned}
\eeq
Taking the second derivative of \eqref{eq:tildeQorth} with respect to $m$ and setting $m=0$, we obtain the term that gives $G_J$ \eqref{eq:GJQJ} and two additional terms:
\begin{align}\label{eq:threetermsQQ}
&\left.\partial_m^2 \left(\int d\tilde{\mu}(m)\,\tilde{Q}_J(a)\tilde{Q}_J(a)\right)\right|_{m=0}=\\
&\left.\partial_m^2 \left(\int d\tilde{\mu}(m)\,Q_J(a)Q_J(a)\right)\right|_{m=0}+\left.2\int d\mu \left(\partial_m\tilde{Q}_J(a)\right)^2\right|_{m=0}+\left.2\int d\mu \,Q_{J}(a)\partial_m^2\tilde{Q}_J(a)\right|_{m=0}\period\nn
\end{align}
However, since $Z_{\text{1-loop}}$ and $Z_{\rm inst}$ are even functions of $m$ (see \eqref{eq:evenZ1}), the second term in \eqref{eq:threetermsQQ} vanishes. Furthermore, since $\partial_{m}^2\tilde{Q}_J(a)$ is a polynomial of degree less than $J$, the last term also vanishes owing to the orthogonality \eqref{eq:orthogonal}. We thus obtain\footnote{In the second equality, we used the fact that the localization integrand is an even function of $m$, \eqref{eq:evenZ1}.} the following simple expression for $G_J$:
\beq
G_J=\left.\frac{\partial_{m}^2\left(\int d\tilde{\mu}(m)\tilde{Q}_{J}(a)\tilde{Q}_{J}(a)\right)}{\int d\tilde{\mu}(m)\tilde{Q}_J(a)\tilde{Q}_J(a)}\right|_{m=0}=\left.\left(\frac{\partial_m^2 \tilde{D}_{J+2}}{\tilde{D}_{J+2}}-\frac{\partial_m^2 \tilde{D}_{J}}{\tilde{D}_{J}}\right)\right|_{m=0}\,.
\eeq
Using \eqref{eq:matrixmodeldeformed}, we arrive at the formula that expresses the integrated correlator as an expectation value of a ``single-trace operator'' in the matrix model
\beq\label{eq:GJasmatrix}
G_{J}=\left<{\rm tr}\left[f\left(M\right)\right]\right>_{J+2}-\left<{\rm tr}\left[f\left(M\right)\right]\right>_{J}\comma
\eeq
where the function $f(a)$ is given by
\beq\label{eq:fadefini}
f(x)\equiv \left.\partial_{m}^2 \mathcal{F}(\sqrt{x})\right|_{m=0}=\left.\partial_{m}^2\left(\log Z_{\text{1-loop}}(m,\sqrt{x},\tau)+\log |Z_{\rm inst}(m,\sqrt{x},\tau)|^2\right)\right|_{m=0}\comma
\eeq
and $\left<\bullet\right>_J$ is the expectation value in the Wishart-Laguerre matrix model of size $J/2$,
\beq
\left<{\rm tr}\left[f\left(M\right)\right]\right>_{J}=\frac{\int_{0}^{\infty}\prod_{i=1}^{\frac{J}{2}}\left(dx_i \sqrt{x_i}e^{-4\pi {\rm Im}\tau x_i}\right)\prod_{i<j}(x_i-x_j)^2\sum_{k=1}^{\frac{J}{2}}f(x_k)}{\int_{0}^{\infty}\prod_{i=1}^{\frac{J}{2}}\left(dx_i \sqrt{x_i}e^{-4\pi {\rm Im}\tau x_i}\right)\prod_{i<j}(x_i-x_j)^2}\,.
\eeq

The formula \eqref{eq:GJasmatrix} is valid at finite $\tau$ and $J$, but it simplifies in the large charge 't Hooft limit as the limit coincides with the standard 't Hooft limit of this matrix model. 
In this case, the (anti)instanton contributions are non-perturbatively suppressed in $J$ and we can set $|Z_{\rm inst}|=1$.
Then, using the following integral formula\fn{This can be derived by using the following two formulae:
\beq
\begin{aligned}
\log H(x)=\sum_{n=1}^{\infty}\frac{(-1)^{n}}{n+1}\zeta(2n+1)x^{2n+2}\comma\qquad \zeta(2n+1)=\frac{2^{2n}}{\Gamma(2n+2)}\int_0^{\infty}dw\frac{w^{2n+1}}{\sinh^2(w)}\period\nn
\end{aligned}
\eeq
} given in (A.12) of \cite{Beccaria:2022ypy},
\beq\label{eq:newHformula}
\log H(x)=\int^{\infty}_{0}\frac{dw}{w}\frac{1-2w^2x^2-\cos (2wx)}{2\sinh^2 (w)}\comma
\eeq
we can rewrite \eqref{eq:fadefini} with 
\ie 
f(x) =   8 \int_0^{\infty} dw \frac{\sin^2 (2\sqrt{x} w)}{\sinh^2 (w)} \,.
\label{logF}
\fe
To the leading large $J$ limit, the matrix model is 
solved by the distribution \cite{Grassi:2019txd},
\ie\label{MPdist}
\rho_0(y)={1\over 2\pi }\sqrt{4-y\over y}\,, \quad y\in [0,4]\,,
\fe
where $y$ is related to the eigenvalue $x$ by $x=\frac{\lambda_J y}{16\pi^2}$.
Consequently, \eqref{eq:GJasmatrix} can be evaluated in the large $J$ limit by
\ie 
G_J=\int_0^4 d y  \rho_0(y) (\lambda_J\partial_{\lambda_J}+1) f\left (\frac{\lambda_J y}{16\pi^2}\right)\,.
\fe
Starting from the expression \eqref{logF} for $f(a)$ and integrating over $x$ first, using
\ie 
\int_0^4 dy \,\rho_0(y) \sin^2\left(\sqrt{\frac{\lambda_J y}{4\pi^2}}w\right)=\frac{1}{2}-\frac{\pi J_{1}\left(\frac{2w\sqrt{\lambda_J}}{\pi}\right)}{2w\sqrt{\lambda_J}}\,,
\fe
we obtain
\ie \label{eq:n2integralfinal}
G_J=4\int_0^\infty dw\frac{w}{\sinh^2(w)}\left(1-J_0\left(\frac{2w\sqrt{\lambda_J}}{\pi}\right)\right) \,.
\fe
It is straightforward to obtain the weak coupling (small $\lambda_J$) expansion from \eqref{eq:n2integralfinal} by expanding the Bessel function and performing the integral. The result reads
\beq
G_J=\sum_{n=1}^\infty\frac{(-1)^{n-1} 2^{2-2 n} (2 n+1)!  \zeta (2 n+1)}{ \pi^{2n}(n!)^2} \lambda_J ^n\,\period
\label{GJweakexp}
\eeq
This agrees\footnote{There is a difference of an overall factor of $4$ as compared to the result in \cite{Paul:2023rka} owing to the difference in the normalization of operators. See e.g.~(3.29) in \cite{Paul:2023rka}. \label{ft:factorfour}} with the results obtained in recent papers \cite{Paul:2023rka} and \cite{Brown:2023why}, which used different methods.\footnote{Note that the definitions of the large charge 't Hooft coupling in \cite{Paul:2023rka,Brown:2023why} are slightly different: $\lambda_J^{\text{\cite{Paul:2023rka,Brown:2023why}}}=g_{\rm YM}^2J$.}
It is easy to see that the weak coupling expansion has a finite radius of convergence $|\lambda_J|<\pi^2$.
This coincides with the radius of convergence for the structure constant of the Konishi operator \eqref{weakradiusconv}. We will discuss their common physical origin in Section~\ref{subsubsec:convergence}.
\paragraph{Comparison with the direct large charge computation.} The result \eqref{GJweakexp} can also be reproduced by a direct integration of the four-point function at large charge, which we computed in Section~\ref{subsec:HHLL}.

The first step is to note the following relation between the HHLL four-point function $t(z,\bar z)$ in \eqref{tintegralform} and the un-integrated four-point function $\cT_J(U,V)$ in 
\eqref{eq:integratedexplicitly},
\beq
\mathcal{T}_J(U,V)=\frac{4}{z\bar{z}}\left(t(z,\bar{z})^2-1\right)\period
\eeq
The prefactor $1/z\bar{z}$ comes from $\Theta (U,V,Y_{ij})$ (see \cite{Binder:2019jwn}) and the overall numerical factor $4$ is to account for the correct normalization. The subtraction $-1$ is needed to remove the tree-level contribution. We next use the crossing symmetry under $z,\bar{z}\to 1/z,1/\bar{z}$ and restrict the range of integration of $r$ to $[0,1]$:
\beq
G_J= -\frac{4}{\pi}\int_0^{1}  dr\int_0^{\pi} d\varphi \left. {r^3 \sin^2\varphi \over U^2}\cT_J(U,V) \right|_{U=1+r^2-2r \cos\varphi,V=r^2}\,.
\eeq
We then express $t(z,\bar{z})$ as a sum over magnons \eqref{eq:tsumbound} and perform the integral explicitly. The result reads
\beq\label{eq:integratedintegrabilitylike}
G_J=4\sum_{a=1}^{\infty}\left(\frac{1}{a}-\frac{a^2}{(a^2+16g^2)^{3/2}}\right)\period
\eeq
It is then straightforward to check that the weak-coupling expansion of \eqref{eq:integratedintegrabilitylike} is in precise agreement with the result from the matrix model \eqref{GJweakexp}.
\paragraph{A trick to evaluate the integral at strong coupling.}
To obtain the strong coupling (large $\lambda_J$) expansion, we evaluate the remaining $w$ integral  by contour deformation. A similar trick gives a proof of the conjectured formula (4.21) in \cite{Grassi:2019txd}.

 We first rewrite the integral as a contour integral over $\C$ just above the real line defined by
\ie 
\C=[-\infty,-\ep]\cup C_\ep \cup  [\ep,\infty]\,,
\fe
where $C_\ep$ is an infinitesimal semicircle of radius $\ep$ centered at the origin in the upper half-plane. Using the relation between the Bessel function $J_0(x)$ and the Hankel function $H_0^{(1)}(x)$,
\ie 
H^{(1)}_0(x)-H^{(1)}_0(-x)=2J_0(x)\,,
\fe
we have
\ie 
2\int_\ep^\infty dw  {w\over \sinh^2 w} J_0\left( 2w \sqrt{\lambda_J }\over \pi \right)
=
\left(\int_{-\infty}^{-\ep} dw+\int_\ep^\infty dw\right)    {w\over \sinh^2 w} H_0^{(1)}\left( 2w \sqrt{\lambda_J }\over \pi \right)\,.
\label{contour1}
\fe
Furthermore, the Hankel function $H_0^{(1)}(x)$ has a branch point at $x=0$ and behaves at small $x$ as
\ie 
H_0^{(1)}(x)=1+  {2i\over \pi} \left(\log {x\over 2}+\gamma_E \right) +\cO(x)\,.
\fe
Consequently
\ie 
2\int_\ep^\infty dw    \,\frac{w}{\sinh^2(w)}+\int_{C_\ep} dw   {w\over\sinh^2w} H_0^{(1)}\left( 2w \sqrt{\lambda_J }\over \pi \right)=2(1+\gamma_{\rm E})+\log {\lambda_J\over 4\pi ^2}\,.
\label{contour2}
\fe
Putting together \eqref{contour1} and \eqref{contour2}, we obtain
\ie 
G_J=2\log {\lambda_J\over 4\pi ^2}+4(1+\gamma_{\rm E})-2\int_\C dw  {w\over \sinh^2 (w) }H_0^{(1)}\left( 2w \sqrt{\lambda_J }\over \pi \right)\,.
\fe
Deforming the contour upwards, one picks up residues at $w=n\pi i$ for positive integer $n$. From the relation between Hankel function on the positive imaginary axis and Bessel functions,
\ie 
H_t^{(1)}(ix)={2\over \pi i^{t+1}}K_t(x)\,, 
\fe
and the recurrent relations for Bessel functions, we thus obtain the following explicit expression for the integrated four-point function
\ie 
G_J=2\log{\lambda_J\over 4\pi^2} 
+4(1+\C_E)+8\sum_{n=1}^\infty \left(2n \sqrt{\lambda_J}  K_1(2n\sqrt{\lambda_J})
-K_0(2n\sqrt{\lambda_J}) \right)\,.
\label{GJstrongexp}
\fe
Here again, the Bessel functions on the RHS are exponentially suppressed for large $\lambda_J$ and correspond to the worldline instanton contributions to the integrated correlator, as is expected from the form of the un-integrated four-point function $t(z,\bar z)$ given in \eqref{eq:massivestrong}.

Using the
  following Mellin integral expression for the Bessel functions $K_t(2y)$ with $t>0$,
\ie 
K_t(2 y)={1\over 8\pi i}\int_{\ep-i\infty}^{\ep+i\infty} ds \,y^{-s}   \Gamma \left({s+t\over 2}\right)\Gamma \left({s-t\over 2}\right)\,,
\fe
where $\ep >t$ and $y=n\sqrt{\lambda_J/2}$,
we can equivalently write \eqref{GJstrongexp} in the following way,
\ie 
G_J=2\log{\lambda_J\over 4\pi^2} 
+4(1+\C_E)+
\int_{0^+ -i\infty}^{0^+ +i\infty} {ds\over 2\pi i}\lambda^{-{s\over 2}}
  {2 (s-1) }   \zeta (s) \Gamma \left(\frac{s}{2}\right)^2\,.
  \label{GJalternate}
\fe
Note that  the weak coupling expansion \eqref{GJweakexp} immediately follows from \eqref{GJalternate} by deforming the contour to the left and picking up the residues.
\subsubsection{Radius of convergence and massless magnon point}\label{subsubsec:convergence} One interesting outcome of our analysis is that both the three-point function of the Konishi operator and the integrated four-point function have the same (finite) radius of convergence $|\lambda_J|=\pi^2$ and they become singular at $\lambda_J=-\pi^2$. Precisely at this point, the mass of the lightest massive magnon,
\beq
\sqrt{1+16g^2}=\sqrt{1+\frac{\lambda_J}{\pi^2}}\comma
\eeq
vanishes. It is therefore natural to conjecture that the physical origin of the  singularity is a tachyonic instability induced by this magnon.

A few comments are in order. First, this relation between singularity and the emergence of massless particles is reminiscent of the structure of the Coulomb branch moduli space in $\mathcal{N}=2$ supersymmetric gauge theories, which also exhibits singularity when BPS particles become massless \cite{Seiberg:1994rs,Seiberg:1994aj}. In fact, using the dictionary between the large $R$-charge and the Coulomb branch moduli \eqref{eq:chargeVEV}, one can map the massless magnon point at large charge to the massless W-boson point.

Second, in perturbation around a trivial vacuum $\langle \phi\rangle_{\rm cl}=0$, the theory becomes unstable as soon as the coupling $g_{\rm YM}^2$ becomes negative and the perturbation series has a zero radius of convergence. What we found implies that the large charge 't Hooft vacuum is more stable than the trivial vacuum owing to the background $R$-charges that delays the onset of the instability. One interesting question for the future is to take other quantum numbers (in particular the Lorentz spin) to be large and study the stability.

Third, the radius of convergence is finite also in the standard large $N_c$ 't Hooft limit. This often allows one to take the {\it double-scaled large-$N$ limit} in which one zooms in the singularity. The limit corresponds to the continuum limit of Feynman graphs and the critical behavior near the singularity is governed by (gravitationally dressed) conformal field theory \cite{Brezin:1990rb,Douglas:1989ve,Gross:1989vs}. It would be interesting to take a similar double-scaled limit around $\lambda_J=-\pi^2$ and study it from the point of view of Feynman diagrams. Another interesting question is to find an effective (conformal) field theory\fn{A natural guess is a theory that contains both the Goldstone boson around the large charge vacuum and the magnon that becomes massless at the singularity.} governing the critical behavior around $\lambda_J=-\pi^2$. 

\subsubsection{Generalization to ${\rm SU}(N)$ gauge theories}We now generalize the result to $\mathcal{N}=4$ SYM with the SU$(N)$ gauge group. In $\mathcal{N}=2$ SCFTs with higher-rank gauge groups, there are multiple chiral operators with the identical $R$-charge, and in general they all mix with each other when the theory is placed on ${\rm S}^4$. This makes the Gram-Schmidt procedure and the derivation of the large-charge matrix model significantly more complicated. However, in $\mathcal{N}=4$ SYM, it was shown in \cite{Gerchkovitz:2016gxx} that operators organize into a set of ``towers'' and  the mixing occurs only within each tower. This is a special property of $\mathcal{N}=4$ SYM. See Section 3.2.1 of \cite{Gerchkovitz:2016gxx} for more details.

To define the towers of operators, we consider the set of operators made out of single-trace operators of length greater than 2, i.e.~$\prod_{k=3}^{N}\left(\mathcal{S}_{k}\right)^{n_k}$ and order them such that their charges are non-decreasing; $B_n$ with $J_n\leq J_{n+1}$. We then construct an orthogonal basis $\mathcal{O}_{0}^{(n)}$ by the following recursive procedure,
\beq
\tilde{\mathcal{O}}_{0}^{(n)}\equiv B_n -\sum_{\substack{n^{\prime}<n\\J_n-J_{n^{\prime}}\in 2\mathbb{Z}}}\frac{\left< B_n \,,\,\tilde{\mathcal{O}}_{0}^{(n^{\prime})}\right>}{\left<\tilde{\mathcal{O}}_{\frac{J_n-J_{n^{\prime}}}{2}}^{(n^{\prime})}\,,\,\tilde{\mathcal{O}}_{0}^{(n^{\prime})}\right>}\tilde{\mathcal{O}}_{\frac{J_n-J_{n^{\prime}}}{2}}^{(n^{\prime})}\comma
\eeq
with
\beq
\tilde{\mathcal{O}}_{k}^{(n)}\equiv \left(\mathcal{S}_2\right)^{k}\tilde{\mathcal{O}}_{0}^{(n)}\period
\eeq
For the first few $n$, $\tilde{\mathcal{O}}_{0}^{(n)}$ takes the following form:
\beq\label{eq:tildeOdefi}
\tilde{\mathcal{O}}_{0}^{(0)}= 1 \comma\quad \tilde{\mathcal{O}}_{0}^{(1)}= \mathcal{S}_{3}\comma\quad \tilde{\mathcal{O}}_{0}^{(2)}=\mathcal{S}_4-\frac{\langle \mathcal{S}_4\rangle}{\langle(\mathcal{S}_2)^2\rangle}(\mathcal{S}_2)^2\period
\eeq 
Unlike the Gram-Schmidt procedure we used for the SU$(2)$ gauge theory, the operators $\tilde{\mathcal{O}}_{0}^{(n)}$'s only contain $B_n$'s with the same charge. Consequently, there is still a residual mixing among operators with different charges. A special property of $\mathcal{N}=4$ SYM is that the residual mixing occurs only among operators in the same tower, $\{\tilde{\mathcal{O}}_{0}^{(n)},\tilde{\mathcal{O}}_{1}^{(n)},\cdots\}$. Within each tower, there is only a single operator with a given $R$-charge and therefore the mixing can be resolved in the same way as in the ${\rm SU}(2)$  theory;
\beq
\mathcal{O}_{k}^{(n)}=\tilde{\mathcal{O}}_{k}^{(n)}+\sum_{K=1}^{k}u_{k,k-K}^{(n)}\tilde{\mathcal{O}}_{k-K}^{(n)}\period
\eeq

\paragraph{General formula.} The ${\rm S}^{4}$ partition function of the mass-deformed ${\rm SU}(N)$ SYM is \cite{Pestun:2007rz},
\beq
Z(\tau,m)={1\over N!}\int \left(\prod_{i=1}^{N}da_i\right) \left(\prod_{i<j}(a_i-a_j)^2\right)\delta ({\rm tr}({\bm a}))e^{-2\pi {\rm Im}\tau {\rm tr}({\bm a}^2)}Z_{\text{1-loop}}(m,{\bm a},\tau)\left|Z_{\text{inst}}(m,{\bm a},\tau)\right|^2\comma
\eeq
with ${\bm a}={\rm diag}(a_1,\ldots, a_N)$ and
\beq
\begin{aligned}
&Z_{\text{1-loop}}(m,{\bm a},\tau)={1\over H(m)^{N-1}}\prod_{i\neq j}\frac{H(a_i-a_j)}{H(a_i-a_j+m)}\period
\end{aligned}
\eeq
The operators $\tilde{\mathcal{O}}_{k}^{(n)}$ and $\mathcal{O}_{k}^{(n)}$ correspond to the following insertions in the localization integral
\beq
\tilde{\mathcal{O}}_{k}^{(n)}\,\,\mapsto \,\, g^{(n)}({\bm a})\left({\rm tr}\left({\bm a}^2\right)\right)^{k}\comma\qquad \mathcal{O}_{k}^{(n)}\,\,\mapsto \,\, g^{(n)}({\bm a})Q_k^{(n)}\left({\rm tr}\left({\bm a}^2\right)\right)\comma
\eeq
where $g^{(n)}({\bm a})$ is an insertion corresponding to the operator $\tilde{\mathcal{O}}_{0}^{(n)}$, whose form can be deduced from its structure \eqref{eq:tildeOdefi}, and $Q_k^{(n)}(x)$ is a polynomial of degree $k$ defined by\footnote{Note that the normalization of $x$ is slightly different from the one we used in the ${\rm SU}(2)$ gauge theory. In terms of the eigenvalue $a$ for the ${\rm SU}(2)$ gauge theory, $x$ here is given by $2a^2$ while $x$ in the previous analysis was given by $a^2$.\label{ft:xnorm}}
\beq
Q_k^{(n)}(x)=x^{k}+\sum_{K=1}^{k}u_{k,k-K}^{(n)}x^{k-K}\period
\eeq
We then follow \cite{Beccaria:2020azj} and separate the integral into the integral along the ``radial'' coordinate $x={\rm tr}({\bm a}^2)$ and the integral of ``angular''  coordinates $\int_{S^{N-1}}d\Omega$ with the traceless constraint,
\beq
\int \left(\prod_{i=1}^{N}da_i\right)\left(\prod_{i<j}(a_i-a_j)^2\right)\delta ({\rm tr}({\bm a}))=\int_{0}^{\infty} dx \,  x^{\frac{N^2-3}{2}}\int_{S^{N-1}} d\Omega \,J(\Omega)\comma
\eeq
where $J(\Omega)$ is a Jacobian for the change of coordinates. For later convenience we define the constant $C_N=\int d\Omega \,J(\Omega)$ and the normalized angular integration
\ie 
\llangle  f({\bm a}) \rrangle = \frac{1}{C_N}\int_{S^{N-1}} d\Omega J(\Omega)  f({\bm a})\,,
\fe
which produces a function of the radial variable $x={\rm tr}({\bm a}^2)$. By construction, the angular integration is closely related to the full matrix integral for the hermitian Gaussian matrix model,
\ie 
\la f({\bm a})\ra_{\rm GMM}\equiv  {1\over \cN}\int \left(\prod_{i=1}^{N}da_i\right)\left(\prod_{i<j}(a_i-a_j)^2\right) f({\bm a}) \D(\tr {\bm a}) e^{-\tr ({\bm a}^2)}\,,
\fe
where $\cN$ is a normalization factor such that $\la 1\ra_{\rm GMM}=1$. For example if $f$ is homogeneous of degree $\Delta$ \cite{Beccaria:2020azj}, 
\ie 
\llangle f({\bm a}) \rrangle = {x^{\frac{\Delta}{2}}\Gamma ({N^2-1\over 2})\over \Gamma ({N^2-1+\Delta\over 2}  )}
\la f({\bm a})\ra_{\rm GMM}  
\,,
\label{angularint}
\fe
which will be useful for us later.

 The polynomials $Q_{k}^{(n)}$ are orthogonal to each other under the radial measure $d\mu^{(n)}$
\beq
\begin{aligned}\label{radialmeasure}
&\int d\mu^{(n)} Q_{k}^{(n)}(x)Q_{k^{\prime}}^{(n)}(x)=\frac{D^{(n)}_{k+1}}{D^{(n)}_k}\delta_{k,k^{\prime}}\,,\quad d\mu^{(n)}\equiv dx\,x^{\frac{N^2-3}{2}}e^{-2\pi {\rm Im}\tau x}\llangle \left(g^{(n)}({\bm a})\right)^2\rrangle\,,
\end{aligned}
\eeq
where $D^{(n)}_k$ admits a `matrix model' representation
\beq
D^{(n)}_k=\frac{1}{k!}\int_0^{\infty}\left(\prod_{i=1}^{k}d\mu^{(n)}(x_i)\right) \prod_{i<j}(x_i-x_j)^2\period
\eeq

To derive a matrix model for the integrated correlators, we consider the mass-deformed measure $d\tilde{\mu}^{(n)}\equiv dx\,x^{\frac{N^2-3}{2}}e^{-2\pi {\rm Im}\tau x} \llangle  \left(g^{(n)}({\bm a})\right)^2 e^{\cF({\bm a})}\rrangle$ 
where $\mathcal{F}=\log Z_{\text{1-loop}}+\log |Z_{\rm inst}|^2$.
Following the derivation for the ${\rm SU}(2)$ gauge theory, we arrive at the following representation for the integrated correlator of the operators $\mathcal{O}_{k}^{(n)}$
\beq\label{eq:matrixmodelgeneral}
G_{k}^{(n)}=\left<{\rm tr}\left[f^{(n)}\left(M\right)\right]\right>_{k+1}-\left<{\rm tr}\left[f^{(n)}\left(M\right)\right]\right>_{k}\comma
\eeq
where $\langle \bullet\rangle_k$ is the expectation value in the following matrix model of size $k$:
\beq\label{eq:WishartSUN}
\left<{\rm tr}\left[f^{(n)}\left(M\right)\right]\right>_{k}=\frac{\int_{0}^{\infty}\prod_{i=1}^{k}\left(d\mu^{(n)}(x_i)\right)\prod_{i<j}(x_i-x_j)^2\sum_{j=1}^{k}f^{(n)}(x_j)}{\int_{0}^{\infty}\prod_{j=1}^{k}\left(d\mu^{(n)}(x_i)\right)\prod_{i<j}(x_i-x_j)^2}\comma
\eeq
and the function $f^{(n)}(x)$ is defined by the angular integral
\beq\label{eq:fndefine}
f^{(n)}(x)\equiv  \frac{\llangle \left(g^{(n)}({\bm a})\right)^2\partial_{m}^2 \left.\mathcal{F}({\bm a})\right|_{m=0} \rrangle}{\llangle  \left(g^{(n)}({\bm a})\right)^2\rrangle} \comma
\eeq
As in the case of the ${\rm SU}(2)$ theory, the formula \eqref{eq:matrixmodelgeneral} is valid at finite $\tau$, $N$ and $J$. 

\paragraph{Result for maximal-trace operators.}The matrix integral \eqref{eq:WishartSUN} takes a particularly simple form for $\mathcal{O}_k^{(0)}$'s, which are often called the ``maximal-trace'' operators. In this case $g^{(0)}({\bm a})=1$ and the radial measure \eqref{radialmeasure} coincides with the measure for the Wishart-Laguerre (WL) matrix model
\beq\label{eq:measurewishartN}
d\mu^{(0)}= dx \, x^{\frac{N^2-3}{2}}e^{-2\pi {\rm Im}\tau x}\period
\eeq 
Note that, even for $n=0$, the observable $f^{(n)}(x)$ takes a rather complicated form since it is defined by an angular integral \eqref{eq:fndefine}. 

We 
use the integral representation of $\log H(x)$ \eqref{eq:newHformula} and obtain 
\ie 
\left.\partial_m^2\mathcal{F}({\bm a})\right|_{m=0}= \int_0^\infty dw  \frac{w {\cal I}(w {\bm a} )}{\sinh^2 (w)}\,,\quad {\cal I}({\bm a})\equiv 8\sum_{i<j}\sin^2 (a_i-a_j)\,.
\label{d2mF}
\fe 
The first step is to evaluate the angular integral of ${\cal I}(wa)$, which we can write by expanding in $w$ as,
\ie 
\llangle {\cal I}(w{\bm a}) \rrangle = \sum_{s=1}^\infty x^s w^{2s} b(s,N)\,.
\fe
We will come back to the explicit form of $b(s,N)$ shortly. Let us proceed by performing the remaining rank $k$ WL matrix integral and the $w$ integral in \eqref{d2mF}. The latter is straightforward with the following identity
\ie 
\int_0^\infty dw {w^{2s+1}\over \sinh^2(w)}=2^{-2s} \Gamma(2s+2)\zeta(2s+1)\,.
\label{wintegralid}
\fe
The WL matrix integral with measure \eqref{eq:measurewishartN} is different from the $N=2$ case considered in the previous section. However, to the leading order in the large $k$ limit, the eigenvalue distribution is $N$ independent (see e.g. \cite{Grassi:2019txd} Appendix A) and given by
\eqref{MPdist}. The relation between the $y$ variable in \eqref{MPdist} and the WL eigenvalue $x$ is $x={\lambda_J y\over 8\pi^2}$ here, where a factor 2 originates from the different normalization conventions for general $N$ as explained in footnote~\ref{ft:xnorm}. 
From the integral identity
\ie 
\int_0^4dy \rho_0(y) y^s= {1\over s+1}{(2s)!\over (s!)^2}\,,
\fe
together with \eqref{wintegralid}, 
we then obtain the following formula for the integrated correlator of the maximal trace operators with charge $J$, $\cO_{J/2}^{(0)}$,
\ie 
G_{J/2}^{(0)}
=\sum_{s=1}^\infty 2^{-s}\pi^{-1-2s}{(2s+1)} \zeta(2s+1)\Gamma\left(s+{1\over 2}\right)^2 b(s,N) \lambda_J^s\,. 
\fe

We now come back to the explicit form of $b(s,N)$.
As a consequence of \eqref{angularint}, the coefficients $b(s,N)$ are related to those from the Gaussian matrix model integral,
\ie 
\la {\cal I}(w{\bm a})\ra_{\rm GMM}=\sum_{s=1}^\infty  2^{s} w^{2s} \tilde b(s,N)\,, \quad b(s,N)=2^{s}\tilde b(s,N) {\Gamma ({N^2-1\over 2})\over \Gamma ({N^2-1\over 2}+s  )}\,.
\label{IvevGMM}
\fe
The integral $\la {\cal I}(w{\bm a})\ra_{\rm GMM}$ has been studied previously (see for example 
\cite{Chester:2019pvm,Hatsuda:2022enx}) and it determines completely the integrated correlator of the stress energy tensor multiplet in the ${\rm SU}(N)$ SYM \cite{Dorigoni:2021bvj,Dorigoni:2021guq,Collier:2022emf}. In particular, the coefficient $\tilde b(s,N)$ admits a meromorphic extension to complex $s$, captures the spectral decomposition of the $SL(2,\mZ)$ invariant correlator into non-holomorphic Eisenstein series
and obey a three-term recursion relation in $N$. Here we see that via \eqref{IvevGMM}, the same quantity completely determines the integrated correlator of large charge maximal-trace operators. 

By now there are several equivalent expressions for $\la {\cal I}(w{\bm a})\ra_{\rm GMM}$ which can be derived by standard resolvent methods. Here we use the following expression which can be deduced from formulae in Appendix A \cite{Hatsuda:2022enx}, 
\ie
\cK(t)= e^{-t} L_{N-1}^{(1)}(t)^2  
+
N  \int_0^t dt' e^{-t'}\left(L_{N-1}^{(1)}(t')L_{N-1}(t')-L_{N-2}^{(1)}(t')L_N(t')\right) -N^2 \,,
\fe
where $L_n^{(\A)}(t)$ are the generalized Laguerre polynomials and 
$\la {\cal I}(w{\bm a})\ra_{\rm GMM}=
-2\cK(2w^2)$. The constant is fixed by $\cK(0)=0$.

The Mellin transform of $\cK(t)$ easily follows from the orthogonality of the Laguerre polynomials and integration by part as in \cite{Hatsuda:2022enx},\footnote{This is related to the function $c(s,N)$ in \cite{Hatsuda:2022enx} by an overall factor of ${1\over 4}s(s-1)(1-2s)$.
The function $c(s,N)$ determines the integrated correlator of the stress tensor multiplet in \cite{Binder:2019jwn} completely as shown in \cite{Dorigoni:2021bvj,Dorigoni:2021guq,Collier:2022emf}.
}
\ie 
 \int_0^\infty dt \,{t^{s-1}\over \Gamma(s)} \cK(t)
=\,{N(N-1)(1-2s) \,} {}_3F_2(2-N,1-s,s;2,3;1)\,.
\label{MellinofK}
\fe
On the other hand, the coefficients $\tilde b(s,N)$ in \eqref{IvevGMM} are picked out by the residue at origin,
\ie 
\tilde b(s,N)=-2\oint {dt\over 2\pi i }t^{-s-1} \cK(t)\,.
\fe
Nonetheless, by considering a contour deformation in \eqref{MellinofK} with the new contour $\C'$ that comes in from $\infty$ just below the positive real axis, circles the origin by a small semicircle $C'_\ep$ and goes back to the infty just above the positive real axis,
\ie 
\C'=[-i\ep+\infty,-i\ep] \cup C'_\ep \cup [i\ep,i\ep+\infty] \,,
\fe
it is easy to obtain the following relation
\ie 
\tilde b(s,N)={-2e^{-\pi i s}\over \Gamma(s+1)}\int_0^\infty dt {t^{-s-1}\over \Gamma(-s)} \cK(t)\,,
\fe
and consequently from \eqref{MellinofK},
\ie 
\tilde b(s,N) =
{-2(-1)^s (1+2s)\over \Gamma(s+1)} N(N-1){}_3F_2(2-N,1+s,-s\,;\,2,3\,;\,1)\,.
\label{tbform}
\fe
We thus arrive at the final formula by combining \eqref{IvevGMM} and \eqref{tbform},
\ie 
G_{J/2}^{(0)}
=&\,-2N(N-1)\sum_{s=1}^\infty{(-1)^s((2s+1)!)^2\over (s!)^3} \zeta (2 s+1) \left(\frac{\lambda_J}{16 \pi ^2}\right)^s 
\\&
  {\times \,}_3F_2(2-N,-s,s+1;2,3;1) {\Gamma({N^2-1\over 2})\over\Gamma({N^2-1\over 2}+s)}    \,,
  \label{finalgenNweak}
\fe
which reduces to \eqref{GJweakexp} for $N=2$ and agrees with the result in the recent paper \cite{Paul:2023rka} derived from recursion relations in both $N$ and $J$ \cite{Paul:2022piq} (up to a factor of 4 as noted in Footnote~\ref{ft:factorfour}). 
The radius of convergence in \eqref{finalgenNweak} is $N$ independent and coincide with \eqref{weakradiusconv}.

We expect the matrix model approach described here to be useful to study more general large charge integrated correlators that involve operators $\mathcal{O}_k^{(n)}$ that are not of the maximal trace type (i.e. $n\neq 0$).
It would also be interesting to analyze 
the non-planar contributions in the large charge limit,
as well as 
other scaling limits such as the large $N$-large $J$ limit\fn{In the limit $J\propto N^2$ with fixed $\tau$, $x^{\frac{N^2-3}{2}}$ in \eqref{eq:measurewishartN} becomes dominant and the integral \eqref{eq:WishartSUN} can be approximated by a matrix model with a logarithmic potential.} $J\propto N^2$ discussed in \cite{Paul:2023rka}. We leave the more extensive analysis to future work.

	\section{Conclusions and future directions\label{sec:generalization}}
	The results of this paper highlight the large charge 't Hooft limit as a new and intriguing solvable corner of $\mathcal{N}=4$ SYM that shares striking similarities with the conventional large $N_c$ limit. Through the analysis of the spectrum and correlation functions, we demonstrated the validity of this claim. However, we emphasize that our analysis only scratched the surface of this interesting limit, and there are countless exciting directions for future exploration. 
	\subsection{Direct generalization of the results in this paper}
	\paragraph{Index and partition function at large charge.}In Section~\ref{sec:partition}, we showed using the contour deformation tricks that the superconformal index and the partition function simplify in the large charge limit and can be rewritten as a sum over magnons. The analysis was done only for the ${\rm SU}(2)$ gauge group and for the large charge limit of $1/2$ BPS operators. It would be interesting to generalize it to higher-rank gauge groups and to less supersymmetric states. If a similar rewriting exists for $1/16$ BPS states, it could be relevant for the microstate counting of supersymmetric black holes, about which significant progress has been made recently \cite{Cabo-Bizet:2018ehj,Choi:2018hmj,Benini:2018ywd}.
	\paragraph{Constraints on CFT data from dilaton Ward identity.} In Section~\ref{subsubsec:CBconformal}, we conjectured the implication of the dilaton Ward identity on the CFT data. However the argument presented there was rather heuristic and it is important to make them rigorous and prove them from the conformal bootstrap. This will be the first step towards understanding necessary and sufficient conditions for CFTs to have the vacuum moduli.
	\paragraph{Integrated correlator at large charge.} In Section~\ref{subsec:MM}, we derived an emergent matrix model description for the integrated correlator both for the ${\rm SU}(2)$ gauge group and general ${\rm SU}(N)$ gauge groups. However, for the ${\rm SU}(N)$ gauge groups, we only analyzed the simplest class of large charge operators (i.e.~the maximal trace operators) and it would be interesting to perform explicit computations for other large charge operators. In addition, we expect that the matrix-model reformulation to be useful for other limits such as the combined large $N$-large $J$ limit ($J\propto N^2$) discussed in \cite{Paul:2023rka}. 
	\paragraph{Massless magnon point, double-scaling limit, and EFT.} One interesting outcome of our analyses is that various quantities in the large charge 't Hooft limit become singular at $\lambda_J=-\pi^2$. As discussed in Section~\ref{subsec:MM}, this singularity signals a tachyonic instability of the lightest magnon. It would be interesting to see if one can define a ``double-scaling limit'' near this instability, much like the double-scaling limit of matrix integrals studied in the past \cite{Brezin:1990rb,Douglas:1989ve,Gross:1989vs}. Based on the analogy with the matrix integrals, we expect that this limit is dominated by ``dense'' Feynman diagrams at large charge. It would be interesting to identify diagrams that dominate in the limit and come up with a ``continuum'' description. A related question is to understand the effective (conformal) field theory describing such a limit.
	\paragraph{Exact worldline instanton.} In \eqref{eq:massivestrong}, we derived an exact strong coupling expansion of the resummed conformal ladder integral and interpreted it as a sum over worldline instantons. It would be interesting to derive this expression from the first-quantized worldline formalism, generalizing the analysis in \cite{Dondi:2021buw}. 
	\subsection{Other important future directions}
	\paragraph{Less supersymmetric states.}In this paper, we analyzed the fluctuations around the $1/2$ BPS large charge states. One interesting future direction is to generalize the analysis to less supersymmetric states, most notably to the $1/16$ BPS states, which are dual to supersymmetric black holes in the large $N_c$ limit. Concrete questions in this direction are to perform the semiclassical analysis around the $1/16$ BPS solutions analyzed in \cite{Grant:2008sk,Yokoyama:2014qwa} and to understand the residual (centrally-extended) symmetry around the background.
	
	In addition, in a recent paper \cite{Cuomo:2022kio}, the authors analyzed operators in generic (nonsupersymmetric) CFTs that have both large spin and large charge, and identified a new phase called the ``giant vortex'' phase. The $1/16$ BPS states are interesting also in this regard since they can carry nonzero Lorentz spin unlike the $1/2$ BPS states discussed in this paper, and might provide a supersymmetric analog of the giant vortex phase.
	\paragraph{Higher-rank gauge groups.} Most of the analyses performed in this paper are for the ${\rm SU}(2)$ gauge group and a natural next step is to generalize it to higher-rank gauge groups. In higher-rank theories, there are multiple $1/2$ BPS operators with a given R-charge, and correspondingly, there are different semiclassical backgrounds depending on how one distributes the charge among the Cartan directions (see below for $SU(3)$);
	\beq
	\langle \phi \rangle_{\rm cl}\sim \left(\begin{array}{ccc} \phi^{0}_{11}&&\\&\phi^{0}_{22}&\\&&\phi^{0}_{33}\end{array}\right) \qquad \phi^{0}_{11}+\phi^{0}_{22}+\phi^{0}_{33}=0\period
	\eeq
	Even in these cases, the centrally-extended $\mathfrak{psu}(2|2)^2$ still governs the spectrum but the values of the central charges for the magnons depend on how one distributes the charges (i.e.~$\phi_{ii}^{0}-\phi_{jj}^{0}$). In particular, we expect that the charge distribution along the Cartan to become continuous in the large $N_c$ limit and the magnons can take an arbitrary continuous value of the central charge. This would be a potential way to relate the analysis in this paper to the centrally-extended symmetry in the planar limit \cite{Beisert:2005tm,Beisert:2006qh}.
	
	\paragraph{Three-point functions.} In Section~\ref{subsec:Konishi}, we computed the structure constant of the Konishi operator and two large charge operators. A possible generalization would be to extend the analysis to other non-BPS operators. Even more interesting would be to consider the structure constants of three heavy non-BPS operators which are obtained by small deformation of large charge $1/2$ BPS operators.\footnote{The paper \cite{Cuomo:2021ygt} analyzed the correlation function of three large charge operators using the semi-classics. It might be possible to perform a similar analysis for $\mathcal{N}=4$ SYM. However for the non-extremal three-point function, it seems difficult to even find a semiclassical saddle point at large charge. (The configuration discussed in \cite{Cuomo:2021ygt} corresponds to the extremal correlator in $\mathcal{N}=4$ SYM.)} In the planar limit, it was shown in \cite{Basso:2015zoa} that the (non-extremal) three-point function of $1/2$ BPS operators preserves a single copy of the centrally-extended $\mathfrak{psu}(2|2)$ and one can use it to constrain the structure constants of non-BPS operators. The same symmetry should survive also in the large charge 't Hooft limit, and it would be interesting to understand the constraints thereof.
	\paragraph{Large spin 't Hooft limit.} Yet another direction is to see if there exists an analog of the large charge 't Hooft limit for the Lorentz spin $S$. Since the anomalous dimension of large spin operators scales as $g_{\rm YM}^2 \log S$ at weak coupling, one candidate limit is to send $S$ to infinity keeping $g_{\rm YM}^2\log S$ fixed. To understand the physics of this limit, it is useful to employ the coordinate transformation used in \cite{Alday:2007mf} and map a large spin operator to a state in ${\rm AdS}_3 \times {\rm S}^1$. As discussed in Section 2.2 of \cite{Alday:2007mf}, the large spin operator corresponds to a fluxtube state in this geometry even at finite $N_c$ as long as the coupling $g_{\rm YM}$ is small,\footnote{This is because the probability of the fluxtube to break is exponentially suppressed $e^{-1/g_{\rm YM}^2}$ in the limit.} and one can study it by (Kaluza-Klein) reducing $\mathcal{N}=4$ SYM to an effective two-dimensional Yang-Mills with higher derivative corrections. It would be interesting to compute the anomalous dimensions in this limit using the perturbative dilatation operator and compare them against the effective $2d$ YM description.
	
	Relatedly, there have been interesting works \cite{Olivucci:2021pss,Olivucci:2022aza} which identified a double-scaled lightlike limit of correlation functions at large $N_c$, where the 't Hooft coupling $\lambda$ is sent to zero with the ``cusp times'' $t_i^2 =\lambda \log x_{i-1,i}^2\log x_{i,i+1}^2$ held fixed.  Physically this limit is controlled by operators with large spin and is similar to the large spin limit discussed above. To make a more concrete connection, it would be interesting to study a similar double-scaling limit (with $\lambda$ replaced with $g_{\rm YM}^2$) for the correlation functions in the ${\rm SU}(2)$ theory. 
	
	In addition, the results in \cite{Olivucci:2022aza} are given by determinants of Bessel functions, which also show up in the analysis of the large $N_c$ 2d YM. One interesting question is if the correlation functions studied in \cite{Olivucci:2022aza} can be mapped to some observables\footnote{The correlation function in the light-like limit is known to be related to a null polygonal Wilson loop \cite{Alday:2010zy}. In the planar limit, the null polygonal Wilson loop at finite $\lambda$ can be studied using the so-called Pentagon decomposition \cite{Basso:2013vsa}, in which the basic building block is the pentagon form factor which can be interpreted as a form factor of a branch-point twist operator \cite{Cardy:2007mb} on the two-dimensional worldsheet. It is therefore tempting to conjecture that the relevant observable in this limit is a branch-point twist operator in $2d$ YM.} in the effective 2d YM.
	
	\paragraph{Combining large $N$ and large $J$ and black holes.} For the application to (standard) holography, it is important to study the combined large $N$-large $J$ limits. One difficulty in this case is that the semiclassical analysis cannot be easily justified. To understand this point, let us recall the basics of the standard 't Hooft limit. The action of $\mathcal{N}=4$ SYM is proportional to $N/\lambda$, which is infinitely large in the 't Hooft limit,
	\beq
	S\sim \frac{N}{\lambda}\int d^4 x {\rm Tr}\left[\cdots\right]\period
	\eeq
	One might therefore think that the limit can be studied using the saddle point approximation. As is well-known, this conclusion is incorrect for the following reason: if we approximate the path integral by its saddle point, the saddle point action scales like
	\beq\label{eq:estimateS}
	S_{\rm saddle}\sim \frac{N}{\lambda} \left.{\rm Tr}\left[\cdots \right]\right|_{\rm saddle}\sim \frac{N^2}{\lambda}\period
	\eeq
	However, since there are $\sim N^2$ dynamical degrees of freedom, the one-loop correction to the saddle point contribution also scales like
	\beq
	S_{\text{1-loop}}\sim N^2\period
	\eeq
 Thus, unless we take $\lambda$ to be small, the saddle-point approximation cannot be justified.
	
	One potential way to overcome this difficulty is to combine it with the large $J$ limit. For instance, when the saddle point configuration scales like\footnote{The scaling with respect to $g_{\rm YM}$ and $J$ is fixed by the fact that the Noether charge is quadratic in fields and proportional to $g_{\rm YM}^2$; namely $J\sim \frac{1}{g_{\rm YM}^2}{\rm Tr}[\Phi_{\rm cl}^2]$. The exponent $\alpha$ depends on how the charge is distributed among different components of the ${\rm SU}(N)$ adjoint field. For instance, if the semiclassical configuration only has a single non-vanishing diagonal entry, we expect $\alpha=0$ while if it is distributed equally among $N$ diagonal entries, we expect $\alpha=1$.} $g_{\rm YM}\sqrt{J}/N^{\alpha}$, \eqref{eq:estimateS} will be enhanced to
	\beq
	S_{\rm saddle}\sim \frac{N^2}{\lambda}\frac{g_{\rm YM}^2J}{N^{2\alpha}}\comma
	\eeq
	while the scaling for $S_{\text{1-loop}}$ remains the same.  By taking the ratio $g_{\rm YM}\sqrt{J}/N^{\alpha}\sim \sqrt{\lambda J/N}/N^{\alpha}$ to be large, we can then suppress the 1-loop correction parametrically. Of course the estimate we gave here is rather crude and it is important to perform detailed analysis for concrete cases.
	
	It would be interesting to study double-scaling limits also on the gravity side. When the 't Hooft coupling $\lambda$ is infinite, the bulk is described by supergravity. As we decrease $\lambda$, it receives higher derivative corrections proportional to some powers of $1/\lambda$. On the other hand, the effects of these higher derivative corrections are suppressed for a black hole of a very large size. Since the size of the black hole with a large quantum number $J$ is controlled roughly by the parameter $J/N^2$, the effective strength of higher derivative corrections is parametrized by $\frac{N^2}{\lambda^{\beta}J }$ with some positive exponent $\beta$. This offers the possibility to study the limit $\frac{N^2}{\lambda^{\beta}J }\ll 1$ both at weak and strong couplings. At strong coupling, it is described by  supergravity as discussed above, while we can try to reach this limit from the weak coupling by first taking the double-scaling limit in which $\lambda\to 0$ with $\tilde{\lambda}\equiv \lambda^{\beta} J/N^2$ fixed and then later taking the limit $\tilde{\lambda}\to \infty$, much like what we did in this paper for the strong coupling limit of $\lambda_J$. Note that there is no guarantee that different ways of taking the limits give the same answer and more detailed analysis is certainly necessary. 
	
	\paragraph{$\mathcal{N}=2$ SCFT.}  Much of our analysis can be generalized to study large charge operators and their correlation functions in the double-scaling limit for general $\cN=2$ SCFTs with a conformal manifold such as the superconformal QCD (SQCD) described by $\cN=2$ ${\rm SU}(N)$ SYM coupled to $2N$ fundamental hypermultiplets. Previous works on large charge limits of these $\cN=2$ theories have focused on the two-point function of BPS operators \cite{Grassi:2019txd,Hellerman:2021duh,Hellerman:2021yqz} which contain information about the extremal correlator sector of the SCFT \cite{Gerchkovitz:2016gxx} and are nontrivial due to the reduced supersymmetry. In particular, at strong coupling $\lambda_J\gg 1$, these two-point functions exhibit an infinite tower of worldline instanton contributions similar to what we have seen here for more general observables in the $\cN=4$ SYM. 
 	 Nonetheless, it would be much more interesting to extend the $\cN=2$ large charge analysis to non-BPS operators and also higher-point functions by
	exploiting our methods here. 
	
	Indeed, one copy of the maximally-centrally-extended $\mf{psu}(2|2)$ symmetry that played an important role in our ``bootstrap'' analysis of the non-BPS large charge operator spectrum survives in general $\cN=2$ theories. In this case, the superconformal R-symmetry is ${\mf u}(1)_R\times \mf{su}(2)_R$ and the relevant large charge operators are Coulomb branch chiral primary operators $\cO_J$ that carry charge $J>0$ under the ${\mf u}(1)_R$ and are singlet under $\mf{su}(2)_R$. The BPS condition reads $\Delta=J$ and thus as before, the operator inserted at origin is preserved by the superconformal generator $C\equiv {\hat D-\hat J\over 2}$.  Together with the rotation symmetry $\mf{so}(4)$ and the $\mf{su}(2)_R$ symmetry, the full superconformal subalgebra preserved by the operator insertion $\cO_J(0)$ is
	\ie 
	\mf{su}(2) \times \mf{psu}(2|2) \ltimes \mR\,,
	\fe
	compared to \eqref{centralsym} for the $\cN=4$ case (see around \eqref{eq:susygens}). In the large charge 't Hooft limit, we expect this symmetry to be further centrally extended, which includes a maximally-centrally-extended $\mf{psu}(2|2) \ltimes \mR^3$ factor, that would govern the physics of excitations on the 
	large charge state created by $\cO_J(0)$. It would be interesting to carry out this analysis explicitly, for example in the case of the $\cN=2$ ${\rm SU}(2)$ SQCD, to learn about the non-BPS large charge operators. 
	
	Another potentially intriguing direction is to develop a better understanding of the large charge EFT in the large charge 't Hooft limit  which shares many similarities with the conventional Coulomb branch EFT, as we have commented on in the main text, in general $\cN=2$ theories. 
	It is well-known that the latter EFT has an elegant low energy description in terms of  the Seiberg-Witten curve and the complex geometric data therein, largely thanks to the $\cN=2$ Poincar\'e supersymmetry \cite{Seiberg:1994rs,Seiberg:1994aj}. It is thus tantalizing to speculate a version of the Seiberg-Witten curve for the large charge EFT that follows from the centrally extended $\mf{psu}(2|2)$ symmetry.\footnote{See Section~\ref{subsubsec:Poincare} for comments on the relation between the centrally extended $\mf{psu}(2|2)$ symmetry preserved in the large charge EFT and the Poincar\'e supersymmetry that govern the Coulomb branch EFT.} In particular, it would be interesting to extract the correlation functions of large charge Coulomb branch operators from such an EFT. We leave this investigation to future work.

	\subsection*{Acknowledgement} We thank Gabriel Cuomo, Nicola Dondi, Simeon Hellerman, Zohar Komargodski, Andrew McLeod, Domenico Orlando, Eric Perlmutter, Himanshu Raj, Susanne Reffert, Raffaele Savelli, Kostya Zarembo, Sasha Zhiboedov and Masataka Watanabe for helpful discussions. We in particular thank Eric Perlmutter for communicating to us the results of \cite{Paul:2023rka} prior to the publication. SK thanks the organizers and participants of the workshop ``Large Charge in Les Diablerets'' in SwissMAP Research Station for an opportunity to present this work and have stimulating discussions.
	 The work of YW was
	supported in part by NSF grant PHY-2210420 and by the Simons Junior Faculty Fellows program.
	
	\appendix
	\section{Action \& conventions} \label{eftconv}
	In this appendix, we establish our conventions for the flat space action, which will be used for perturbative computations. We begin with the $\mathcal{N}=4$ SYM action which we write as 
\beq
\mathcal{S}_{\mathcal{N}=4 {\,\rm{SYM}}} =\frac{2}{g^{2}_{\rm{YM}}}\int d^4x  \Tr\left( -\frac{1}{4}F_{\mu \nu} F^{\mu \nu}-\frac{1}{2}D_{\mu}\Phi_{I}D^{\mu}\Phi_{I}  +\frac{1}{4}[\Phi_{I},\Phi_{J}][\Phi_{I},\Phi_{J}] + {\rm{fermions}}\right)\,,
\eeq
where we omit the fermionic part of the action since we will not use it explicitly.

We express the six real scalar fields in terms of complex scalar fields in the usual manner, 
\begin{align}\label{Phi2Z}
\Phi_1 &= \frac{Z+\bar{Z}}{\sqrt{2}} , \quad \Phi_2 = \frac{Z-\bar{Z}}{i\sqrt{2}} , \quad \Phi_3 = \frac{X+\bar{X}}{\sqrt{2}} ,\quad
\Phi_4 = \frac{X-\bar{X}}{i\sqrt{2}} , \quad \Phi_5 = \frac{Y+\bar{Y}}{\sqrt{2}} , \quad \Phi_6 = \frac{Y-\bar{Y}}{i\sqrt{2}} .
\end{align}
Note that each of the complex scalar fields takes values on the Lie algebra of the gauge group SU(2), with the generators of the Lie algebra taken to be Hermitian in our conventions. This allows us to represent the fields explicitly as follows
\beq
\begin{alignedat}{2}
 & Z = \begin{pmatrix}
    z^0 & z^{+} \\
    z^{-} & -z^0
\end{pmatrix}, \quad &&\bar{Z} = \begin{pmatrix}
    \bar{z}^0 & \bar{z}^{-} \\
    \bar{z}^{+} & -\bar{z}^0
\end{pmatrix},\\
&X = \begin{pmatrix}
   x^0 & x^{+} \\
    x^{-} & -x^0
\end{pmatrix}, \quad &&\bar{X} = \begin{pmatrix}
    \bar{x}^0 & \bar{x}^{-} \\
    \bar{x}^{+} & -\bar{x}^0
\end{pmatrix}, \\
&Y = \begin{pmatrix}
   y^0 & y^{+} \\
    y^{-} & -y^0
\end{pmatrix}, \quad &&\bar{Y} = \begin{pmatrix}
    \bar{y}^0 & \bar{y}^{-} \\
    \bar{y}^{+} & -\bar{y}^0
\end{pmatrix}\,.
 \end{alignedat}
 \eeq
Similarly, the gluons, fermions, and ghost fields can be decomposed in the same way
 \beq
 A_{\mu} = \begin{pmatrix}
    a_{_\mu}^0 & a_{_\mu}^{+} \\
    a_{_\mu}^{-} & -a_{_\mu}^0
\end{pmatrix}, \quad
 \psi= \begin{pmatrix}
    \psi^0 & \psi^{+} \\
    \psi^{-} & -\psi^0
\end{pmatrix}, \quad  
 \bar\psi= \begin{pmatrix}
   \bar{\psi}^0 & \bar{\psi}^{-} \\
   \bar{\psi}^{+} & -\bar{\psi}^0
\end{pmatrix},\eeq 
\beq
 c= \begin{pmatrix}
    c^0 & c^{+} \\
    c^{-} & -c^0
\end{pmatrix}, \quad  
 \bar{c}= \begin{pmatrix}
   \bar{c}^0 & \bar{c}^{-} \\
   \bar{c}^{+} & -\bar{c}^0
\end{pmatrix}\,.
 \eeq
Among these fields, $Z$ and $\bar{Z}$ acquire a classical profile in the large charge background
\beq
Z_{\rm{cl}}= \begin{pmatrix}
   z_{{\rm{cl}}}^0 & 0\\
   0 & -z_{{\rm{cl}}}^0
\end{pmatrix},\quad \bar{Z}_{\rm{cl}}= \begin{pmatrix}
 \bar{z}_{{\rm{cl}}}^0 & 0\\
   0 & -\bar{z}_{{\rm{cl}}}^0
\end{pmatrix}\,,
\eeq
around which we expand the action. In order to perform perturbative computations, we first gauge fix the action by adding to it the gauge fixing term
\beq
\mathcal{S}_{\rm gf}=\frac{2}{g^2_{\rm{YM}}}\int d^{4}x\,\left(-\frac{1}{2} \Tr \left(\partial_{\mu} A^{\mu}+i [Z,\bar{Z}_{\rm cl}]+i [\bar{Z},Z_{\rm cl}]\right)^2 \right)
\eeq
and the corresponding ghost term
\beq
\mathcal{S}_{\rm gh}= \frac{2}{g^2_{\rm{YM}}}\int d^{4}x\,\Tr \left(\bar{c}\left(\partial^{\mu} D_{\mu} c - [[c,Z],Z_{\rm cl}]  -[[c,\bar{Z}],Z_{\rm cl}]  -[[c,\bar{Z}_{\rm cl}],Z_{\rm cl}] -[[c,Z_{\rm cl}],\bar{Z}_{\rm cl}]    \right)\right)\,,
\eeq
where $c, \bar{c}$ are the ghost fields.
The resulting effective action can be split as follows
\beq
\mathcal{S}_{\mathcal{N}=4 {\,\rm{SYM}}}+\mathcal{S}_{\rm gf} +\mathcal{S}_{\rm gh} = \mathcal{S}_{\rm{kin}} + \mathcal{S}_{\rm{m,b}}+ \mathcal{S}_{\rm{cubic}}+ \mathcal{S}_{\rm{quartic}}+{\rm{fermions}}\,,
\eeq
where from now on we will omit completely the discussion about the fermionic part.
The kinetic terms piece $\mathcal{S}_{\rm{kin}}$ is given by
\begin{equation}
\begin{aligned}
\mathcal{S}_{\rm{kin}} = \frac{2}{g_{\rm{YM}}^2} \int d^4x \Bigl[a_\mu^{0} \Box (a^{0})^{\mu}+ a_\mu^{+} \Box (a^{-})^{\mu} + 2 \sum_{\Phi\in\{x,y,z,c\}}\bar{\Phi}^{0} \Box \Phi^{0}+ \sum_{\substack{\Phi\in\{x,y,z,c\}\\p=\pm}} \bar{\Phi}^{p} \Box \Phi^{p}\Bigr].
\end{aligned}
\end{equation}
where 
  $\Box=\partial_{\mu} \partial^{\mu}$.
The bosonic mass part of the action $\mathcal{S}_{\rm{m,b}}$ reads
\begin{equation}\label{quadratic}
\mathcal{S}_{\rm{m,b}} = -\frac{16}{g^2_{\rm{YM}}}\int d^{4}x \Bigl[ |z^0_{\rm{cl}}|^2 \Bigl(a^+_{\mu }(a^-)^{\mu }+\sum_{\substack{\Phi \in\{x,y,z\}\\p=\pm}}|\Phi^p|^2\bigr) + \frac{i}{2}\,\partial^{\mu}\bar{z}^0_{\rm{cl}}\bigl(z^+ a^-_{\mu}-z^- a^+_{\mu}\bigr)+\frac{i}{2}\,\partial^{\mu}{z}^0_{\rm{cl}}\bigl(\bar{z}^- a^-_{\mu}-\bar{z}^+ a^+_{\mu}\bigr)\Bigr].
\end{equation}
We note that the classical profile of $z^0_{\rm{cl}}$ and $\bar{z}^0_{\rm{cl}}$ induces a mixing term between the gauge fields and the scalars $z^{\pm},\bar{z}^{\pm}$. 

We now proceed with the cubic part of the action, which due to its length can be sub-split into several parts according to the field content
\beq
\mathcal{S}_{\rm{cubic}} =\mathcal{S}_{\rm{cubic,g}} +\mathcal{S}_{\rm{cubic,s}} +\mathcal{S}_{\rm{cubic,g,s}}
+\mathcal{S}_{\rm{cubic,s,gh}}\,.
\eeq
Each one of these parts read as
\beq
\mathcal{S}_{\mathrm{cubic,g}} = \frac{2}{g_{\mathrm{YM}}^2} \int d^4x \,2i \Big[(a_{\mu}^- a_{\nu}^0 - a_{\nu}^- a_{\mu}^0) \partial^{\nu}(a^{\mu})^+ + (a_{\nu}^+ a_{\mu}^0 - a_{\mu}^+ a_{\nu}^0) \partial^{\nu}(a^{\mu})^- + (a_{\mu}^+ a_{\nu}^- - a_{\mu}^- a_{\nu}^+) \partial^{\nu}(a^0)^{\mu} \Big]\,,
\eeq
\begin{equation}
\begin{aligned}
\mathcal{S}_{\rm{cubic,s}} = \frac{2}{g^2_{\rm{YM}}} \int d^{4}x \,& 4 \Biggl[
\sum_{\Phi \in \{x,y\}} \Bigl(\bar{\Phi}^0\bar{z}^0_{\rm{cl}}(\Phi^- z^++\Phi^+ z^-)  +\bar{\Phi}^0z^0_{\rm{cl}} (\Phi^- \bar{z}^-+\Phi^+ \bar{z}^+) -2\, \bar{z}^0 z_{\rm{cl}}^0(|\Phi^{+}|^2+|\Phi^{-}|^2)\Bigr)   \\
&-\, \bigl(|z^-|^2+|z^+|^2\bigr) \bar{z}^0 z_{\rm{cl}}^0+ 2\, z^- z^+ \bar{z}^0 \bar{z}^0_{\rm{cl}} +h.c. \Biggr]\,,
\end{aligned}
\end{equation}
where  $h.c.$ means adding the conjugate of all terms of the Lagrangian where unbarred fields become barred and vice-versa. The remaining cubic terms read

\begin{equation}
\begin{aligned}
\mathcal{S}_{\text{cubic,g,s}} = \frac{4}{g_{\text{YM}}^2} \int d^4x &\Biggl[ -2 (a^-)^\mu a_\mu^0 (z^+ \bar{z}_{\text{cl}}^0 + \bar{z}^- z_{\text{cl}}^0) - 2 (a^+)^\mu a_\mu^0 (z^- \bar{z}_{\text{cl}}^0 + \bar{z}^+ z_{\text{cl}}^0) \\
&- 4 (a^-)^\mu a_\mu^+ (\bar{z}^0 z_{\text{cl}}^0 + z^0 \bar{z}_{\text{cl}}^0) + \sum_{\Phi \in \{x,y,z\}} \biggl( i\partial^\mu\Phi^- (\bar{\Phi}^- a_\mu^0 - \bar{\Phi}^0 a_\mu^+) \\
&\quad+ i\partial^\mu\Phi^+ (\bar{\Phi}^0 a_\mu^- - \bar{\Phi}^+ a_\mu^0) + i\partial^\mu\Phi^0 (\bar{\Phi}^+ a_\mu^+ - \bar{\Phi}^- a_\mu^-) + h.c. \biggr) \Biggr]\,,\\
\mathcal{S}_{\rm{cubic,s,gh}}  = \frac{2}{g^2_{\rm{YM}}}\int d^{4}x & \,  \Biggl[ \bar{z}^0 z^0_{\rm{cl}}\left(2  \bar{c}^-c^-+2  \bar{c}^+c^+\right)+z^0 \bar{z}^0_{\rm{cl}} \left(2  \bar{c}^-c^-+2  \bar{c}^+c^+\right) \\
&- \bar{z}^0_{\rm{cl}} \left(2 z^- \bar{c}^-+2 z^+ \bar{c}^+\right)c^0 - z^0_{\rm{cl}} \left(2 \bar{c}^+ \bar{z}^-+2 \bar{c}^- \bar{z}^+\right)c^0\Biggr]\,.
\end{aligned}
\eeq
Finally, we display explicitly the quartic part of the action which we write as
\beq
\mathcal{S}_{\rm{quartic}} = \mathcal{S}_{\rm{quartic,g}}+ \mathcal{S}_{\rm{quartic,g,s}}+ \mathcal{S}_{\rm{quartic,s}}
\eeq
where
\begin{equation}
\begin{aligned} 
\mathcal{S}_{\rm{quartic,g}}= \frac{1}{g^2_{\rm{YM}}}\int d^{4}x \,& \Biggl[4 (a^0)^{\mu } (a^0)^{\nu } \left(a^+_{\mu } a^-_{\nu }+a^-_{\mu } a^+_{\nu }\right)-2 \,a^-_{\nu } (a^+)^{\nu } \left(a^-_{\mu } (a^+)^{\mu }+2 a^0_{\mu } (a^0)^{\mu }\right) \\
&-4 \,a^-_{\mu } (a^+)^{\mu } a^0_{\nu }(a^0)^{\nu }+a^+_{\mu }\left(a^+\right)^{\mu }  \left(a^-\right)^{\nu }a^-_{\nu }+a^-_{\mu }(a^-)^{\mu } a^+_{\nu } (a^+)^{\nu} \Biggr]\,,
\end{aligned}
\end{equation}

\begin{equation}
\begin{aligned} \label{quartic}
 \mathcal{S}_{\rm{quartic,g,s}}= \frac{2}{g^2_{\rm{YM}}}\sum_{\Phi \in \{x,y,z\}}\int d^{4}x\,& \Biggl[ \,4\, a^-_{\mu } (a^0)^{\mu }\left(\Phi^+ \bar{\Phi}^0+ \bar{\Phi}^-\Phi^0\right)+2\, a^-_{\mu } \left(a^-\right)^{\mu }\left(\Phi^+ \bar{\Phi}^-\right)\\
 &+4\, a^+_{\mu }(a^0)^{\mu } \left(\Phi^- \bar{\Phi}^0+\Phi^0 \bar{\Phi}^+\right)+2\, a^+_{\mu }\left(a^+\right)^{\mu }  \left(\Phi^- \bar{\Phi}^+\right)\\
&-2\, (a^-_{\mu } (a^+)^{\mu }+2 a^0_{\mu }(a^0)^{\mu }) \left(\Phi^- \bar{\Phi}^-+\Phi^+ \bar{\Phi}^+\right)
-8 \, a^-_{\mu } (a^+)^{\mu } \left(\Phi^0 \bar{\Phi}^0\right)\Biggr]\,,
\\
 \mathcal{S}_{\rm{quartic,s}}= \frac{2}{g^2_{\rm{YM}}}\int d^{4}x\,&\sum_{\substack{\Phi \in \{x,y,z\} \\ \Psi \in \{x,y,z\}}} \Biggl[ -4  \bar{\Psi }^- \bar{\Phi }^+ \Psi ^0 \Phi ^0+4 \Psi ^- \bar{\Phi }^- \Phi ^0 \bar{\Psi }^0-8 \Psi ^- \bar{\Psi }^- \Phi ^0 \bar{\Phi }^0\\
 &+4 \Psi ^+ \bar{\Phi }^+ \Phi ^0 \bar{\Psi }^0-8 \Psi ^+ \bar{\Psi }^+ \Phi ^0 \bar{\Phi }^0-\Psi ^- \Phi ^- \bar{\Psi }^- \bar{\Phi }^--\Psi ^+ \Phi ^+ \bar{\Psi }^+ \bar{\Phi }^+\\
 &-2 \Psi ^+ \Phi ^- \bar{\Psi }^+ \bar{\Phi }^-+4 \Psi ^- \Phi ^+ \bar{\Psi }^0 \bar{\Phi }^0+4 \Psi ^+ \Phi ^- \bar{\Psi }^- \bar{\Phi }^+ \Biggr]\,.
\end{aligned}
\eeq

\section{Large charge background and scalar propagators} \label{backlc}
In this section, we aim to find the propagators for the scalar fields in the effective action. We consider a two point function of local operators with a very large charge, which we denote as
\beq
O_{J}(x) = (z^{0}(x))^{J}\,,\quad \bar{O}_{J}(x)=(\bar{z}^{0}(x))^{J}\,.
\eeq
In the limit when $J \rightarrow \infty$, the insertion of such operators in the path integral source a non-trivial profile for $z^0$ and $\bar{z}^0$ which can be obtained as a saddle point of the two-point function given by 
\beq \label{limit}
\langle \bar{O}_{J}(x_1)   O_{J}(x_2) \rangle  \simeq e^{S_{\rm{eff}}^{\ast}} \quad  {\rm{as}} \quad J\rightarrow \infty\; ,\; g_{\rm{YM}}\rightarrow 0 \quad {\rm{with}} \quad  \lambda_{J} = g^2_{\rm{YM}} J/2 \; {\rm{fixed}}\,,
\eeq
where  the corresponding effective action is given by 
\beq
S_{\rm{eff}}^{\ast} = S_{\rm{eff}}\Big\rvert_{z^0,\,\bar{z}^0\rightarrow z^0_{\rm{cl}},\,\bar{z}^0_{\rm{cl}}} \;\; {\rm with} \;\; S_{\rm{eff}} =\frac{2 J}{\lambda_{J}} \int d^{4}x  \, \left[  \bar{z}^{0} \Box z^{0} + \frac{\lambda_{J}}{2} \left( \log (\bar{z}^{0} )\,\delta(x-x_1)+\log (z^{0} )\,\delta(x-x_2) \right) \right]\,.
\eeq
In this expression, $ z^{0}_{\rm{cl}},\, \bar{z}^{0}_{\rm{cl}}$ are the solutions of the saddle point equations that follow from the action $S_{\rm{eff}}$. 
The explicit solutions for $z^{0}_{\rm{cl}}(x)$ and $\bar{z}^{0}_{\rm{cl}}(x)$ are given by
\beq \label{saddlesol}
z^{0}_{\rm{cl}}(x) = \frac{e^{i\phi} \, |x_1-x_2|}{2\pi |x-x_1|^2}\,\sqrt{\frac{\lambda_{J}}{2}}\,,\quad \bar{z}^{0}_{\rm{cl}}(x) = \frac{e^{-i\phi} \, |x_1-x_2|}{2\pi |x-x_2|^2}\,\sqrt{\frac{\lambda_{J}}{2}}\,.
\eeq 
From (\ref{quadratic}), we find that the scalars $z^{\pm}, x^{\pm}$ and $y^{\pm}$  have masses given by
\beq
m^{2}(x) = 8 |z_{\rm{cl}}^0|^2 =\frac{ |x_1-x_2|^2}{ \pi^2 |x-x_1|^2|x-x_2|^2} \,\lambda_{J}\,.
\eeq
The quadratic part of the action for each of these scalars, which we generically denote by $\Phi$, is then given by
\beq
\mathcal{S}_{\rm{quad},\Phi} = \frac{2}{g^2_{\rm{YM}}} \int d^{4}x \left( \bar{\Phi} \Box \Phi- m^{2}(x) |\Phi|^2 \right)\,.
\eeq
The scalar propagator is obtained by solving the equation
\beq
 \quad (-\Box_x + m^{2}(x) )\,G(x,y)  =   \,\delta^{(4)}(x-y) \qquad {\rm{with}}\quad  \, G(x,y) := \frac{2}{g^{2}_{\rm{YM}}}\langle \bar{\Phi}(x) \Phi(y) \rangle  \,.
\eeq
The solution to this equation has already been worked out in  \cite{Giombi:2020enj}. One considers an expansion in powers the 't Hooft coupling $\lambda_{J}$
\beq \label{twoptresum}
G(x,y)=\sum_{k=0}^{\infty} g_{k}(x,y)\,, \quad g_{k}(x,y) \sim \mathcal{O}(\lambda_{J}^{k})\,.
\eeq
The leading order term satisfies the massless propagator equation
\beq
-\Box_{x} g_{0} = \delta^{(4)}(x-y)\quad \Rightarrow \quad g_{0} = \frac{1}{4\pi^2} \frac{1}{|x-y|^2}\,,
\eeq
whereas the remaining terms obey
\beq
-\Box_x g_{k+1}+m^{2}g_{k} = 0\,,
\eeq
where we have omitted the arguments for simplicity. It is simple to obtain iteratively the perturbative solution at any order $k$
\beq
g_{k}(x,y) = (-1)^{k} \left(\prod_{n=1}^{k} \int d^4 z_n\, m^{2}(z_n) g_{0}(z_n,z_{n+1}) \right) g_{0}(x,z_1)\,,
\eeq
where we defined $z_{k+1}\equiv y$. More explicitly, this is equivalent to
\beq
g_{k}(x,y) =  \left( -4 g^2 \right)^k   \frac{(x_{1}-x_{2})^2}{4\pi^2(x-x_{1})^2(y-x_{2})^2}\, F^{(k)} (z,\bar{z})
\eeq 
where $F^{(k)} (z,\bar{z})$ stems from a ladder type integral and can be pictorially represented by:
  \begin{align*}
       & \eqnDiag{\includegraphics[scale=1.2]{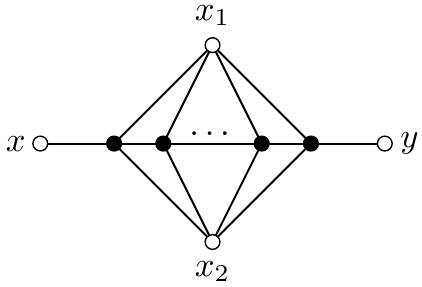}} = \frac{\pi^{2k}}{ (y-x)^2 (x_1-x_2)^{2k}}(1-z)(1-\bar{z})F^{(k)}(z,\bar{z})
  \end{align*}
Here the integration points are represented by the black dots while the white dots are unintegrated external points. The conformal cross-ratios $z,\bar{z}$ are given by
\beq
z\bar{z}\equiv {(y-x_{1})^2 (x-x_2)^2\over (x-x_{1})^2 (y-x_{2})^2}\,,\quad (1-z)(1-\bar{z})\equiv {(x_1-x_{2})^2 (x-y)^2\over (x-x_1)^2(y-x_2)^2}\,.
\eeq 

With the explicit form of the ladder integrals \cite{Usyukina:1993ch,Isaev:2003tk}, it was found in \cite{Broadhurst:2010ds} the resumed expression giving the two-point function (\ref{twoptresum}) 
\beq \label{massprop}
\langle \bar{\Phi}(x) \Phi(y) \rangle  = \frac{g^{2}_{\rm{YM}}}{4\pi^2|x-y|^2}  \int_{0}^{\infty} dt \frac{\sqrt{\frac{z\bar{z}}{(1-z)^2(1-\bar{z})^2}} \sinh(t-\frac{1}{2}\log{z\bar{z}})\, J_{0}\left( \sqrt{\frac{\lambda_{J}}{\pi^2}\, t\,(t-\log z\bar{z})}\right)}{\left(1-\frac{1+z\bar{z}}{(1-z)(1-\bar{z})}+2\sqrt{\frac{z\bar{z}}{(1-z)^2(1-\bar{z})^2}}\cosh\left(t-\frac{1}{2}\log{z\bar{z}}\right)\right)^2}
\eeq
where $J_{k}$ denotes the Bessel function.
The ladder integrals can be represented in an alternative form that is more useful for analyzing the strong coupling behavior of correlation functions \cite{Broadhurst:2010ds,Caron-Huot:2018dsv,Fleury:2016ykk}. It is given by
\beq \label{integlike}
\langle \bar{\Phi}(x)\Phi(y) \rangle =  \frac{g^{2}_{\rm{YM}}}{8\pi^2(x-y)^2}\sum_{n=0}^{\infty} \sum_{a=1}^{\infty}(-4g^2)^n \frac{(1-z)(1-\bar{z})}{z-\bar{z}}\int_{-\infty}^{\infty} \frac{du}{2\pi}\, \frac{2a \sin(a \varphi)e^{-2i u \sigma}}{\left(u^2+\frac{a^{2}}{4}\right)^{n+1}}
\eeq
where $e^{i\varphi} = \sqrt{z/\bar{z}}$ and $e^{-\sigma}=\sqrt{z\bar{z}}$. Since this type of integrals are common in integrability approaches to correlation functions \cite{Fleury:2016ykk}, we will refer to it as the \textit{integrability-like} representation of the ladder integral.

Of particular importance are the following  limits of this formula. When $ \delta^2:=|x-  x_1|^2  \rightarrow 0$ we obtain  \cite{Giombi:2020enj}
\beq
\langle \bar{\Phi}(x_1) \Phi(y) \rangle= \frac{4g^{2}}{J}\frac{1}{|x_1-y|^2\sqrt{1+16g^2}}\left( \frac{\delta |x_2-y|} {|x_1-x_2||x_1-y|}\right)^{\sqrt{1+16g^2}-1}
\eeq
and when $  |y-  x_2|^2  \rightarrow 0$ 
\beq
\langle \bar{\Phi}(x) \Phi(x_2) \rangle= \frac{4g^{2}}{J}\frac{1}{|x_2-x|^2\sqrt{1+16g^2}}\left( \frac{\delta |x_1-x|} {|x_1-x_2||x_2-x|}\right)^{\sqrt{1+16g^2}-1}\,.
\eeq

\section{Perturbative computation} \label{perturbapp}

In this appendix, we set up the perturbative computation of one-loop diagrams from the effective action derived in the Appendix~\ref{eftconv} around the background field determined in Appendix~\ref{backlc}. We will be computing the diagrams listed below:
  \begin{align*}
       & \scalebox{0.88}{$G_1:=$}\eqnDiag{\includegraphics[scale=0.9]{4scalarvertex}} \quad\quad \scalebox{0.88}{$G_2:=$} \eqnDiag{\includegraphics[scale=0.9]{4scalarvertex2}} \quad\quad  \scalebox{0.88}{$G^{\pm}_3:=$} \eqnDiag{\includegraphics[scale=0.9]{gluonexchange}}\,.
  \end{align*}
  The double line corresponds to the massive scalar propagator and the coiled line represents the propagator of the massless mode of the gluon field.  

\paragraph{Diagram $G_1$.} After factoring out the tree level contribution and using the part of the action denoted by $\mathcal{S}_{\rm{quartic,s}}$ in (\ref{quartic}), we obtain the following one-loop result:
\begin{equation}
G_{1}=-\frac{32 g^2}{J} \frac{1}{1+16g^2}\,\pi^2\,(2\pi)^4\,|x_1-x_2|^4 X_{1122}\,,
\end{equation}
which is independent of the sign of $y$. We consider the simultaneous point-splitting $x_{3}^{\mu} = x_{1}^{\mu}+\epsilon^{\mu}$ and $x_{4}^{\mu} = x_{2}^{\mu} +\epsilon^{\mu}$ with $\epsilon^{\mu} \rightarrow 0$ of the integral $X_{1234}$ given by
\begin{equation}
X_{1234} = \int \frac{d^4z}{(2\pi)^8} \frac{1}{|x_1-z|^2|x_2-z|^2|x_3-z|^2|x_4-z|^2},.
\end{equation}
The result is given by \cite{Caetano:2014gwa}
\begin{equation}
X_{1122} \equiv \lim_{\substack{x_3 \rightarrow x_1 \\ x_4\rightarrow x_2}} X_{1234} =\frac{1}{8\pi^2} \left( 1- \log\left(\frac{\epsilon^2}{|x_1-x_2|^2} \right)\right) \times \frac{1}{(2\pi)^4 |x_1-x_2|^4}.
\end{equation}
We conclude that the logarithmic divergence of such a diagram is
\begin{equation}
G_1\big\rvert_{\log} = \frac{1}{J} \frac{4 g^2}{1+16 g^2} \simeq \frac{1}{J}\left( 4 g^2 - 64 g^4+\dots \right).
\end{equation}

\paragraph{Diagram $G_2$.} The diagram $G_2$ is similar to $G_1$ except with a different combinatorial factor. We get
\begin{equation}
G_{2}=\frac{64 g^2}{J} \frac{1}{1+16g^2}\,\pi^2\,(2\pi)^4\,|x_1-x_2|^4 X_{1122}\,.
\end{equation}
This gives the following logarithmic divergence:
\begin{equation}
G_2\big\rvert_{\log} = -\frac{1}{J} \frac{8 g^2}{1+16 g^2} \simeq \frac{1}{J}\left( - 8 g^2 + 128 g^4+\dots \right).
\end{equation}

\paragraph{Diagram $G_3^{\pm}$.}The gluon exchange diagram is derived from the part of the action denoted by $ \mathcal{S}_{\rm{quartic,g,s}}$ in (\ref{quartic}) and  produces the following result
\beq
G_3^{\pm}=\mp \frac{64 g^2}{J}\frac{1}{1+16g^2} (2\pi)^{6}|x_1-x_2|^4\lim_{\substack{x_3\rightarrow x_1 \\ x_4 \rightarrow x_2}} \partial_{2} \cdot \partial_{4} H_{12,34}
\eeq
with the two-fold integral $H_{12,34}$ given by
\beq
H_{12,34} = \int \frac{d^4z d^{4}w}{(2\pi)^{10}} \frac{1}{|x_1-z|^2|x_2-z|^2|w-z|^2|x_3-w|^2|x_4-w|^2}\,.
\eeq
We use the result for its derivatives  that can be found for example in \cite{Caetano:2014gwa} and reads
\beq
\lim_{\substack{x_3\rightarrow x_1 \\ x_4 \rightarrow x_2}} \partial_{2} \cdot \partial_{4} H_{12,34} =\frac{1}{16\pi^2} \left(2- \log\left(\frac{ \epsilon^2}{|x_1-x_2|^2}\right)\right)\times \frac{1}{ (2\pi)^4|x_1-x_2|^4}\,.
\eeq
When extracting the logarithmic divergence we now obtain
\beq
G_3^{\pm}\big\rvert_{\log}= \pm \frac{1}{J}\frac{16 g^2}{1+16g^2} \simeq \pm\frac{1}{J} \left(16 g^2-256 g^4+\dots \right)\,.
\eeq
This concludes the perturbative one-loop computation.

\pdfbookmark[1]{\refname}{draftLC}
\bibliographystyle{JHEP}
\bibliography{draftLC}

	\end{document}